\documentclass[useAMS,usenatbib,psfig]{mn2e}
\usepackage{graphicx}

\def\gtrsim{\mathrel{\hbox{\rlap{\hbox{\lower4pt\hbox{$\sim$}}}\hbox{$>$}}}}

\newcommand{\kms}{{\rm km\,s^{-1}}}
\newcommand{\dm}{{\Delta}m_{12}}
\newcommand{\fsub}{f_{\rm sub}}
\newcommand{\kpc}{\rm kpc}
\newcommand{\Mpc}{\rm Mpc}

\newcommand{\keV}{\rm keV}

\newcommand{\ergs}{{\rm erg\,s^{-1}}}

\newcommand{\mic}{{\rm \mu m}}

\newcommand{\secs}{{\rm sec}}
\newcommand{\dex}{{\rm dex}}
\def\Msol{\mathrel{\rm M_{\odot}}}
\def\ls{\mathrel{\hbox{\rlap{\hbox{\lower4pt\hbox{$\sim$}}}\hbox{$<$}}}}
\def\gs{\mathrel{\hbox{\rlap{\hbox{\lower4pt\hbox{$\sim$}}}\hbox{$>$}}}}

\begin{document}

\title[LoCuSS: Strong-lensing Clusters]{LoCuSS: First Results from
  Strong-lensing Analysis of 20 Massive Galaxy Clusters at z=0.2}

\author[Richard et al.]{
  Johan Richard,$\!^{1\star}$ 
  Graham P.\ Smith,$\!^{2}$
  Jean-Paul Kneib,$\!^{3}$
  Richard S.\ Ellis,$\!^{4}$\newauthor
  A.\ J.\ R.\ Sanderson,$\!^{2}$
  L.\ Pei,$\!^{4}$
  T.\ A.\ Targett,$\!^{5}$
  D.\ J.\ Sand,$\!^{6,7}$
  A.\ M.\ Swinbank,$\!^{1}$\newauthor
  H.\ Dannerbauer,$\!^{8}$
  P.\ Mazzotta,$\!^{9}$
  M.\ Limousin,$\!^{3,10}$
  E.\ Egami,$\!^{11}$
  E.\ Jullo,$\!^{12}$\newauthor
  V.\ Hamilton-Morris,$\!^{2}$
  S.\ M.\ Moran,$\!^{13}$ 
  \vspace{1mm}\\
  $^{1}$ Institute for Computational Cosmology, Department of Physics,
  Durham University, South Road, Durham, DH1 3LE, England \\
  $^{2}$ School of Physics and Astronomy, University of Birmingham,
  Edgbaston, Birmingham, B15 2TT, England\\ 
  $^{3}$ Laboratoire d'Astrophysique de Marseille, CNRS- Universit\'e
  Aix-Marseille, 38 rue Fr\'ed\'eric Joliot-Curie, 13388
  Marseille Cedex 13, France\\ 
  $^{4}$ California Institute of Technology, Mail Code 105--24,
  Pasadena, CA 91125, USA \\  
  $^{5}$ University of British Columbia, Department of physics and
  astronomy, 6224 Agricultural Rd., Vancouver, B.C., V6T 1Z1,
  Canada \\ 
  $^{6}$ Harvard Center for Astrophysics and Las Cumbres Observatory
  Global Telescope Network Fellow\\ 
  $^{7}$ Harvard-Smithsonian Center for Astrophysics, 60 Garden
  Street, Cambridge MA02138, USA\\ 
  $^{8}$ Max-Planck-Institut f\"ur Astronomie,K\"onigstuhl 17, 69117
  Heidelberg, Germany\\ 
  $^{9}$ Department of Physics, Universit\'a di Roma Tor Vergata, via
  della ricerca scientifica, 1, 00133 Roma, Italy\\ 
  $^{10}$ Dark Cosmology Centre, Niels Bohr Institute, University of 
  Copenhagen, Juliane Maries Vej 30, 2100 Copenhagen, Denmark\\
  $^{11}$ Steward Observatory, University of Arizona, 933 North Cherry
  Avenue, Tucson, AZ 85721, USA\\ 
  $^{12}$ Jet Propulsion Laboratory, Caltech, MS 169-327, Oak Grove
  Dr, Pasadena CA 91109, USA\\ 
  $^{13}$ Johns Hopkins Dept. of Physics and Astronomy, Baltimore, MD
  21218, USA\\ 
  $^{\star}$ Marie-Curie fellow. E-mail: johan.richard@durham.ac.uk\\
}

\date{Accepted. Received; in original form }

\pagerange{\pageref{firstpage}--\pageref{lastpage}} \pubyear{2002}

\maketitle

\label{firstpage}

\begin{abstract} We present a statistical analysis of a sample of 20
  strong lensing clusters drawn from the Local Cluster Substructure
  Survey (LoCuSS), based on high resolution \emph{Hubble Space
    Telescope (HST)} imaging of the cluster cores and follow-up
  spectroscopic observations using the Keck-I telescope.  We use
  detailed parameterized models of the mass distribution in the
  cluster cores, to measure the total cluster mass and fraction of
  that mass associated with substructures within $R\le250\kpc$.  These
  measurements are compared with the distribution of baryons in the
  cores, as traced by the old stellar populations and the X-ray
  emitting intracluster medium.  Our main results include: (i) the
  distribution of Einstein radii is log-normal, with a peak and
  $1\sigma$ width of
  $\langle\log_{10}\theta_E(z=2)\rangle=1.16\pm0.28$; (ii) we detect
  an X-ray/lensing mass discrepancy of $\langle M_{SL}/M_X\rangle=1.3$
  at $3\sigma$ significance -- clusters with larger substructure
  fractions displaying greater mass discrepancies, and thus greater
  departures from hydrostatic equilibrium; (iii) cluster substructure
  fraction is also correlated with the slope of the gas density
  profile on small scales, implying a connection between
  cluster-cluster mergers and gas cooling.  Overall our results are
  consistent with the view that cluster-cluster mergers play a
  prominent role in shaping the properties of cluster cores, in
  particular causing departures from hydrostatic equilibrium, and
  possibly disturbing cool cores.  Our results do not support recent
  claims that large Einstein radius clusters present a challenge to
  the CDM paradigm.
\end{abstract}


\section{Introduction}

The evolution of galaxy clusters with cosmic time is an important
cosmological probe, as it traces the gravitational growth of dark
matter on large scales. In particular, the cluster mass function can
be directly tested against cosmological models, as it is related to
fundamental parameters such as the matter density $\Omega_m$, the
normalisation of the power spectrum $\sigma_8$
\citep[e.g.][]{Schuecker,Smith03}, and the dark energy equation of
state parameter $w$.

Commonly used probes of the mass of clusters at $z\sim0.1-0.5$ include
the $K$-band luminosity \citep{lin}, X-ray luminosity and temperature
\citep[e.g.][]{Vikhlinin}, Sunayev-Zeldovich effect
\citep[e.g.][]{Nagai,Bonamente}, cluster kinematics
\citep[e.g.][]{Blindert}, and gravitational lensing
\citep[e.g.][]{Bardeau,Smith05,Okabe}.  Systematic errors in these
various mass probes can be calibrated by an inter-comparison of
results from the various methods.  The pursuit of the most robust
calibrations based on low redshift cluster samples is essential to
achieve reliable results from future studies at higher redshift, and
to control systematic uncertanties in cosmological experiments
\citep{Albrecht06}.

Gravitational lensing plays a central role in this effort, as it does
not rely on assumptions about the symmetry or equilibrium properties
of the cluster mass distribution, although parametrized lensed model implicitly 
assume, for instance, elliptical symmetry.  Indeed, the highest quality
measurements of the dark matter mass and substructure in low redshift
clusters have been obtained through the combination of strong and weak
gravitational lensing.  The identification of background galaxies
forming strong lensing arcs in cluster cores is a direct measurement
of the enclosed mass within the Einstein radius $\theta_E$, providing
an accurate normalisation of cluster mass models both in the core, and
extending to larger radii \citep[e.g.][]{Kneib03}.  The main caveats
on lensing studies are that the lensing signal is sensitive to the
mass distribution projected along the line of sight through the
cluster, that the commonly used parametric models must by design
assume a parametric form of the mass distribution, and that
gravitational lensing may prove to be impractical in samples of more
distant ($z>1$) cluster samples that are starting to be discovered
\citep[e.g.][]{Rosati1, Rosati2}.

Early joint lensing/X-ray cluster studies identified large
discrepancies between total cluster masses obtained from the two
methods \citep[e.g.][]{Jordi95}.  These discrepancies were
subsequently eliminated, albeit within quite large uncertainties, in
cool core clusters provided that a two phase gas model was used
(\citealt{Allen98}; see also \citealt{Smail97}).  At a similar time,
\emph{HST} data started to became available for some clusters,
allowing more precise strong lensing models to be constructed
\citep[e.g.][]{Kneib96,Tyson98}.  More recent strong lensing studies
of cluster mass distributions have generally continued to concentrate
on detailed studies of spectacular individual strong lensing clusters
based on deep multi-filter \emph{HST} observations
\citep[e.g.][]{Broadhurst05,Limousin07,Richard09}.  The interpretation
of results from such studies, e.g.\ size of Einstein radii and shape
of density profile, in the context of the general cluster population
is inevitably problematic.

In contrast, \citet[][hereafter Sm05]{Smith05} studied 10 X-ray
selected clusters ($L_{X,0.2-2.4\keV}>4.1\times10^{44}\ergs$) at
$z\simeq0.2$ as a first step towards building large statistical
samples of strong lensing clusters.  Sm05 combined strong and weak
lensing constraints from moderate depth \emph{HST} observations with
X-ray spectro-imaging from \emph{Chandra}, and near-infrared
photometry of cluster galaxies from UKIRT.  The main results included
the first mass-observable scaling relations to employ lensing-based
mass measurements for a well-defined sample; the only previous example
employed a compilation of 6 clusters from the literature
\citep{Hjorth98}.  The main limiting factor on Sm05's results were
the small sample size, with just 5 of the 10 clusters containing
spectroscopically confirmed strongly lensed galaxies. 

The Local Cluster Substructure Survey (LoCuSS, PI: G.\ P.\ Smith) is
an all-sky systematic survey of 165 X-ray luminous clusters at
$0.15<z<0.30$ selected from the \emph{ROSAT} All-sky Survey catalogs
\citep{Ebeling98,Ebeling00,Boehringer04}.  In addition to seeking to
improve the statistical precision by enlarging the sample size of
studies such as Sm05 and \citet{Bardeau}, LoCuSS aims to incorporate
new constraints on cluster baryons, most notably from observations of
the Sunyaev-Zeldovich Effect \citep{Carlstrom02} to deliver a
definitive local calibration of mass-observable scaling relations that
will be useful for cluster cosmology, interpretation of high-redshift
cluster samples, and probing the physics of gas heating and cooling in
merging clusters at low redshift.  More generally, LoCuSS aims to
constrain the scatter in the baryonic properties of the local cluster
population and to correlate the scatter with the recent hierarchical
assembly history of the clusters, the latter being constrained by the
lensing-based mass maps \citep{Taylor}.  A wide variety of studies are
therefore underway, including wide-field weak-lensing analysis with
Suprime-Cam on Subaru \citep{Okabe}, the first lensing-based
mass-$Y_{SZ}$ relation \citep{Marrone09}, and multiwavelength studies
supported by space-based (\emph{HST}, \textit{Herschel},
\textit{Chandra}, GALEX, \textit{Spitzer}) as well as ground-based
(Keck, VLT, Gemini, MMT, CTIO, KPNO, Palomar) facilities
\citep[e.g.][]{Zhang,Haines09b,Haines09a,Sanderson09a,Smith09b}.

This paper presents a four-fold increase in the number of clusters
with spectroscopically confirmed strong lensing clusters on Sm05.
Such constraints remain critical within the overall context of the
studies outlined above because strong-lensing offers precise
constraints on the mass distribution in cluster cores ($R\sim100\kpc$)
which for example provide an invaluable constraint on small scales
when trying to measure the shape of cluster density profiles in
conjunction with wide-field weak-lensing data from Subaru.  We compare
the details of the central mass distributions (integrated mass and
substructure fraction) obtained from the strong-lensing constraints
with near-infrared and X-ray probes of the cluster baryons.  We
describe the data in \S\ref{data}, and present the strong-lensing
analysis and models in \S\ref{sl}.  The main results on the mass and
structure of the cluster cores are presented in \S\ref{results} and
summarized in \S\ref{conc}.  Throughout the paper, we use magnitudes
quoted in the AB system, and a standard $\Lambda$-CDM model with
$\Omega_m=0.3$, $\Omega_\Lambda=0.7$, and $H_0=70\kms\Mpc^{-1}$,
whenever necessary.  In this cosmology, 1\arcsec\ is equivalent to
$3.3\kpc$ at $z=0.2$.  We adopt the definition $e=1-b/a$ of the
ellipticity, where $a$ and $b$ are the semi-major and semi-minor axis
of the ellipse, respectively.

\section{Observations and data reduction}
\label{data}

\begin{table*}
\caption{\label{images}Optical and  near-infrared imaging observations  of  the new strong-lensing LoCuSS clusters.}
\begin{tabular}{lrrcrlclrrr}
Cluster & $\alpha_{\rm BCG}$ & $\delta_{\rm BCG}$ & $z$ & \multispan3{\dotfill\emph{HST}\dotfill} & \multispan4{\dotfill Near-infrared\dotfill} \\
        &                   &                   &    & PID        & Camera/Filter   & Depth ($Z\sigma$)   & Camera  & $J(Z\sigma)$& $K_S(Z\sigma)$  & $K_S^\star$ \\
\hline 
A\,521$^{(a)}$  &  73.528753 & -10.223605      & 0.2475 & 11312    & WFPC2/F606W    & 27.0 & WIRC & 21.5 & 21.0 & 17.60 \\
A\,611         & 120.236680 &  36.056725      & 0.2850 & 9270     & ACS/F606W      & 27.7 & WIRC & 21.0 & 20.9 & 17.85\\
A\,773         & 139.472660 & 51.727024       & 0.2170 & 8249     & WFPC2/F702W    & 27.6 & WIRC & 21.7 & 21.5 & 17.36\\
A\,868          & 146.359960 & -8.651994      & 0.1535 & 8203     & WFPC2/F606W    & 27.5 & ISPI  & 21.9  & 21.6 & 16.70\\
Z2701$^{(b)}$  & 148.204560 & 51.885143        & 0.2140 & 9270     & ACS/F606W      & 27.7 & WIRC & 21.5 & 21.5 & 17.33\\
A\,1413        & 178.824510 & 23.404451       & 0.1427 & 9292     & ACS/F775W      & 27.3 & WIRC & 23.0  & 22.3 & 16.55\\
               &            &                 &        & 9292     & ACS/F850LP     & 26.7& \\
A\,1835        & 210.258860 & 2.878532        & 0.2528 & 8249     & WFPC2/F702W    & 27.7 & ISPI  & 22.0  & 21.3 & 17.64\\
               &            &                 &        & 10154     & ACS/F850LP    & 27.3      & \\
A\,2204        & 248.195540 & 5.575825        & 0.1524 & 8301      & WFPC2/F606W   & 26.4 & FLAM.\  & 22.1  & 21.9 & 16.68\\
RX\,J1720$^{(c)}$ & 260.041860 & 26.625627       & 0.1640 & 11312     & WFPC2/F606W   & 26.9 & WIRC & 20.6 & 20.7 & 16.82\\
RX\,J2129$^{(d)}$ & 322.416510 & 0.089227        & 0.2350 & 11312     & WFPC2/F606W   & 26.7 & WIRC & 21.7 & 21.7 &17.51\\
\hline
\end{tabular}

$^{(a)}$ also known as RXC\,J0454.1$-$1014 $^{(b)}$ also known as ZwCl\,0949.6+5207 $^{(c)}$ RX\,J1720.1+2638 $^{(d)}$ RX\,J2129.6+0005

\end{table*}

\begin{table*}
\caption{\label{extsample}Extended sample of 10 previously published spectroscopically-confirmed strong-lensing LoCuSS clusters}
\begin{tabular}{lrrcl}
Cluster       & $\alpha_{\rm BCG}$ & $\delta_{\rm BCG}$  & $z$        &  Strong-lensing reference \\
\hline
A\,68     & 9.278626 & 9.156722 & 0.2546  & Sm05, \citet{Richard07,Smith02b} \\
A\,383   & 42.014079 & -3.529040 & 0.1883  & Sm05, \citet{Smith01a,Sand04}\\
A\,963   & 154.264990 & 39.047228& 0.2050  & Sm05, \citet{Ellis91}\\
A\,1201 & 168.227080 & 13.435946& 0.1688  &  \citet{Edge} \\
A\,1689 & 197.872950  & -1.341005 & 0.1832 & \citet{Limousin07} \\
A\,1703 & 198.771971 & 51.817494& 0.2800 & \citet{Richard09}\\
A\,2218 & 248.954604 & 66.212242  & 0.1710 & Sm05, \citet{Kneib96}, \citet{Ardis} \\
A\,2219 & 250.082380 & 46.711561  & 0.2281 & Sm05 \\
A\,2390 & 328.403290 & 17.695740 & 0.2329 & Jullo, PhD thesis \\
A\,2667 & 357.914125 & -26.084375 & 0.2264 & \citet{Covone}\\
\hline
\end{tabular}
\end{table*}

We present here the cluster sample and the relevant datasets used in
our study.  The strong lensing analysis relies on two main
ingredients: (a) the identification of multiply-imaged systems and (b)
measurements of their spectroscopic redshifts.  The former relies on
high angular resolution \emph{HST} imaging, while the latter is
carried out using sensitive optical multi-object spectrographs on
large aperture ground-based telescopes, in this case LRIS on the
Keck-I 10-m telescope.  We also use near-infrared photometry to select
likely cluster galaxies based on the prominent red sequence of cluster
galaxies in the $J-K/K$ color-magnitude diagrams, that is largely
insensitive to spectral type at $z\simeq0.2$.  These photometric
catalogs are also used in the paper during the construction of the
lens models (\S\ref{slsub}).  Finally, we summarize the analysis of
X-ray data available for the majority of these clusters, and which we
latter use as a comparison with the strong-lensing results (see
\S\ref{results}).

\subsection{Cluster sample}

The sample in this article comprises the 10 clusters that we have
spectroscopically confirmed as strong-lenses during our Keck observing
campaign 2004--2008 (Table~\ref{images}), plus a further 10 clusters
within the LoCuSS sample that have been previously published (Table
\ref{extsample}).  To qualify for observations with Keck, a cluster
must have been previously imaged with \emph{HST} (79 clusters), and
lie at $\delta>-25^\circ$.  We note that some of the clusters with new
strong-lensing models had previously been identified as candidate
strong-lensing clusters:

  \begin{itemize}
  \item{A\,773 and A\,1835 were included in Sm05's sample however in
      the absence of spectroscopic confirmation, only weak-lensing
      constraints were used to construct the mass models.}
  \item{A\,521 was studied by \citet{Maurogordato} using ground-based
      images.  A giant arc feature was observed but spectroscopic
      follow-up attempts were unable to confirm its strong lensing
      origin.}
  \item{A\,773, A\,868, A\,1835, A\,2204 and Z2701 were part of the
      \citet{Sand05} sample, who performed a systematic search for
      radial and tangential arc features in archival \emph{HST} images
      of massive clusters. They identified radial arcs in A\,773,
      A\,1835, and A\,2204.}
  \end{itemize} 

\subsection{Imaging data}

\subsubsection{\emph{HST} imaging data}

High resolution imaging data taken with the ACS or WFPC2 instrument on
\emph{HST} are available for each selected cluster in one or two
bands, either through our dedicated LoCuSS program (GO-DD 11312, PI:
G.\ P.\ Smith), or from the archive.  These data are summarized in
Table~\ref{images}.  In every case, one of the observed filters is
located between the $V$ and the $I$ band, the most common being the
F606W filter.

Each \emph{HST} observation consists of several independent frames
arranged in a dither pattern.  These images were combined using the
{\tt multidrizzle} software with a final plate scale of $0.05\arcsec$
(ACS) or $0.1\arcsec$ (WFPC2).  In the case of A\,868, the
observations comprise a mosaic of 6 WFPC2 pointings taken at different
epochs. This mosaic was combined using the relative astrometric shifts
measured from bright objects in common between two adjacent pointings.

A careful visual inspection was performed on the reduced \emph{HST}
images by two of the authors (JR and GPS) to search for strong lensing
features.  A catalog of candidate multiple images was compiled by
selecting highly sheared arc-like features, as well as objects
displaying a very similar morphology and the characteristic symmetric
effect of strong lensing (e.g.,Fig. \ref{cluster}).  The majority of
these multiple images does not appear like arcs, but have a clumpy
morphology, with multiple knots of star-formation.  These lensed
sources are not detected by automatic arc finding techniques
(e.g. {\tt Arcfinder}, \citealt{Seidel}), and therefore justify the
use of a visual inspection of the images.

\subsubsection{Near-infrared imaging data}

\begin{figure*}
  \caption{\label{cm}$J-K_S$ vs $K_S$ colour-magnitude diagrams for each
    observed cluster, showing the red sequence selection of cluster
    members (red dots).  The photometric selection used is shown as
    delimited by the solid box (see text for details).  The last two
    panels show the red sequence parameters (intercept at $K_S=18$ and
    slope $\kappa_{JK_S}$) as a function of cluster redshift.  The solid
    line gives the expected colours of an elliptical galaxy, and the
    dashed line marks the average slope $\kappa_{JK_S}$ for the
    sample.}
\begin{minipage}{5.cm}
\centerline{A\,1413 $z=0.1427$\\}
\centerline{\mbox{\includegraphics[height=0.95\textwidth,angle=270]{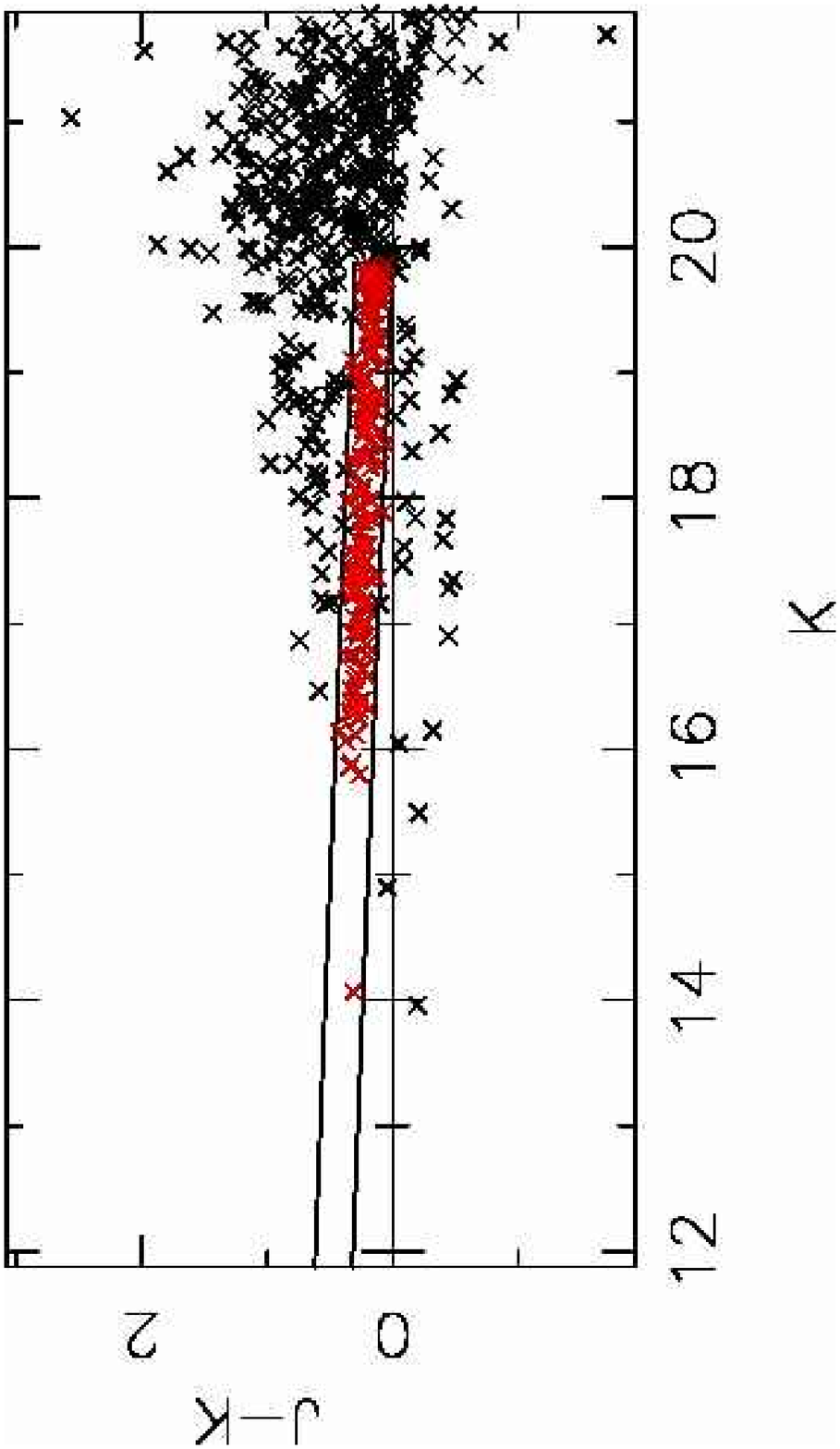}}}
\end{minipage}
\begin{minipage}{5.cm}
\centerline{A\,2204 $z=0.1524$\\}
\centerline{\mbox{\includegraphics[height=0.95\textwidth,angle=270]{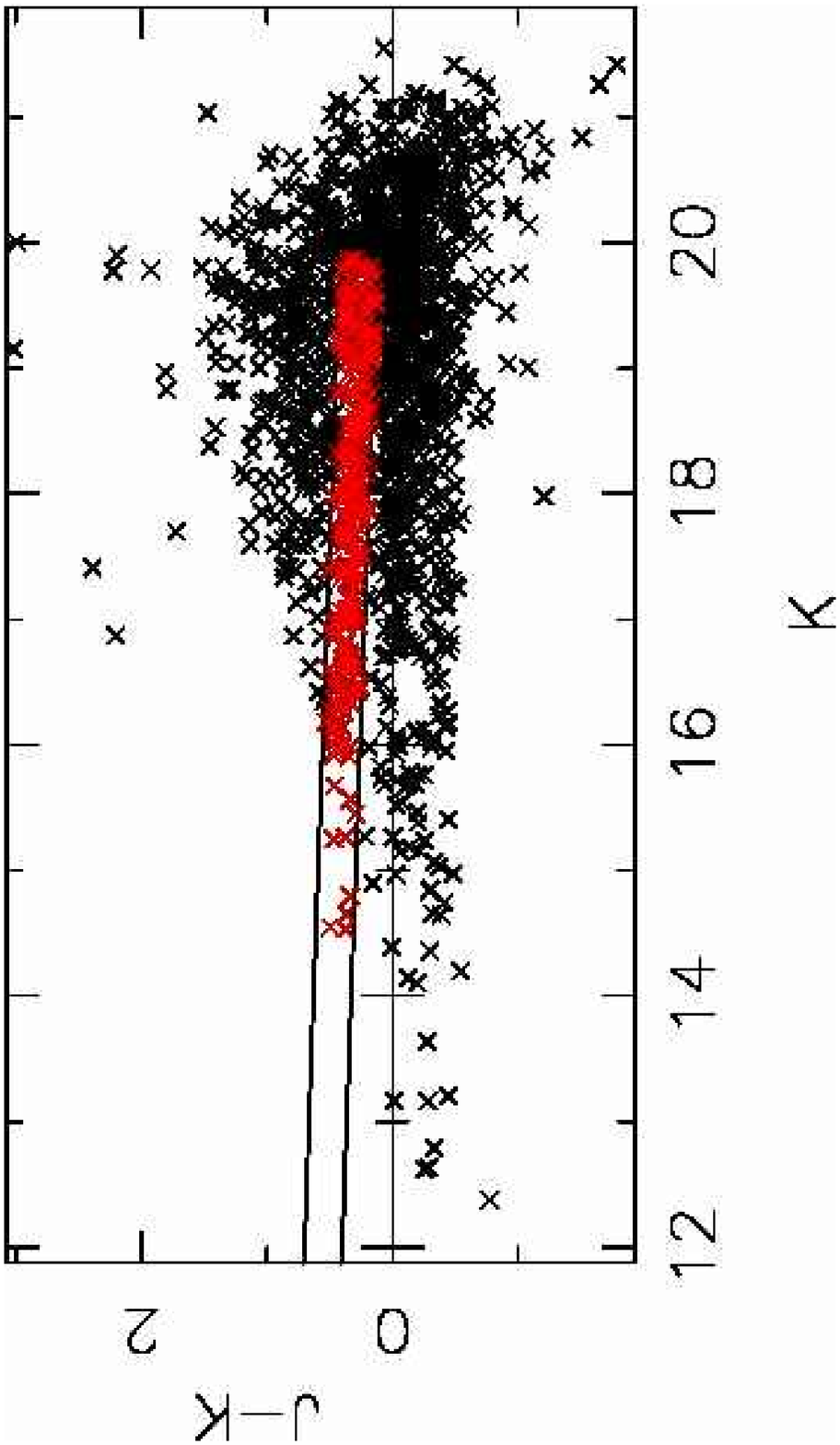}}}
\end{minipage}
\begin{minipage}{5.cm}
\centerline{A\,868 $z=0.1535$\\}
\centerline{\mbox{\includegraphics[height=0.95\textwidth,angle=270]{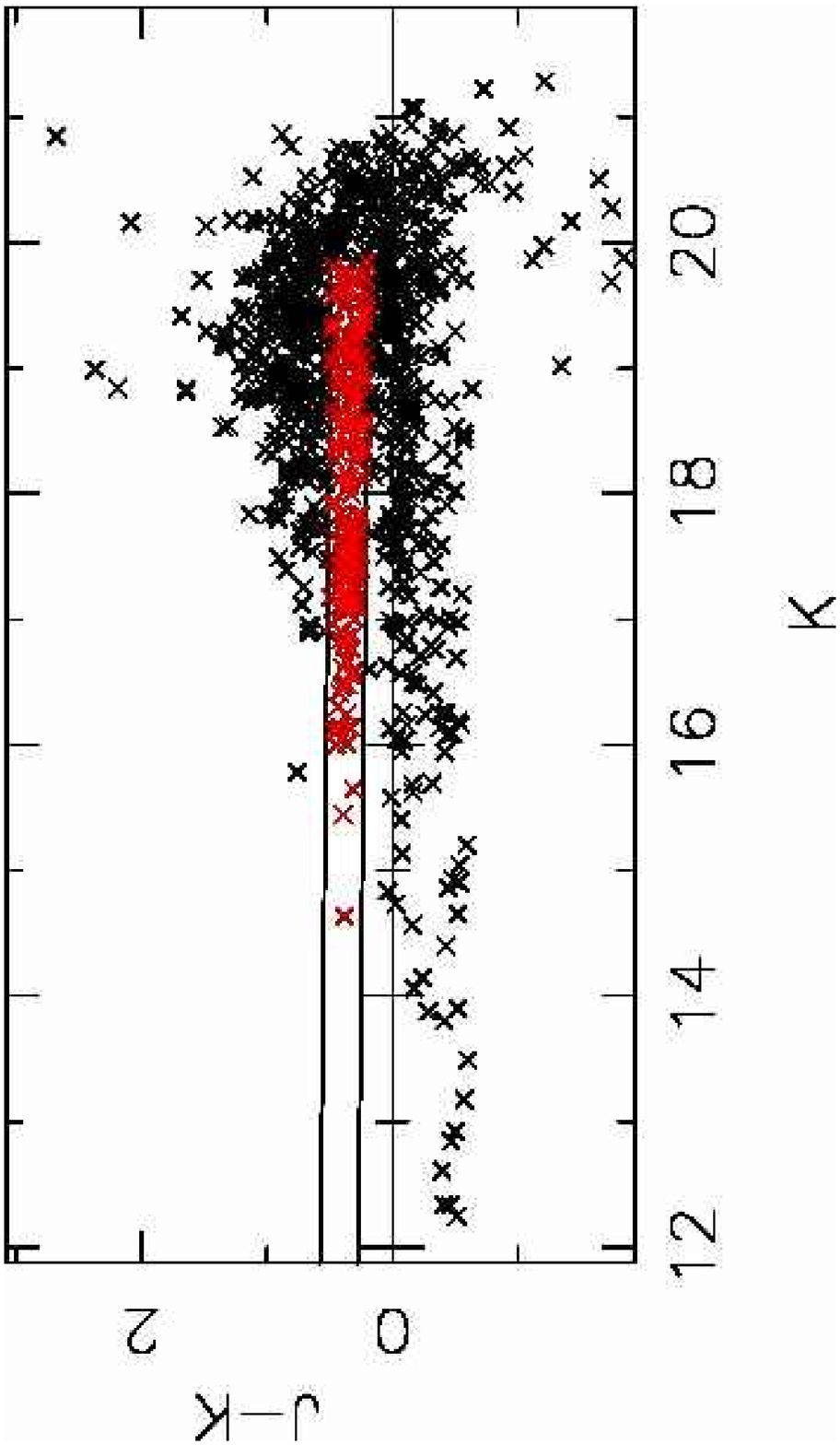}}}
\end{minipage}
\medskip\par
\begin{minipage}{5.cm}
\centerline{RXJ1720 $z=0.1640$ \\}
\centerline{\mbox{\includegraphics[height=0.95\textwidth,angle=270]{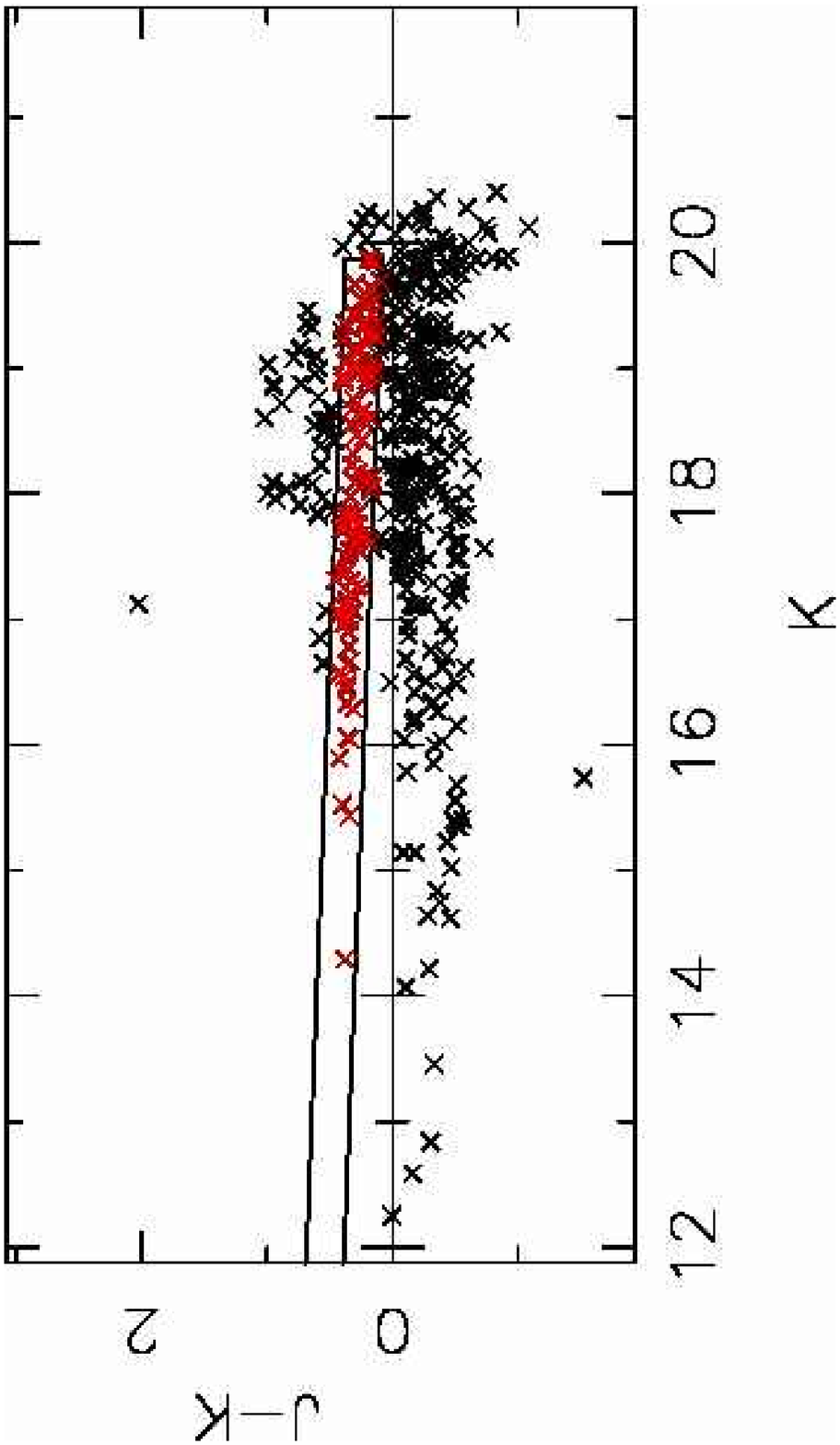}}}
\end{minipage}
\begin{minipage}{5.cm}
\centerline{Z2701 $z=0.2140$ \\}
\centerline{\mbox{\includegraphics[height=0.95\textwidth,angle=270]{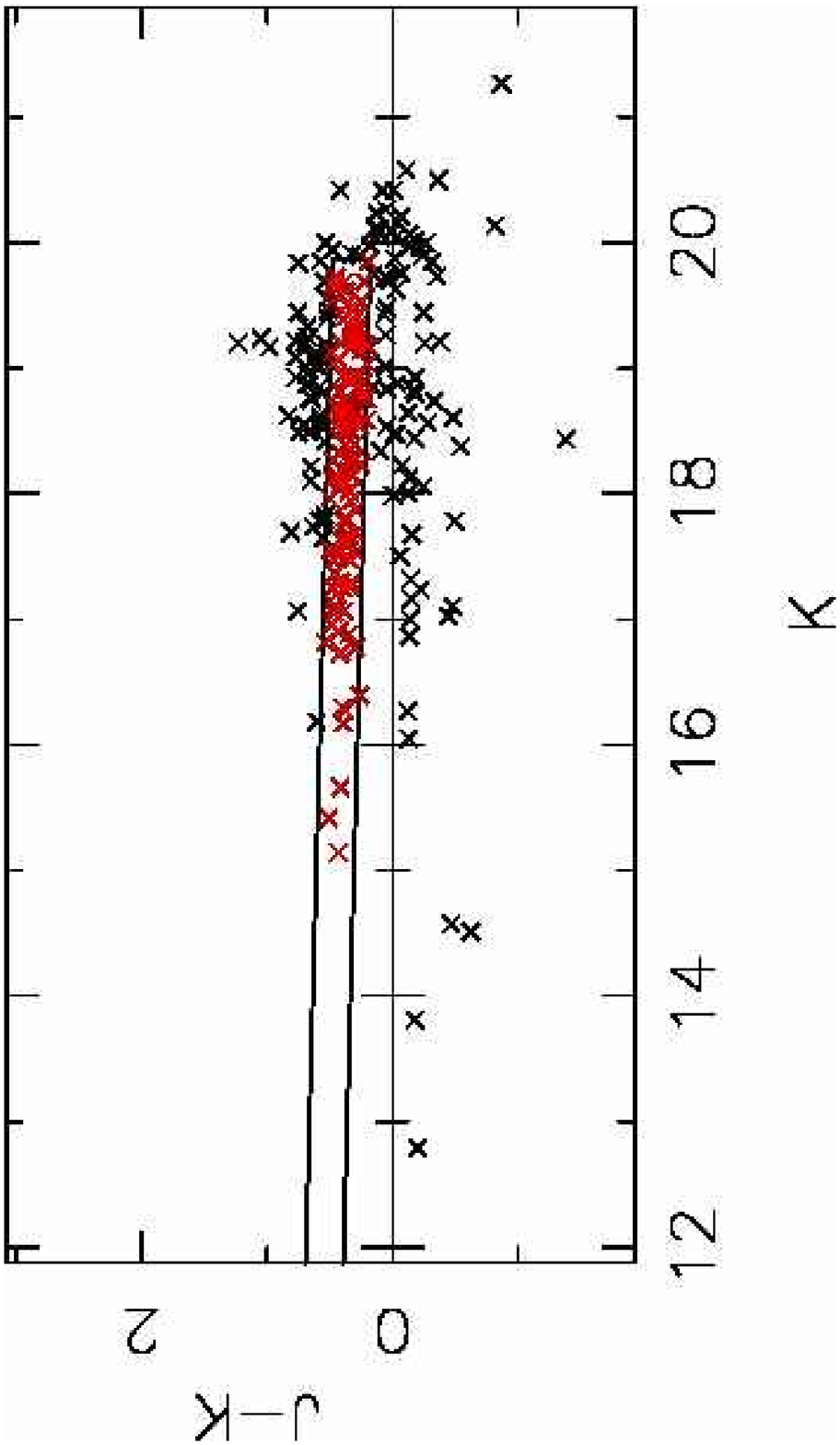}}}
\end{minipage}
\begin{minipage}{5.cm}
\centerline{A\,773 $z=0.2170$ \\}
\centerline{\mbox{\includegraphics[height=0.95\textwidth,angle=270]{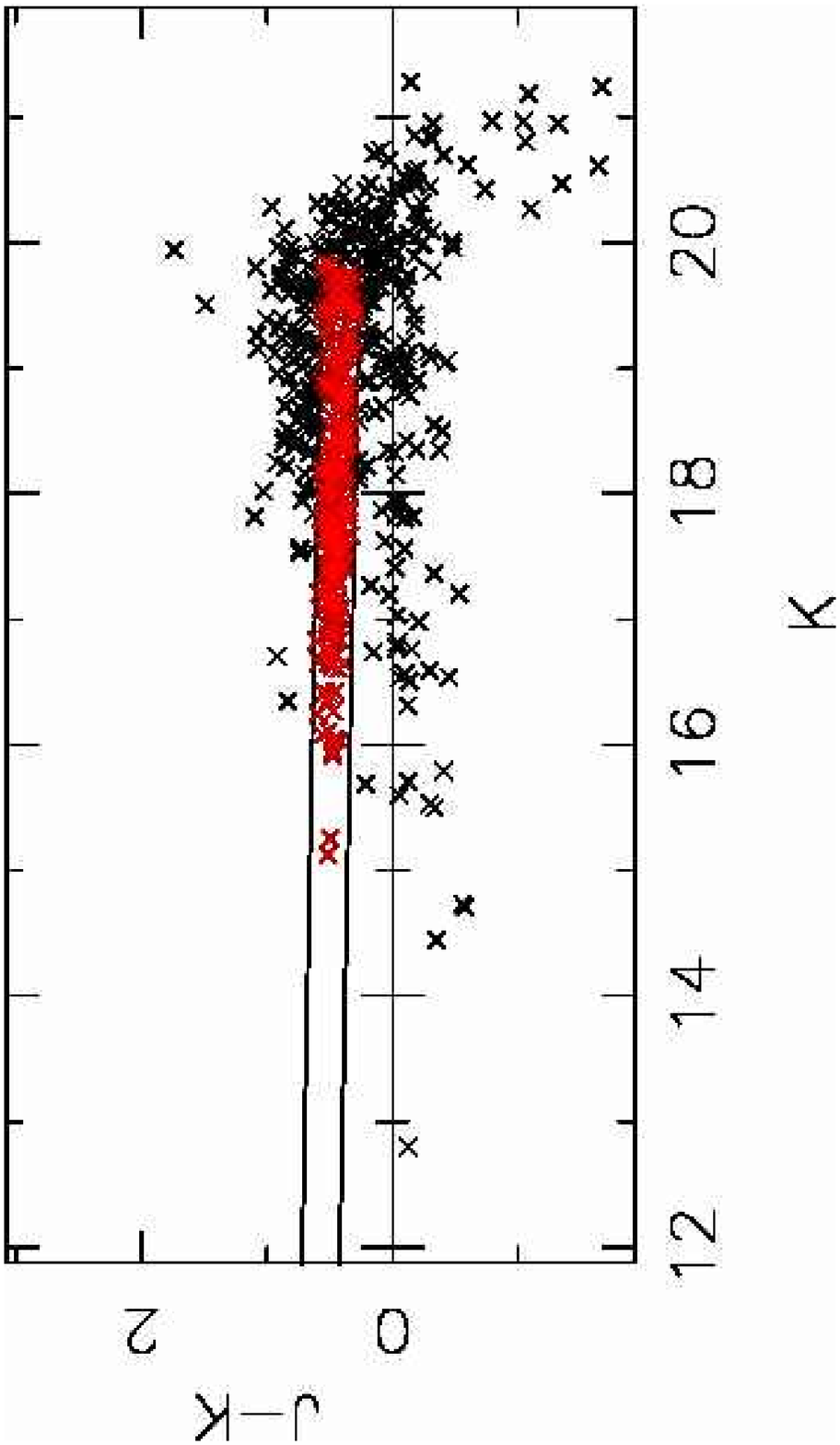}}}
\end{minipage}
\medskip\par
\begin{minipage}{5.cm}
\centerline{RXJ2129 $z=0.2350$ \\}
\centerline{\mbox{\includegraphics[height=0.95\textwidth,angle=270]{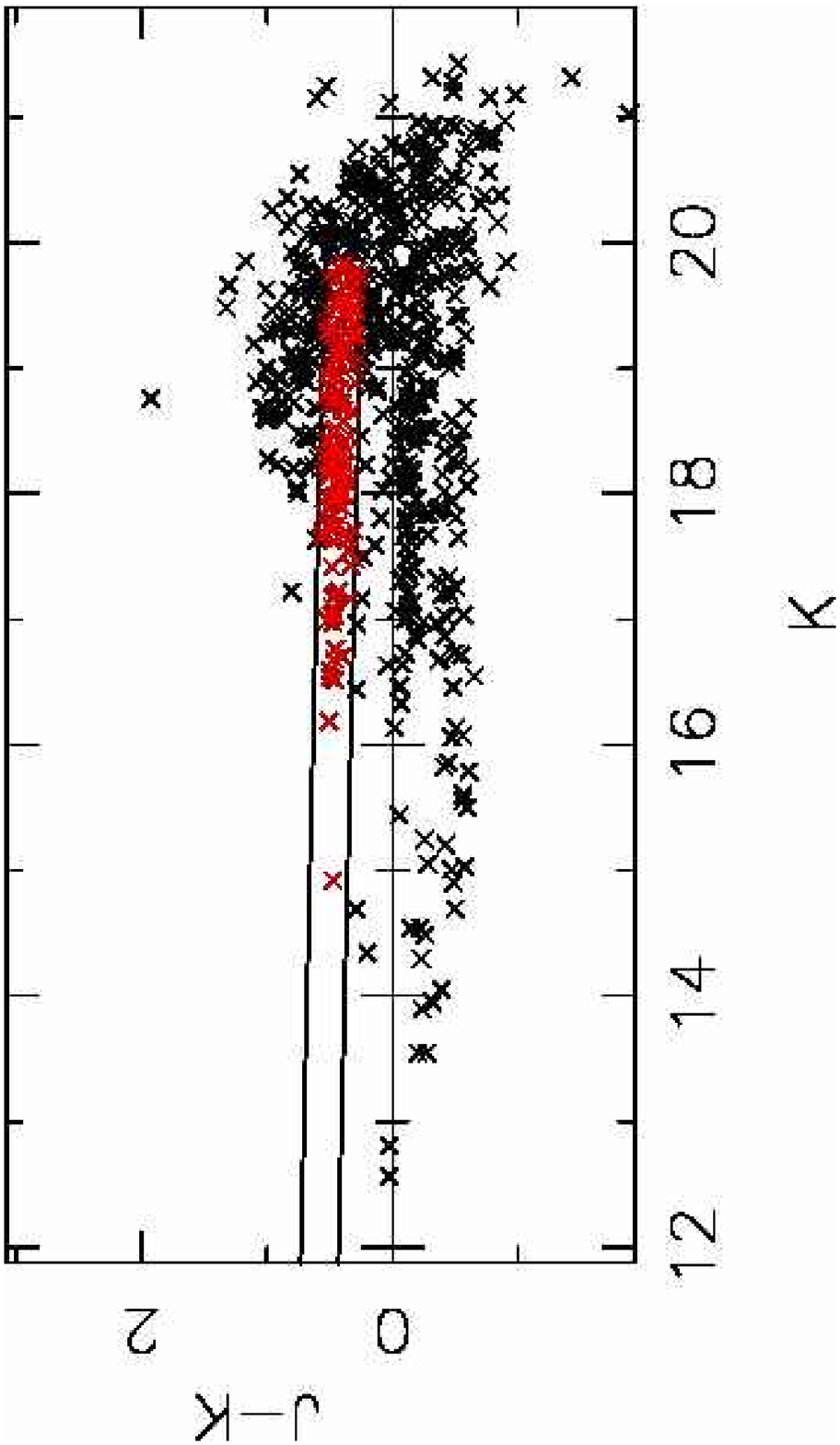}}}
\end{minipage}
\begin{minipage}{5.cm}
\centerline{A\,521 $z=0.2475$\\}
\centerline{\mbox{\includegraphics[height=0.95\textwidth,angle=270]{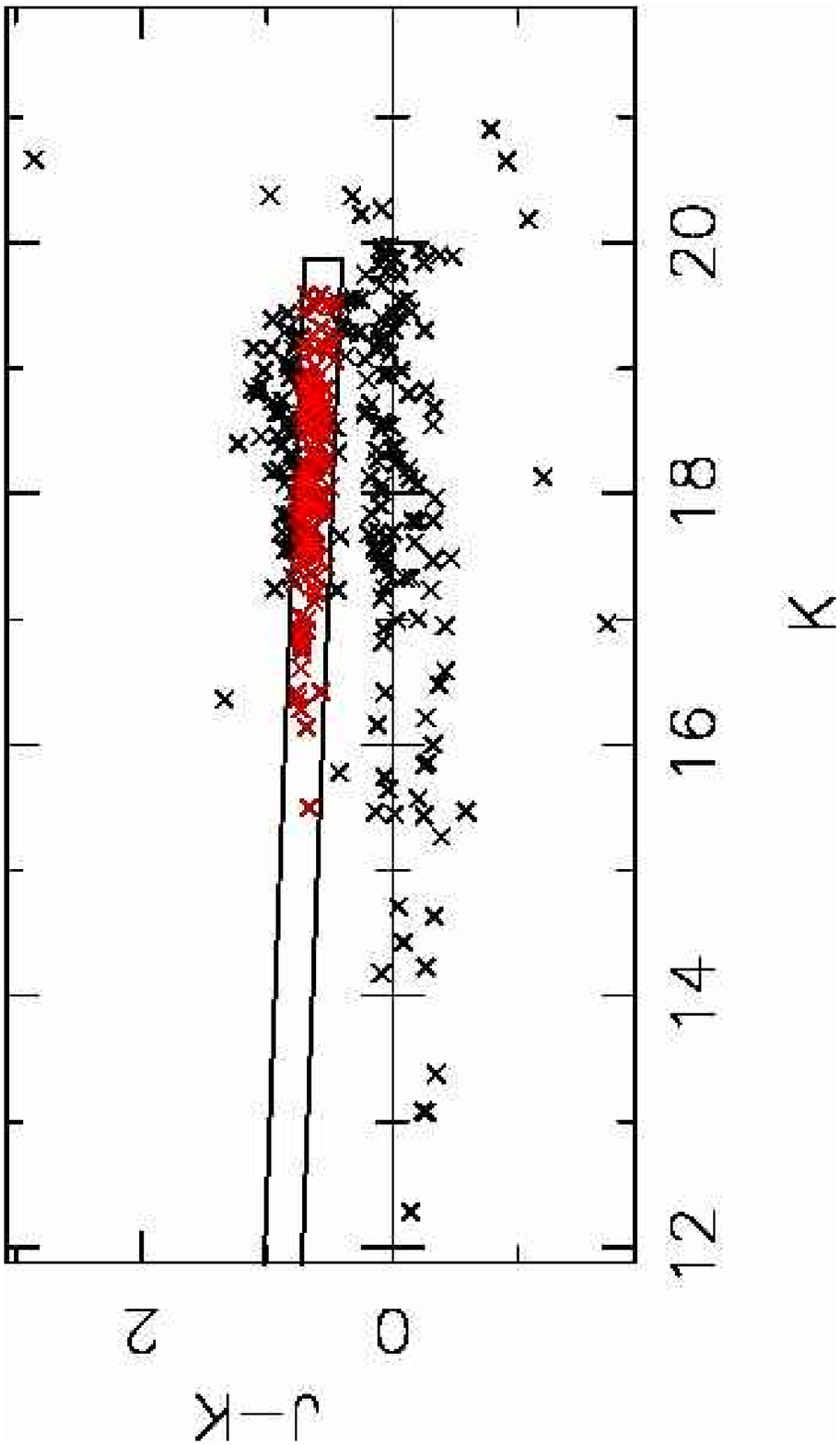}}}
\end{minipage}
\begin{minipage}{5.cm}
\centerline{A\,1835 $z=0.2528$ \\}
\centerline{\mbox{\includegraphics[height=0.95\textwidth,angle=270]{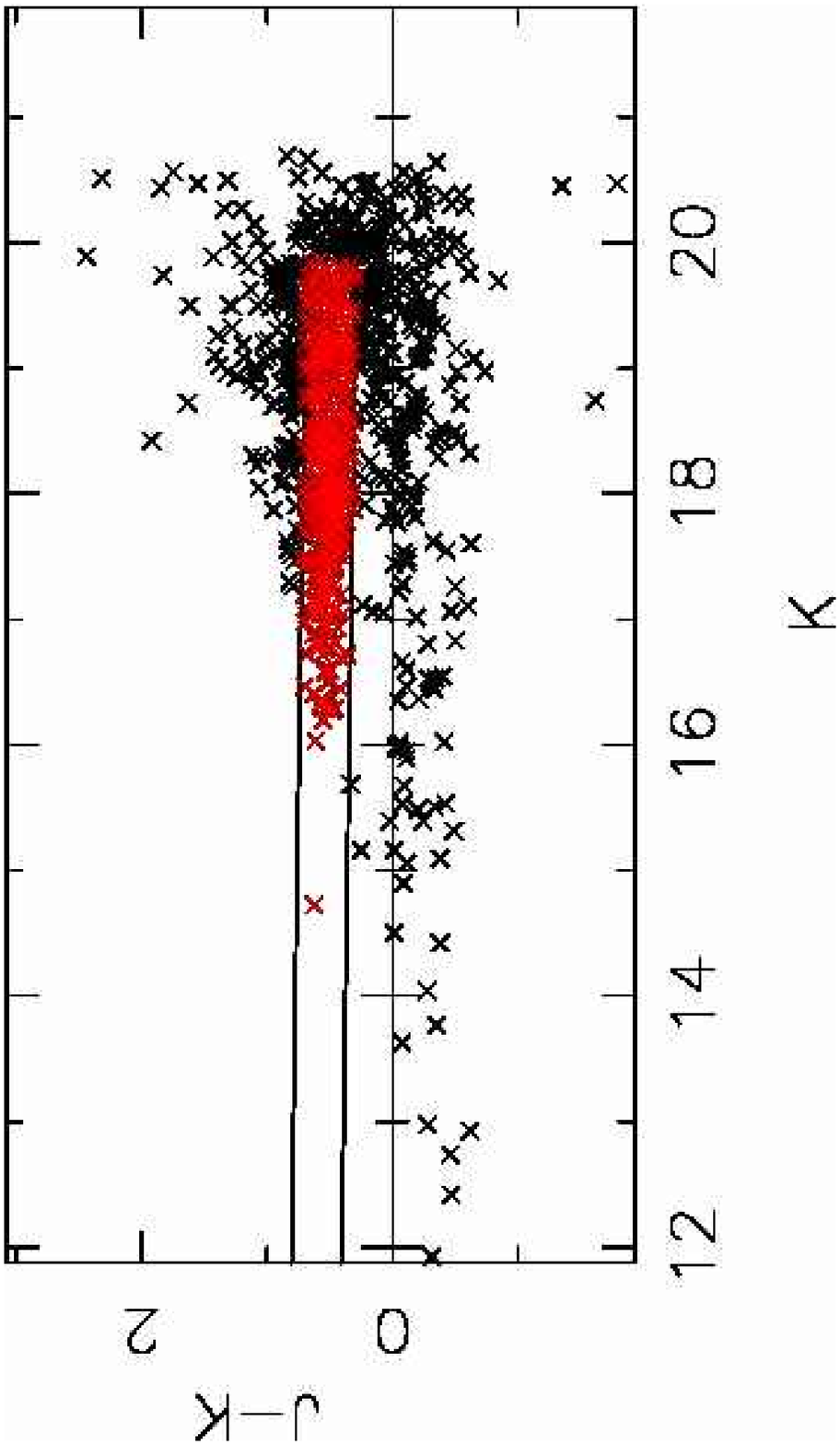}}}
\end{minipage}
\medskip\par
\begin{minipage}{5.cm}
\centerline{A\,611 $z=0.2850$\\}
\centerline{\mbox{\includegraphics[height=0.95\textwidth,angle=270]{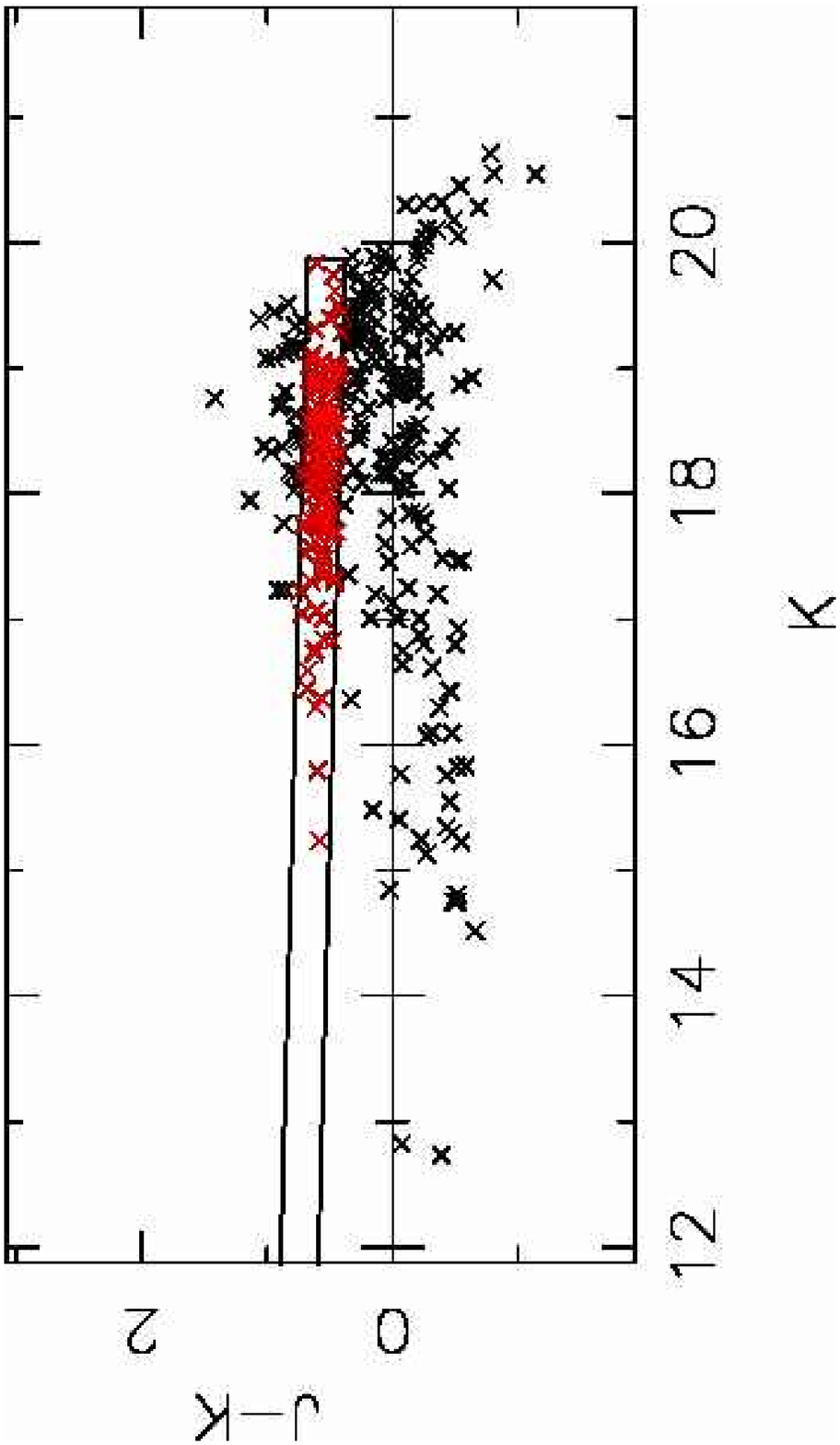}}}
\end{minipage}
\begin{minipage}{5.cm}
\centerline{$J-K_S$ colour at $K_S$=18\\}
\centerline{\mbox{\includegraphics[height=0.75\textwidth,angle=270]{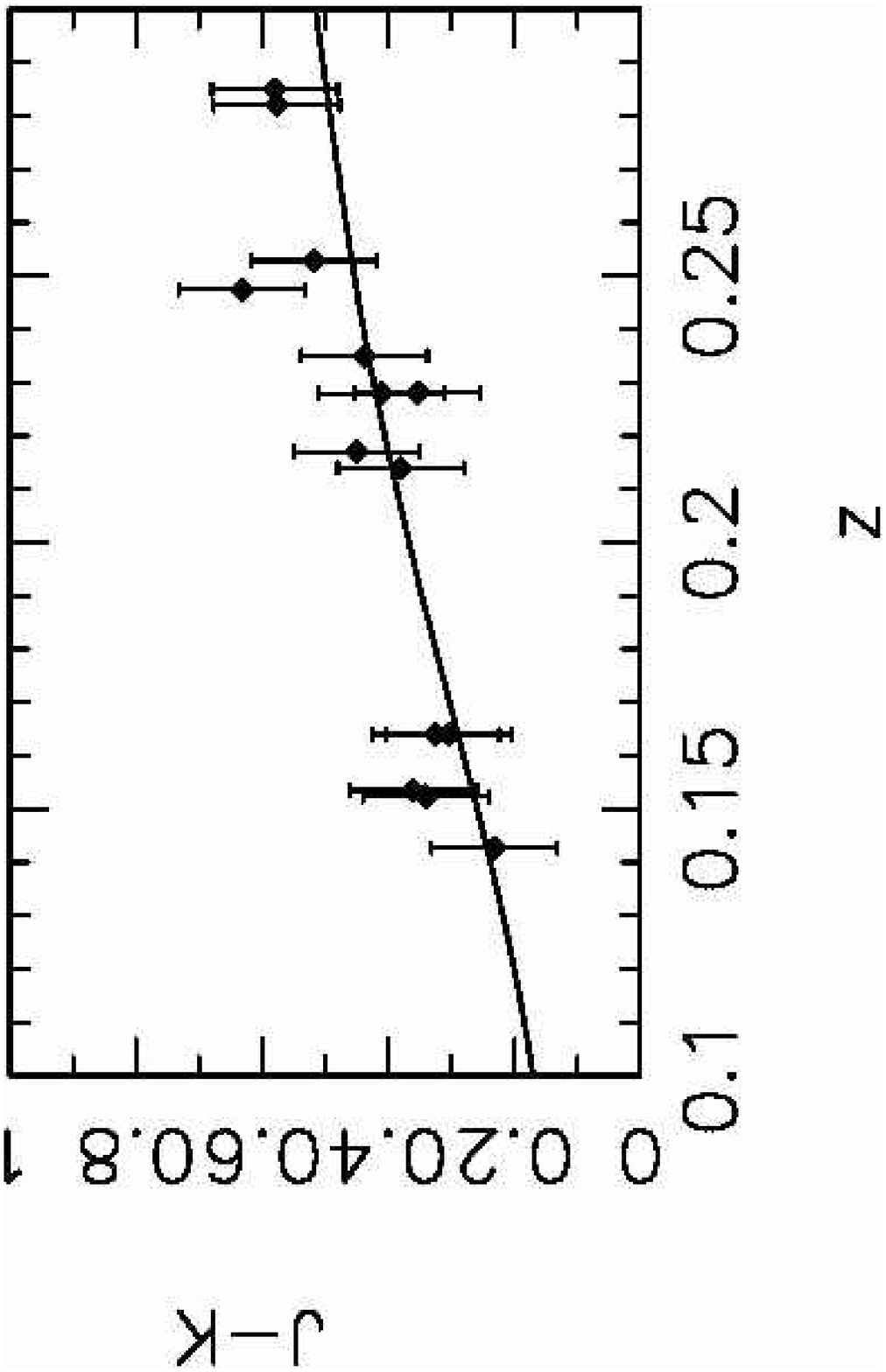}}}
\end{minipage}
\begin{minipage}{5.cm}
\centerline{$\kappa_{JK}$ colour-magnitude slope\\}
\centerline{\mbox{\includegraphics[height=0.75\textwidth,angle=270]{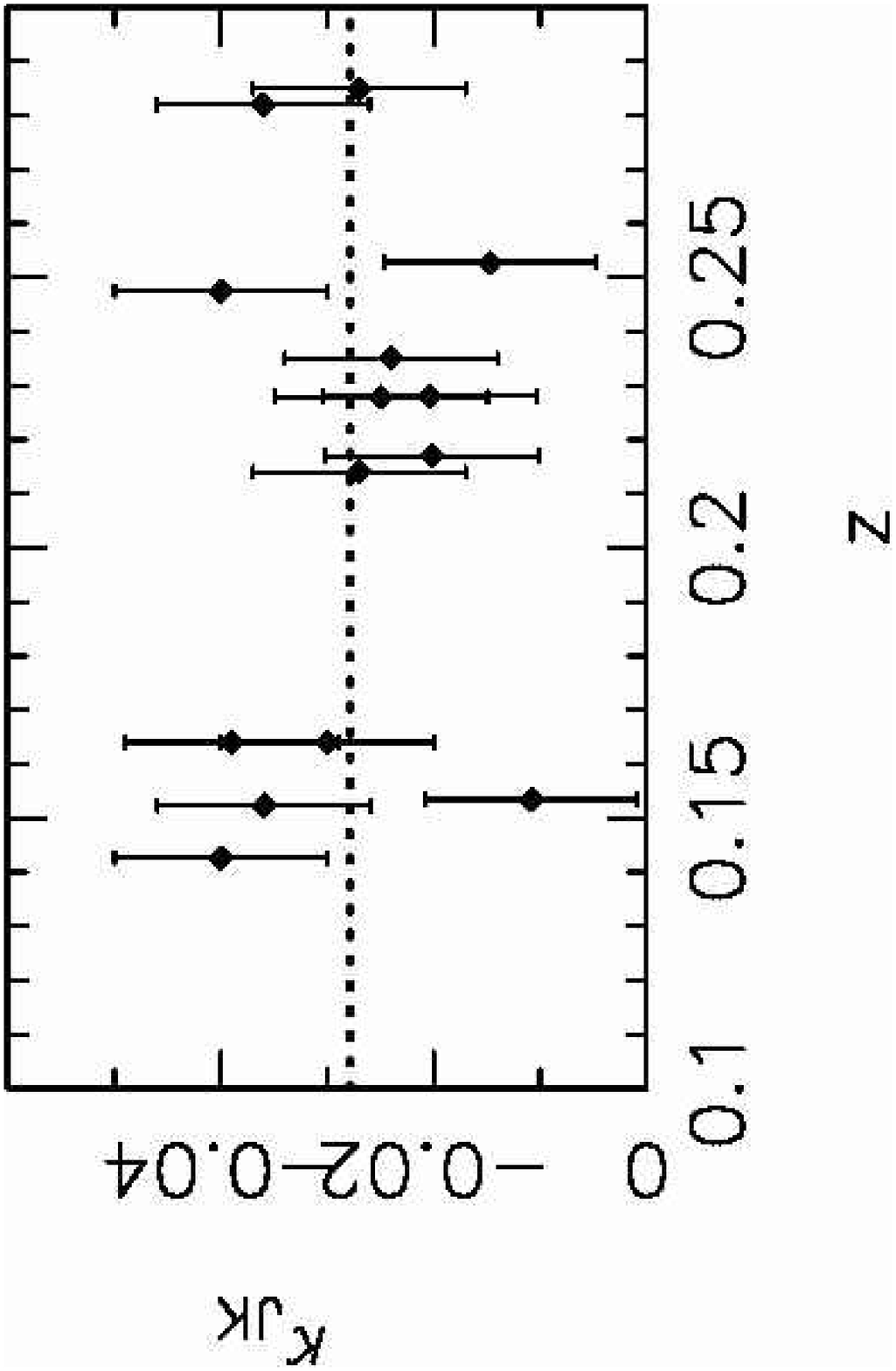}}}
\end{minipage}
\end{figure*}

$J$- and $K_S$-band data were obtained between March 2003 and April
2007 on the following near-infrared instruments: WIRC on the
Palomar-200in telescope, ISPI on the CTIO Blanco 4-m telescope,
and FLAMINGOS on the Kitt Peak (KPNO) 4-m telescope.  The properties
of the data are summarized in Table~\ref{images}.  Observations in
each filter comprised multiple frames with individual exposure times in
the range $30-120\secs$.  These images were combined using standard
IRAF reduction techniques, full details will be provided in Smith et
al. (in preparation), and flux-calibrated using the 2MASS source
catalog. A dithering box of 80\arcsec\ width was used in order to
improve the photometry of the extended envelope of the BCGs. The field
of view of the final dithered images is larger than 8$\times$8
arcmin$^2$, centred on the Brightest Cluster Galaxy (BCG). This
entirely covers the central $R<600$ kpc of each cluster (as measured
from the BCG). We used the SExtractor \citep{SExtractor} software to
derive a photometric catalog of {\tt MAGAUTO} magnitudes in the $K_S$
band, as well as the $J-K_S$ colours inside 4.0-\arcsec diameter
apertures.

\subsubsection{Photometric catalogs of cluster galaxies}
\label{nir}

In order to include galaxy-scale components in the strong-lens models
(\S\ref{slsub}), it is important to perform a systematic selection of
cluster members with their geometrical parameters (centroid
$\alpha_c$, $\delta_c$, ellipticity $e_c$ and position angle
$\theta_c$) and total photometry in the $K_S$ band. Indeed, $K_S$
magnitudes are a more accurate proxy for stellar mass when making the
assumption that light traces mass.

Cluster galaxies were selected from the near-infrared catalog using
the red sequence technique on a $J-K_S/K_S$ colour-magnitude diagram
\citep{Visvanathan}.  We adopted a liming magnitude of $K_S=20$ ({\tt
  MAGAUTO}) and a colour width of 0.3 magnitudes for the sequence
(above 3 times the photometric uncertainties).  The corresponding
diagrams and colour selection regions are presented in
Fig. \ref{cm}.  A final catalog of cluster members is obtained by
correlating the \emph{HST} and near-infrared photometric catalogs: in
the central region, the geometrical parameters ($\alpha_c$,
$\delta_c$, $e_c$, $\theta_c$) of cluster members are replaced by
their more accurate measurements from the \emph{HST} image.

As a sanity check, we plot the red sequence parameters used in our
photometric selection ($J-K_S$ colour at fixed $K_S$-band magnitude, and
linear slope $\kappa_{JK_S}$ of the colour-magnitude diagrams) as a
function of the cluster redshift (Fig. \ref{cm}, last 2 panels). The
correlation between $J-K_S$ at $K_S=18$ follows the expected colour of
elliptical galaxies (using the empirical model from \citet{cww},
while the slope $\kappa_{JK_S}$ is scattered around an average value
$\kappa_{JK_S}=-0.028$, compatible with similar measurements obtained
by \citet{Stott09} in this range of cluster redshifts.

We estimated $K_S^\star$, the $K_S$ band magnitude of an $L^\star$
galaxy, following the results from \citet{lin}.  They show that the
evolution of $K_S^\star$ with cluster redshift is best reproduced by a
single burst model at $z_{form}=1.5$ from \citet{BC03}, assuming a
Salpeter initial mass function and solar metallicity.  Their best fit
$K_S^\star(z)$ values are reported in the final column of Table
\ref{images}.  Our current selection down to $K_S=20$ enables us to
select cluster members down to $\sim0.1L^\star$.  Since the fraction
of fainter cluster members not included in our selection contribute to
less than 1\% of the total $K$-band luminosity, this selection does
not affect our results on the $K$-band properties in the entire
sample.

\subsection{Spectroscopic data}

\begin{table*}
\caption{\label{runs}Summary of the Keck/LRIS spectroscopic data.}
\begin{tabular}{lllcllcccl}
Cluster & Date & Config & $N_{\rm slits}$ &Grism & Grating$[\lambda_c]$ & Dichroic & $T_{\rm exp}$ & Seeing & Standard star\\
        &      &        &                &     &                      &          & (ks)  \\
\hline

A\,521  & Feb 2007 & MOS  & 18 & 400/3400 & 900/5500 [6320]& 560 & 3.8  & 1.1\arcsec & Feige 92 \\
                    &                   &            & & 400/3400 & 831/8200 [8100] & 560 & 3.6  & 1.1\arcsec & Feige 92 \\    
A\,611  &  Nov 2006 & LS      & 1 & 600/4000 & 400/8500 [7970] & 560 & 4.5  & 0.9\arcsec & Feige 110\\
                    &  Jan 2007 &  LS     &  1 & 600/4000& 400/8500 [7970] & 560 & 1.8  &  1.3\arcsec & Feige 34 \\
A\,773  &  Feb 2007& MOS  & 26 & 400/3400 &900/5500 [6400] &560 & 3.6  &1.0\arcsec & Feige 92\\
                   &                    &            & & 400/3400 &831/8200 [8070] &560 & 3.6  &1.0\arcsec & Feige 92\\
A\,868  & May 2008 & MOS & 36 & 300/5000 & 600/7500 [8100]& 680 & 5.4 & 0.7\arcsec & BD+33-2642\\
Z2701        & May 2008 & MOS & 22 & 300/5000 & 600/7500 [8100]& 680 & 3.6 & 1.0\arcsec & BD+33-2642\\
A\,1413 & Feb 2004 & LS       & 1 & 600/4000 & 400/8500 [7640]& 560 & 5.4  & 0.9\arcsec & \\
A\,1835 & March 2005 & MOS & 36& 400/3400 & 600/7500 [6850] & 560 & 7.2 & 0.7\arcsec & G138-31 \\
A\,2204 & May 2008 & MOS & 37 &300/5000 & 600/7500 [8100]& 680 & 3.6 & 0.7\arcsec & BD+33-2642\\
RXJ1720   & May 2008 & MOS & 21 &300/5000 & 600/7500 [8100]& 680 & 5.4  & 0.9\arcsec & BD+33-2642\\
                   &                     &           &       &400/3400 & 600/7500 [8100]& 680 & 5.4  & 0.9\arcsec & BD+33-2642\\
RXJ2129   & May 2008 & MOS &  22 &300/5000 & 600/7500 [8100]& 680 & 5.4& 1.1\arcsec & BD+33-2642\\
\hline
\end{tabular}
\end{table*}

\subsubsection{Observing strategy}

We used the Low Resolution Imager and Spectrograph (LRIS,
\citealt{lris}) on the Keck-I telescope to perform long-slit and
multi-slit observations of the clusters.  The spectroscopic data used
in the current paper is the outcome of 6 different observing runs
between 2004 and 2008, which are summarized in Table~\ref{runs}.

We designed multi-slit masks containing $\sim30$ slits of $1.0\arcsec$
width to include as many of the candidate multiple systems as
possible, with a few tilted slits following the geometry of long
arcs. In the case of fainter or less reliable identifications, several
images of the same system were observed in separate slits. Additional
slits in the mask included faint galaxies with unknown redshifts,
bright infrared sources selected at $24\mic$, or cluster members.

The dichroic/grism/grating combinations listed in Table~\ref{runs} to
ensure a high throughput over the wavelength range 3500 to 9500 \AA\
and good spectral resolution (Table~\ref{runs}) in the red to resolve
OH skylines as well as $[{\rm OII}]$ emission line doublets in the
redshift range $0.85-1.50$. The total exposure time varied between 1
and 3 hours per mask, depending on the seeing conditions and the
detection of strong emission lines when looking over a first reduction
of the data at the telescope.

\subsubsection{Data reduction}
\label{zmeasure}

\begin{table*}
\caption{\label{mult}Properties of the multiple imaged systems. Redshift values quoted with brackets are predictions from 
the lensing model. The last two columns give the observed \emph{HST} magnitudes and linear magnification factor predicted by the lens model.}
\begin{tabular}{lrrcllllllll}
Cluster / Multiple & $\alpha$ & $\delta$ & $z$ & Reference & \emph{HST} & $\mu$ (mags) & $\mu$ (linear) \\
\hline 
A\,521        &           &            &                &         & (F606W)       &\\
1.1$^{a}$        & 73.526932 & -10.223421 & 1.043$\pm$0.01 & $[OII],CII,FeII$ &  $21.49\pm0.04$ & 1.76$\pm$0.10  & 5.1$\pm$0.5 \\
1.2              & 73.527544 & -10.222524 &  1.043$\pm$0.01   &   $CII,FeII$      &  $21.32\pm0.04$ &  1.71$\pm$0.10\ & 4.8$\pm$0.4\\
1.3              & 73.529391 & -10.221431 & 1.043$\pm$0.01 & $[OII]$ &  $21.57\pm0.20$ &  0.98$\pm$0.10 & 2.5$\pm$0.2
\smallskip\\
A\,611        &   &                    &     & &   (F606W) & \\
1.1$^{a}$        & 120.23219 &  36.061593 &       &           & $24.58\pm0.03$ & 1.38$\pm$0.11 & 3.6$\pm$0.4\\
1.2              & 120.24171 &  36.055210 &  $2.06\pm0.02$ & $CIII,CIV$& $23.87\pm0.05$ & 2.52$\pm$0.30 & 10.2$\pm$2.8\\
1.3              & 120.24103 &  36.058273 &  $2.06\pm0.02$ & $CIII,CIV$& $22.80\pm0.10$ & 1.95$\pm$0.18 & 6.0$\pm$1.0\\
1.4              & 120.23555 &  36.054247 &       &           & $24.98\pm0.04$ &  0.79$\pm$0.06 & 2.1$\pm$0.1\\
1.5              & 120.23589 &  36.054887 &       &           & $25.46\pm0.05$  & 1.0$\pm$0.07  & 2.5$\pm$0.2 
\smallskip\\
2.1              & 120.23716 &  36.061183 &  $0.908\pm0.005$ & $[OII]$ & $23.26\pm0.02$ &  2.47$\pm$0.29 & 9.7$\pm$2.6\\
2.2              & 120.24046 &  36.059825 &  $0.908\pm0.005$ & $[OII]$ & $23.26\pm0.09$ & 1.10$\pm$0.08 & 2.8$\pm$0.2\\
2.3              & 120.24206 &  36.057539 &  $0.908\pm0.005$ & $[OII]$ & $23.17\pm0.14$ &  2.48$\pm$0.29 & 9.8$\pm$2.6
\smallskip\\
3.1              & 120.23554 &  36.060897 &  $[2.5^{+0.5}_{-0.2}]$ & & $24.42\pm0.03$ & 4.08$\pm$1.27 & 42$\pm$15\\
3.2              & 120.23739 &  36.060653 &       & & $25.05\pm0.06$ &  2.04$\pm$0.19 & 6.5$\pm$1.2\\
3.3              & 120.24311 &  36.053638 &       & & $25.94\pm0.08$ &  1.38$\pm$0.11 & 3.6$\pm$0.4\\
3.4              & 120.23407 &  36.055807 &       & & $25.63\pm0.07$ &  1.39$\pm$0.11 & 3.6$\pm$0.4
\smallskip\\
4.1              & 120.24182 &  36.056242 & $2.59\pm0.01$ & $Ly_\alpha$ & $27.14\pm0.19$ &  4.35$\pm$1.62 & 55$\pm$17\\
4.2              & 120.23193 &  36.062088 &      &             & $27.67\pm0.19$ &  1.26$\pm$0.09 & 3.2$\pm$0.2 
\smallskip\\
A\,773        &   &                    &     & &   (F702W) &  \\
1.1              & 139.48925 &  51.729656 & 2.300$\pm0.005$ & $Ly_\alpha,SiII,SiIV,CIV$ & $22.95\pm0.02$ & 2.69$\pm$0.35 & 11.9$\pm$3.8 \\
1.2              & 139.48920 &  51.724870 & 2.300$\pm0.005$ & & $23.03\pm0.02$ &  3.0$\pm$0.5 & 16.0$\pm$6.9\\
1.3              & 139.48351 &  51.741593 &       & & $23.47\pm 0.02$ & 1.57$\pm$0.13 & 4.2$\pm$0.5
\smallskip\\
2.1              & 139.46920 &  51.732350 & 3.84$\pm$0.01  & $Ly_\alpha$ & $23.85\pm0.06$ & 2.42$\pm$0.27 & 9.3$\pm$2.3\\
2.2              & 139.48762 &  51.716864 &       &             & $23.58\pm0.06$ & 1.18$\pm$0.09 & 3.0$\pm$0.3 
\smallskip\\
3.1              & 139.49535 &  51.728050 & 1.010$\pm$0.005   & $[OII]$ & $23.09\pm0.05$ & 2.0$\pm$0.2 & 6.1$\pm$1.0\\
3.2              & 139.49501 &  51.728985 & 1.010$\pm$0.005  & $[OII]$ & $21.54\pm0.05$ & 2.0$\pm$0.2 & 6.1$\pm$1.0  
\smallskip\\
A\,868               &           &            &       & &   (F606W) &        \\
1.1              & 146.36459 & -8.6481919 & 0.551$\pm0.002$ & $[OII]$,$[OIII]$ & $22.60 \pm 0.03^{(b)}$ & 3.29$\pm$0.61& 36$\pm$7.8\\ 
1.2              & 146.36474 & -8.6490159 & 0.551$\pm0.002$ & $[OII]$,$[OIII]$ & $22.60 \pm 0.03^{(b)}$ & 2.28$\pm$0.24 &19$\pm$4.7\\
1.3              & 146.36554 & -8.6502853 & 0.551$\pm0.002$ &                  & $23.50\pm0.05$ &  2.64$\pm$0.34 & 23.$\pm$4.1
\smallskip\\
Z2701            &           &            &       & &   (F606W) &        \\
1.1              & 148.20940 &  51.885620 & 1.163 & $[OII]$& $23.73\pm0.03$ & 5.32$\pm$0.98 & 20.7$\pm$6.1\\
1.2              & 148.20708 &  51.883841 &       & & $24.22\pm0.03$ &  3.90$\pm$0.88 & 8.2$\pm$1.8\\
1.3              & 148.20087 &  51.882325 & 1.163 & $[OII]$& $24.34\pm0.03$ & 2.62$\pm$0.33 & 11.4$\pm$3.7
\smallskip\\
2.1              & 148.21056 &  51.884031 & $[2.5^{+0.2}_{-0.3}]$  & & $25.05\pm0.08$ &  3.73$\pm$0.92 & 130$\pm$60\\
2.2              & 148.20988 &  51.883375 &   & & $24.99\pm0.08$ &  3.90$\pm$0.88 & 36.$\pm$9\\
2.3              & 148.20083 &  51.880880 &   & & $25.52\pm0.05$ &  2.62$\pm$0.33 & 11.2$\pm$3.4 
\smallskip\\
A\,1413       &   &                    &      & &   (F775W) & \\
1.1              &  178.82496 & 23.412185 & 2.726$\pm$0.003 & $Ly_\alpha,SiII$ & $23.16\pm0.06$ &  3.73$\pm$0.92\ & 31.$\pm$6.\\
1.2              &  178.82328 & 23.412089 & 2.726$\pm$0.003 & $Ly_\alpha,SiII$ & $22.79\pm0.08$ & 5.00$\pm$2.95 & 100$\pm$27
\smallskip\\
2.1              &  178.82855 & 23.399205 & 2.030$\pm$0.004 &  $Ly_\alpha,SiII$ & $22.90\pm0.16$& 2.37$\pm$0.26 & 8.9$\pm$2.1 \\
2.2              &  178.82591 & 23.398715 & 2.030$\pm$0.004 &  $Ly_\alpha,SiII$ & $22.99\pm0.23$& 2.94$\pm$0.44 & 15$\pm$6\\
2.3              &  178.81985 & 23.399441 &      & & $23.82\pm0.09$&  2.50$\pm$0.30 & 10$\pm$2.8
\smallskip\\
3.1              &  178.82933 & 23.406898 & 1.20$\pm$0.01 & $[OII]$ & $23.03\pm0.06$ &  1.28$\pm$0.10 & 3.3$\pm$0.3\\
3.2              &  178.82391 & 23.407396 & 1.20$\pm$0.01 & $[OII]$ & $23.64\pm0.21$ &  1.54$\pm$0.12 & 4.1$\pm$0.5\\
3.3              &  178.81981 & 23.405955 & 1.20$\pm$0.01 & $[OII]$ & $22.84\pm0.13$ &  1.33$\pm$0.10 & 3.4$\pm$0.3
\smallskip\\
4.1              &  178.82940 & 23.409414 & $[2.9^{+0.3}_{-0.2}]$ & & $25.47\pm0.05$&1.30$\pm$0.10& 3.3$\pm$0.3\\
4.2              &  178.82263 & 23.409657 &  & & $24.50\pm0.03$& 2.07$\pm$0.20 & 6.7$\pm$1.2\\
4.3              &  178.82007 & 23.408814 &  & & $24.57\pm0.03$& 1.95$\pm$0.18 & 6.0$\pm$1.0\\
\hline
\end{tabular}

$^{(a)}$ location of the brightest knot in the image.
$^{(b)}$ Photometry is the sum of 1.1 and 1.2 images
\end{table*}

\begin{table*}
\noindent{\bf Table~\ref{mult}} Continued.
\begin{tabular}{lrrclllllll}
Cluster / Multiple & $\alpha$ & $\delta$ & $z$ & Reference & \emph{HST} & $\mu$ (mags) & $\mu$ (linear)\\
\hline 
A\,1835       &   &                     &     & &   (F850LP) & \\
1.1              &  210.26573 & 2.8706065 & $[2.5^{+0.2}_{-0.2}]$ & & $24.28\pm0.12$  & 2.15$\pm$0.22 & 7.2$\pm$1.5\\
1.2              &  210.26403 & 2.8693007 &  & & $23.64\pm0.20$  &4.33$\pm$1.60 & 54$\pm$16\\
1.3              &  210.24740 & 2.8698812 &      & & $25.12\pm0.25$  & 0.83$\pm$0.06 & 2.1$\pm$0.1
\smallskip\\
2.1              &  210.26614 & 2.8741625 & $[1.3^{+0.1}_{-0.1}]$ & & $23.28\pm0.28$  &  1.60$\pm$0.13 & 4.4$\pm$0.5\\
2.2              &  210.26358 & 2.8714121 &         & & $22.87\pm0.29$  & 0.42$\pm$0.04 & 1.5$\pm$0.05\\
2.3              &  210.24824 & 2.8717531 &         &   &   $25.40\pm0.30$ & 0.85$\pm$0.07 & 2.2$\pm$0.1 
\smallskip\\
3.1              &  210.26326 & 2.8849801 & $[3.0^{+0.2}_{-0.1}]$ & & $25.63\pm0.07$  &  2.31$\pm$0.25 & 8.4$\pm$1.9\\
3.2              &  210.26278 & 2.8852175 &         & & $25.28\pm0.06$  & 2.66$\pm$0.34 & 11.6$\pm$3.6\\
3.3              &  210.24374 & 2.8775381 &         & & $25.87\pm0.07$  & 1.87$\pm$0.17 & 5.6$\pm$0.9
\smallskip\\
4.1              &  210.26382 & 2.8846758 & $[2.3^{+0.1}_{-0.1}]$ & & $24.70\pm0.04$  & 1.96$\pm$0.18 & 6.1$\pm$1.0\\
4.2              &  210.26073 & 2.8858319 &         & & $25.41\pm0.07$  & 3.15$\pm$0.54 & 17.4$\pm$8.0\\
4.3              &  210.24435 & 2.8781489 &         & & $25.08\pm0.13$  & 0.83$\pm$0.06 & 2.1$\pm$0.1 
\smallskip\\
5.1              &  210.25984 & 2.8824075 & 
2.6 & $Ly_\alpha$ (a)& $23.44\pm0.03$ & 0.67$\pm$0.05 & 1.8$\pm$0.1\\
5.2              &  210.24484 & 2.8721518 & 
2.6 & $Ly_\alpha$ (a)& $23.90\pm0.02$ & 0.69$\pm$0.06 & 1.9$\pm$0.1\\
5.3              &  210.25921 & 2.8792670 &         & & $26.04\pm0.18$ &
\smallskip\\
6.1              &  210.26165 & 2.8775836 & $[3.6^{+1.2}_{-1.4}]$ & & $25.77\pm0.09$ &  0.99$\pm$0.07 & 2.5$\pm$0.2\\
6.2              &  210.26131 & 2.8777493 &         & & $25.43\pm0.06$ &  0.80$\pm$0.06 & 2.1$\pm$0.2
\smallskip\\
7.1              &  210.25395 & 2.8731863 & 2.070$\pm$0.004 & Ly$_\alpha,SiII, SiIV,CIV$& $22.28\pm0.04$ &  4.86$\pm$2.59& 88$\pm$30\\
7.2              &  210.25421 & 2.8803387 &      &            & $23.96\pm0.05$ & 4.00$\pm$1.18 & 40$\pm$21\\
7.3              &  210.27108 & 2.8801023 &      &            & $25.02\pm0.05$ &  0.86$\pm$0.07 & 2.2$\pm$0.1
\smallskip\\
A\,2204       &            &           &      &       &  (F606W) & \\
1.1              &  248.19665 & 5.5781303 & 1.06$\pm$0.01 & $[OII]$ (b) & $23.60\pm0.12$& 1.25$\pm$0.09 & 3.2$\pm$0.3 \\
1.2              &  248.19457 & 5.5689343 & 1.06$\pm$0.01 & $[OII]$     & $22.61\pm0.03$ &  0.85$\pm$0.06 & 2.2$\pm$0.1\\
1.3              &  248.19603 & 5.5779196 &     &             & $23.70\pm0.15$&  0.53$\pm$0.05 & 1.6$\pm$0.08 
\smallskip\\
RXJ1720          &            &           &       &       &  (F606W) & \\
1.1              &  260.04314 & 26.624351 & 2.136$\pm0.005$ & $MgII$ & $22.74\pm0.05$ (c)& $3.16\pm0.54$ & $18.4\pm9.1$\\
1.2              &  260.04247 & 26.624155 & 2.136$\pm0.005$ & $MgII$ & $22.74\pm0.05$ (c)& 1.54$\pm0.12$ & 4.1$\pm0.5$\\
1.3              &  260.03779 & 26.626930  &       & & $24.85\pm0.08$ & 0.91$\pm$0.07 & 2.3$\pm$0.2\\
1.4              &  260.04285 & 26.626111 &       & & $24.41\pm0.07$ & 1.33$\pm$0.10 & 3.4$\pm$0.3
\smallskip\\
RXJ2129          &            &           &       &       &  (F606W) & \\
1.1              &  322.42040 & 0.088305 & 1.965$\pm$0.005 & $SiIV,SiII,CIV$ & $23.41\pm0.03$ &  4.30$\pm$1.56 & 52$\pm$17\\
1.2              &  322.42018 & 0.089722 & 1.965$\pm$0.005 & $SiIV,SiII,CIV$ & $23.05\pm0.02$ &  2.33$\pm$0.25 & 8.6$\pm$2.0\\
1.3              &  322.41798 & 0.093250 &       & & $24.74\pm0.06$ &  2.23$\pm$0.23 & 7.8$\pm$1.6\\
\hline
\end{tabular}

$^{(a)}$ Dannerbauer et al., in preparation
$^{(b)}$ See also \citet{Wilman}
$^{(c)}$ Photometry is the sum of 1.1 and 1.2 images

\end{table*}

The data were reduced using the Python version of the \citet{Kelson}
reduction scripts, which offer the advantage of processing the images
in their distorted framework. This helps to reduce noise correlations,
in particular for the case of tilted slits. We performed standard
reduction steps for bias removal, flat-field correction, wavelength
calibration, sky subtraction, and cosmic-ray rejection, and used
observations of spectroscopic standard stars obtained on the same
nights to derive the flux calibration.

The extracted spectra of objects used in the present paper are shown
in Fig. \ref{spectra} (appendix).  These spectra show Lyman-$\alpha$
(in absorption or emission) and/or additional ultraviolet absorption
lines, or the resolved $[{\rm OII}]$ doublet.  For some faint multiple
systems observed in different slits, redshift measurements were
derived after stacking the relevant exposures.  The average redshift
value is obtained from the peaks of the main spectral features
identified, while the corresponding error is taken from the spectral
dispersion.  Additional uncertainties generated by the accuracy of the
relative and absolute wavelength calibrations, about 1.1 and 1.5 \AA,
respectively, were quadratically added to yield the final redshift
errors.  The redshifts are listed in Table~\ref{mult}.

\subsection{X-ray observations}
\label{xprop}

X-ray properties of the LoCuSS clusters have been determined for 18/20
clusters from \textit{Chandra} archival observations, which were
reduced and analyzed as described in \citet{Sanderson09b} following
the methods developed by \citet{Sanderson09c}.  Annular gas
temperature and density profiles were derived and fitted with the
phenomenological cluster model of \citet{Ascasibar}, to determine the
total cluster mass profile.  These models are based on the
\citet{Hernquist} profile, which has the advantage of yielding
convergent three- and two-dimensional mass measurements.  The latter
is particularly helpful when comparing lensing and X-ray based
measurements of the projected mass within $R<250\kpc$ (see
\S\ref{xprop}).

The X-ray models were also used to measure the gradient of the
logarithmic gas density profile at $0.04r_{500}$ ($\alpha$;
\citealt{Vikhlinin07}), which quantifies the strength of cooling in
the core \citep{Vikhlinin07,Sanderson09b}.  Core-excluded mean
temperatures for the clusters were also measured in the range
$0.15-0.2r_{500}$, according to the procedure outlined in
\citet{Sanderson09a}.  We classify clusters as hosting a cool core if
the slope of the gas density profile is $\alpha<-0.85$.  This matches
the range of $\alpha$ displayed by clusters that contain an H$\alpha$
emitting BCG in \citet{Sanderson09b}.  Under this definition, 8/18
clusters as cool core clusters, corresponding to $44\pm14\%$ where the
error bar is a binomial uncertainty following \citet{Gehrels86}.
We also classify clusters as ``disturbed'' if the offset between the
X-ray centroid and the BCG is $>0.01r_{500}$, this identifies all
except 3 of the clusters containing an H$\alpha$ emitting BCG in
\citet[][see their Fig.~5]{Sanderson09b}.  This yields a fraction of
$67^{+12}_{-15}\%$ of the sample that are classified as undisturbed.
All cool core clusters are also classified as undisturbed, however 4
non-cool core clusters are also classified as undisturbed.  The
resulting X-ray properties are given in Table~\ref{xray}.

\begin{table*}
\caption{\label{xray} Summary of Chandra X-ray properties for the Keck
  sample (1st half) and the extended sample (2nd half)} 
\begin{tabular}{lrrcrrccrc}
Cluster    &  $L_X^a$ & ID$^b$& Temperature$^c$ & $r_{500}$  & $M_X(R<250\kpc)$ & $\alpha^d$ & Cool & Offset$^e$ & Disturbed?\\
           &        &         & (keV)          & (kpc)     &  $(10^{13}\Msol)$  &            & core? & (kpc) & \\
\hline 
 A\,521    & 9.45 & 901  & $7.08^{+1.15}_{-1.53}$ & $921\pm26$ & $8.5\pm0.5$ & $-0.10\pm0.29$ & No & 37 & Yes\\
 A\,611    & 8.05 & 3194 & $7.94^{+1.07}_{-1.22}$  & $1342\pm140$ & $20.6\pm1.9$ & $-0.70\pm0.04$    & No & 1 & No\\
 A\,773    & 7.74 & 5006 & $7.50^{+0.87}_{-1.01}$ & $1358\pm45$ & $18.0\pm1.1$ & $-0.40\pm0.05$  & No & 20 & Yes\\
 Z\,2701   & 6.32 & 3195 & $5.08^{+0.41}_{-0.43}$  & $1271\pm300$ & $15.0\pm4.1$ & $-0.88\pm0.12$      & Yes & 3 & No\\
 A\,1413   & 7.80 & 5003 & $6.90^{+0.32}_{-0.35}$  & $1296\pm28$ & $18.4\pm0.5$ & $-0.68\pm0.04$     & No & 2 & No\\
 A\,1835   & 22.80& 6880 & $9.82^{+0.43}_{-0.44}$  & $1506\pm57$ & $24.6\pm1.0$ & $-1.17\pm0.05$     & Yes & 5 & No\\
 A\,2204   & 12.57& 7940 & $9.64^{+0.54}_{-0.54}$  & $1420\pm59$ & $23.8\pm1.4$ & $-1.22\pm0.04$     & Yes & 8 & No\\
 RX\,J1720 & 9.54 & 4361 & $7.96^{+0.97}_{-0.82}$  & $1349\pm71$ & $19.2\pm1.1$ & $-1.06\pm0.04$     & Yes & 6 & No\\
 RX\,J2129 & 11.00& 552  & $8.27^{+1.48}_{-1.87}$ & $1155\pm33$ & $16.1\pm1.1$ & $-1.09\pm0.03$     & Yes & 6 & No\\
 A\,868    & 3.46 & N/A &  & & & & & &\\
\hline A68 & 8.81 & 3250 & $8.89^{+1.76}_{-3.37}$ & $941\pm56$ & $11.9\pm1.8$ & $-0.25\pm0.02$ & No & 52 & Yes\\
 A\,383    & 5.27 & 2320 & $5.01^{+0.43}_{-0.47}$ & $1014\pm47$ & $12.1\pm1.0$ & $-1.09\pm0.03$     & Yes & 2 & No\\
 A\,963    & 6.16 & 903  & $6.73^{+0.52}_{-0.77}$ & $1210\pm39$ & $16.9\pm0.7$ & $-0.68\pm0.05$     & No & 6 & No\\
 A\,1201   & 3.72 & 4216 & $5.56^{+0.61}_{-0.8}$ & $1016\pm30$ & $10.7\pm0.6$ & $-0.65\pm0.13$     & No & 11 & Yes\\
 A\,1689   & 16.27& 5004 & $8.86^{+0.68}_{-0.88}$ & $1451\pm23$ & $25.7\pm1.0$ & $-0.77\pm0.03$     & No & 3 & No\\
 A\,2218   & 5.51 & 1666 & $7.17^{+0.51}_{-0.58}$ & $1216\pm34$ & $16.0\pm0.6$ & $-0.33\pm0.03$  & No & 41 & Yes\\
 A\,2219   & 12.07& 896  & $11.52^{+0.8}_{-0.9}$ & $1786\pm149$ & $26.9\pm1.4$ & $-0.31\pm0.03$  & No & 28 & Yes\\
 A\,2390   & 12.69& 4193 & $9.78^{+0.55}_{-0.56}$ & $1437\pm68$ & $20.0\pm1.0$ & $-0.94\pm0.02$      & Yes & 2 & No\\
 A\,2667   & 15.78& 2214 & $5.66^{+0.62}_{-0.85}$ & $1243\pm115$ & $14.1\pm1.6$ & $-0.89\pm0.05$     & Yes & 3 & No\\
 A\,1703   & 8.66 & N/A &  & & & & & &\\
\hline
\end{tabular}

  $^a$~X-ray luminosity in the $0.1-2.4\keV$ band from \emph{ROSAT}
  \citep{Ebeling98,Ebeling00,Boehringer04}.
  $^b$~\textit{Chandra} observation identifier. 
  $^c$~Measured between $0.15$ and $0.2r_{500}$ \citep[see][]{Sanderson09a}. 
  $^d$~Logarithmic slope of the gas density profile at $0.04r_{500}$ \citep[see][]{Vikhlinin07,Sanderson09c}.
  $^e$~X-ray centroid/BCG offset from \citet{Sanderson09b}. Errors are $1\sigma$.

\end{table*}

\section{Strong-lensing analysis}
\label{sl}

We now describe the multiply-imaged galaxies used to constrain mass
models of the 10 new strong-lensing clusters (\S\ref{specz}), and the
methods used to construct the models themselves (\S\ref{slsub}).

\subsection{Multiple images and spectroscopic redshifts}\label{specz}

The multiple-image constraints used for each cluster are described
below, and summarized in Table~\ref{mult}, including their astrometry,
photometry and available spectroscopic redshifts.  The locations of
the multiple images are also marked in Fig.~\ref{cluster}.  The naming
convention for multiple images in a given cluster is the following:
(1.1, 1.2, 1.3, ... ) are individual images in system 1, (2.1, 2.2,
2.3,...) in system 2, and so on.

\subsubsection*{A\,521}

The WFPC2 observations revealed a triply-imaged face-on spiral just
$6\arcsec$ from the BCG, and containing many individual knots of star
formation. The clear symmetry between these knots (marked A to E in
Fig.~\ref{cluster}) confirms a strong-lensing system of 3 images,which
we measure to be at $z=1.034$ from $[OII]$ emission and UV absorption
lines.  We included the 3 brightest knots of each image as 9
individual constraints for the mass modeling.  The system was
previously identified by \citet{Maurogordato96} using CFHT
ground-based imaging.  However the poor angular resolution of their
data prevented them from identifying it as strongly-lensed, and
subsequent spectroscopy was also inconclusive \citep{Maurogordato}.

\subsubsection*{A\,611}
We identify 4 systems, the first of which (1.1 to 1.5) clearly appears
as 5 images of identical clumpy morphology, each of them composed of 5
individual knots (A to E) of star formation.  We obtained a
spectroscopic redshift for 1.2 and 1.3, showing identical features of
a $z=2.06$ galaxy.  The locations of the radial counter-images (1.4
and 1.5) are confirmed by a preliminary mass model.  The second system
(2.1 to 2.3) appears as a giant arc to the North-East of the BCG, for
which we derive $z=0.908$ from strong $[OII]$ emission. This giant arc
is clearly affected by the presence of cluster substructure, with
significant changes of curvature close to individual galaxies.  The
third system (3.1 to 3.4) was identified thanks to the symmetry
between 3.1 and 3.2.  The mass model predicts counter images that we
identify as 3.3 and 3.4.  Finally, a very faint arc (4.1)
serendipitously falling in one of the slits was confirmed at $z=2.54$
with strong Lyman-$\alpha$ emission. The mass model confirmed it is
located on the $z=2.5$ critical line, and predicts the counter image
that we identify as 4.2.

\subsubsection*{A\,773}

We identify 3 multiply imaged systems in this cluster.  System 1 is a
triply-imaged system with a clear symmetry between components 1.1 and
1.2.  We derived a redshift $z=2.3$ for both images and identify the
counter-image 1.3 with help from the mass model. The second system
(2.1 and 2.2) is a radial image and counter-image at $z=3.84$ from strong Ly$_\alpha$
emission. Finally, we found a spectroscopic redshift $z=1.0$ for the third
  system (3.1 and 3.2) curving around one of the cluster galaxies.

\subsubsection*{A\,868}

A giant arc containing three bright images (1.1 to 1.3) is clearly
found around the second brightest galaxy to the north-east of the
BCG. We derive its redshift at $z=0.551$ from the combination of
$[OII]$ and $[OIII]$ emissions. The curvature of this arc suggests it
is strongly affected by a secondary mass component, as well as the
main component centred on the BCG.

\subsubsection*{Z2701}

We identify a first system (1.1 to 1.3) as three compact images
$\sim10\arcsec$ from the BCG, confirmed to be at $z=1.163$.  A second
system is formed by two faint symmetric arcs (2.1 and 2.2), for which
the mass model predicts a counter-image identified as 2.3.

\subsubsection*{A\,1413}

The two-filter ACS data enable 4 multiply-imaged galaxies to be
identified from their colours and morphologies, three of them having
spectroscopic redshifts between $z=1.17$ and $z=2.77$, from $[OII]$ or
Lyman-$\alpha$ in emission.  System 1 is a bright fold arc to the
north, where we identify two bright knots as 1.1 and 1.2. The three
other systems are triply-imaged with clear mirror symmetry.

\subsubsection*{A\,1835}

Several studies have previously identified strongly-lensed arcs in
this cluster \citep[Sm05]{Schmidt01,Sand05} however to date
spectroscopic redshifts have not been measured for any of these
background galaxies, thereby limiting the precision of previous
attempts to model the mass distribution in this cluster.  We report
redshift measurements for two systems: one of them is a triple system
(labelled 5) discovered during a narrow-band search for Lyman-$\alpha$
emitters at $z=2.5$ (Dannerbauer et al., in preparation), as two
objects with similar morphologies and strong Ly-$\alpha$ emission. A
radial counter image (5.3) is predicted close to the BCG.  The second
is a pair of merging images (system 7) measured at $z=2.07$, for which
we identify two counter images 7.2 and 7.3.  A 4th image 7.4 is
predicted very close to the BCG but the light of the BCG prevents a
reliable identification.  Finally, we checked that the submillimetre
sources J1/J2 discussed by \citet{smail05} at $z=2.56$ are indeed
predicted by the mass model to be singly-imaged.

\subsubsection*{A\,2204}

The redshift of a radial arc (1.1 and 1.2) was measured to be $z=1.0$
from strong $[OII]$ emission during an IFU observation of the central
galaxy with the VIMOS instrument \citep{Wilman}.  We confirm this
redshift independently, and find the same redshift for 1.3, which our
mass model predicted to be the counter image.

\subsubsection*{RX\,J1720}

This cluster is constrained by a giant arc (1.1 and 1.2) curving
around the BCG, with a spectroscopic redshift $z=2.136$ from the
magnesium doublet.  We identify two counter images (1.3 and 1.4) with
help from the mass model.

\subsubsection*{RX\,J2129}

Two compact sources with mirror symmetry (1.1 and 1.2) have been
identified $z=1.965$ from UV absorption lines, and the location of the
counter image 1.3 was identified with the mass model.

\begin{figure*}
\begin{minipage}{7cm}
\includegraphics[width=7cm,angle=0]{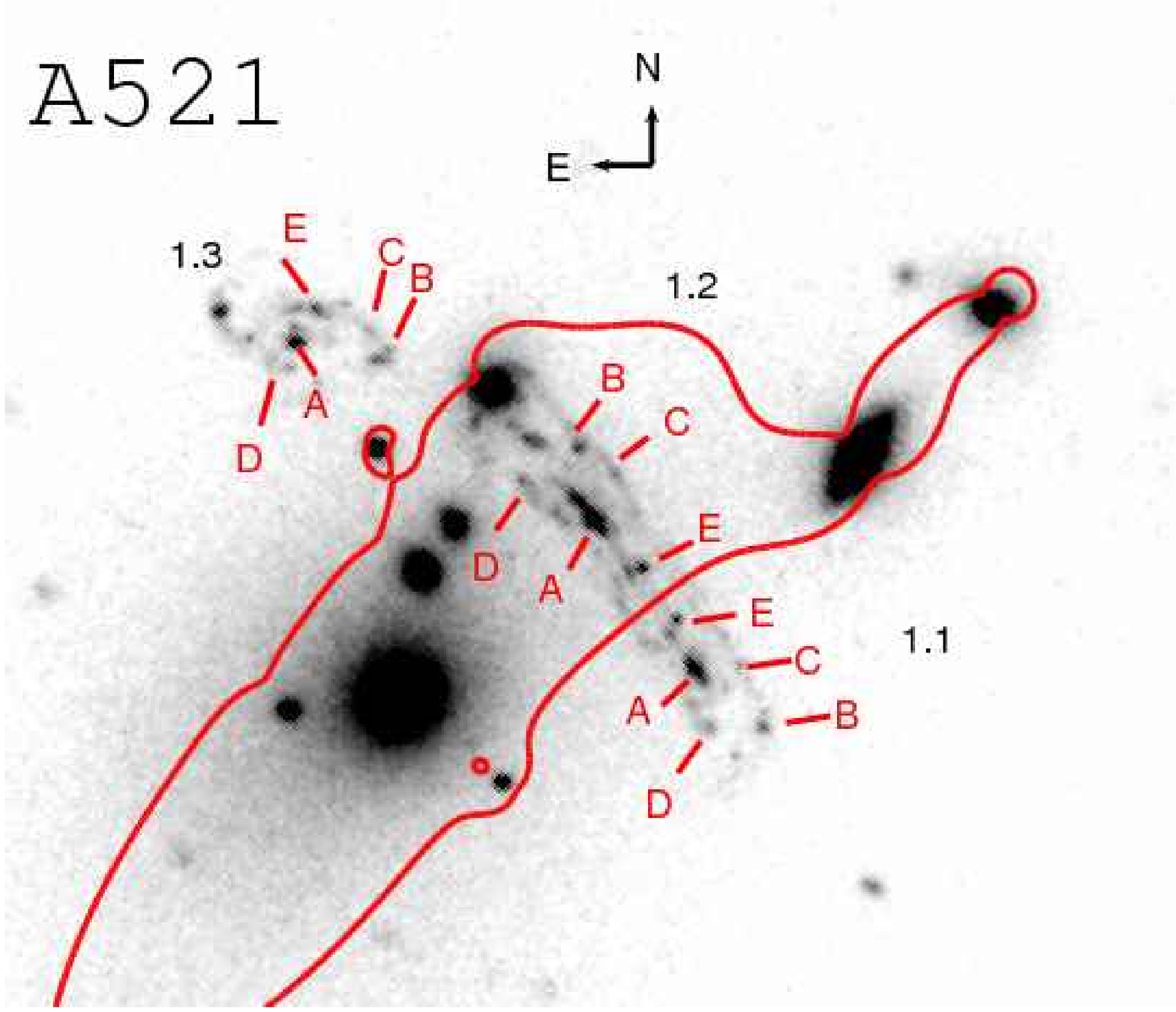}
\end{minipage}
\begin{minipage}{7.5cm}
\includegraphics[width=7.5cm,angle=0]{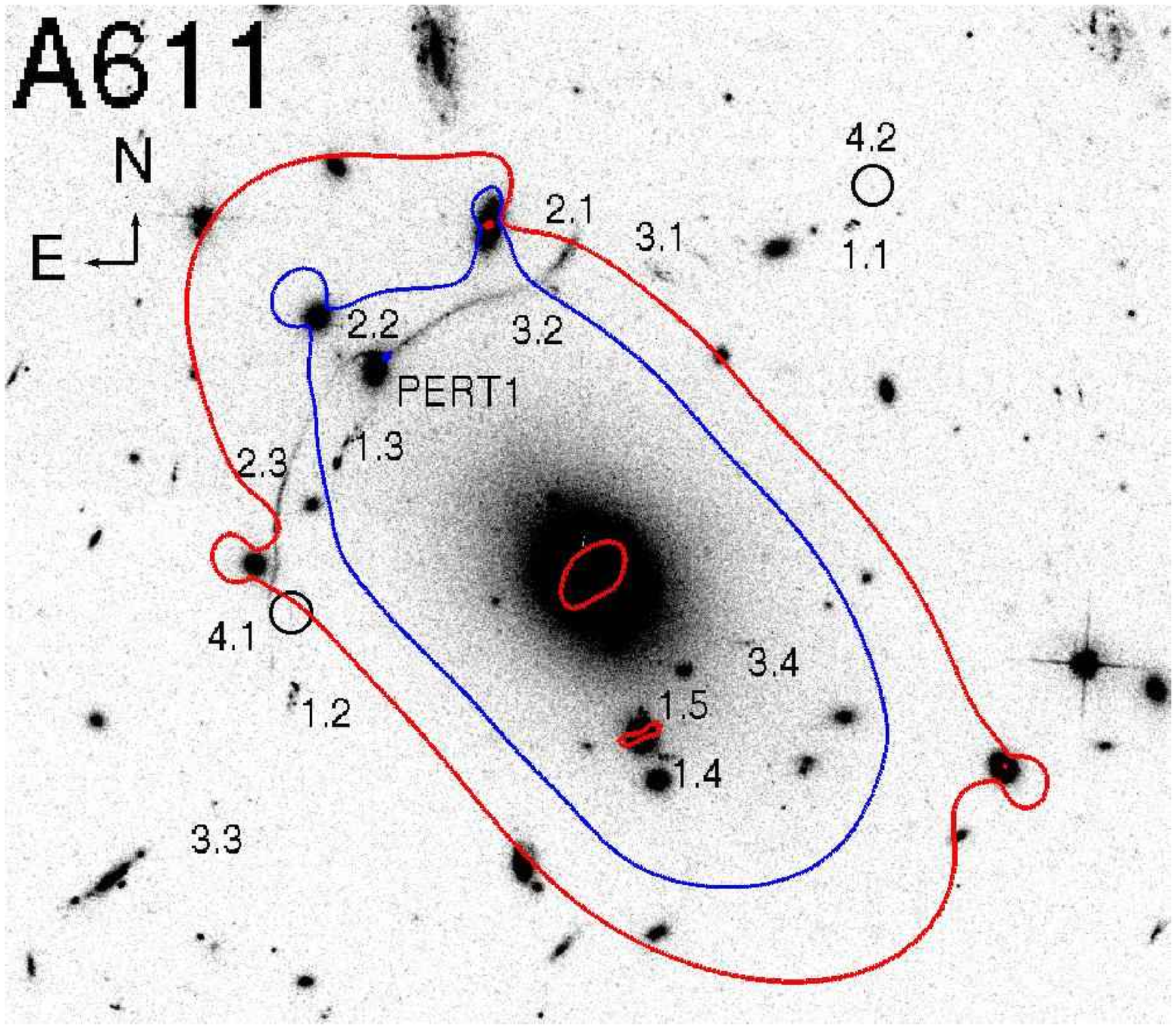}
\end{minipage}
\begin{minipage}{2.8cm}
\includegraphics[height=2.8cm,angle=0]{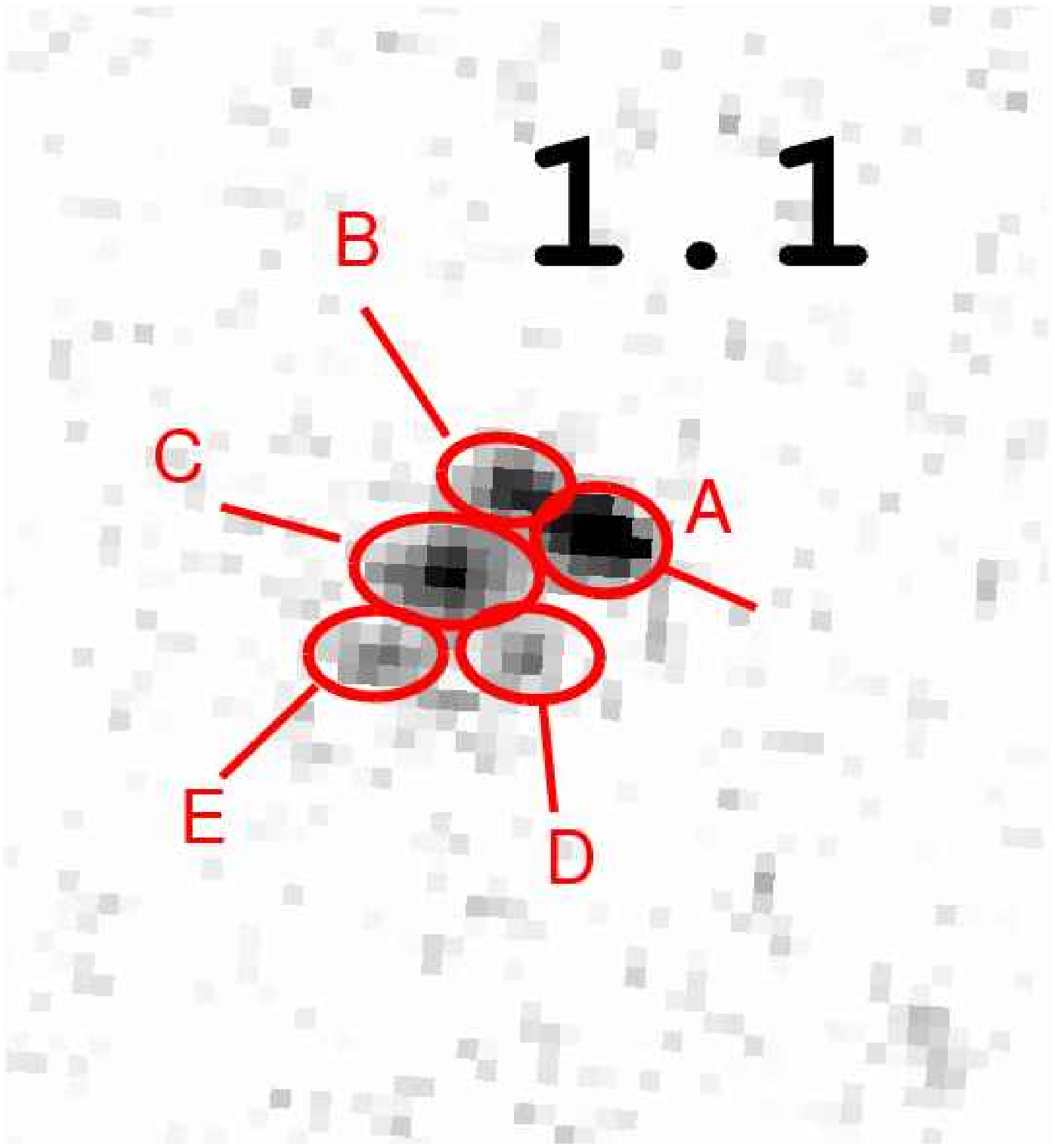}
\includegraphics[height=1.4cm,angle=0]{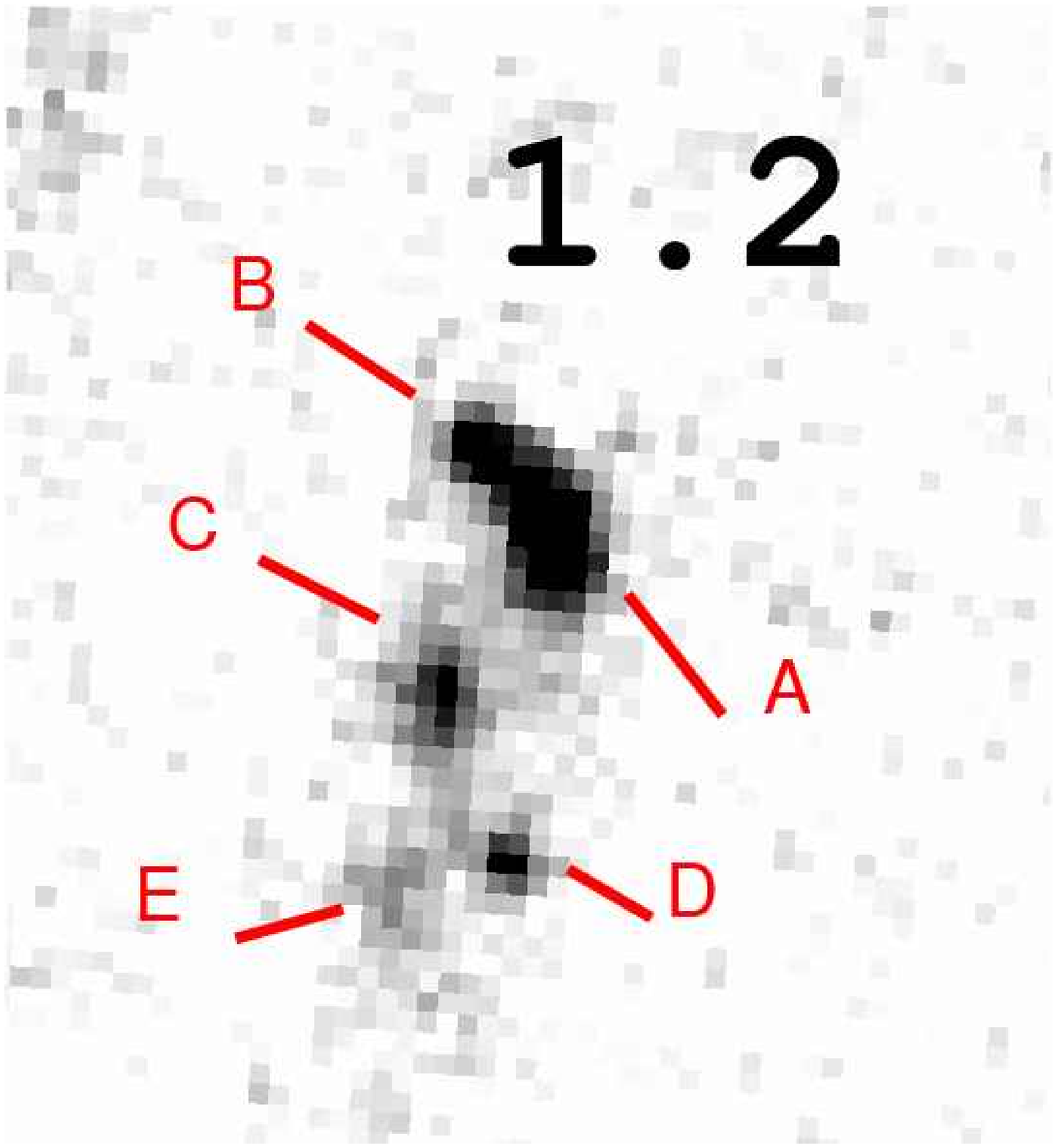}
\includegraphics[height=1.4cm,angle=0]{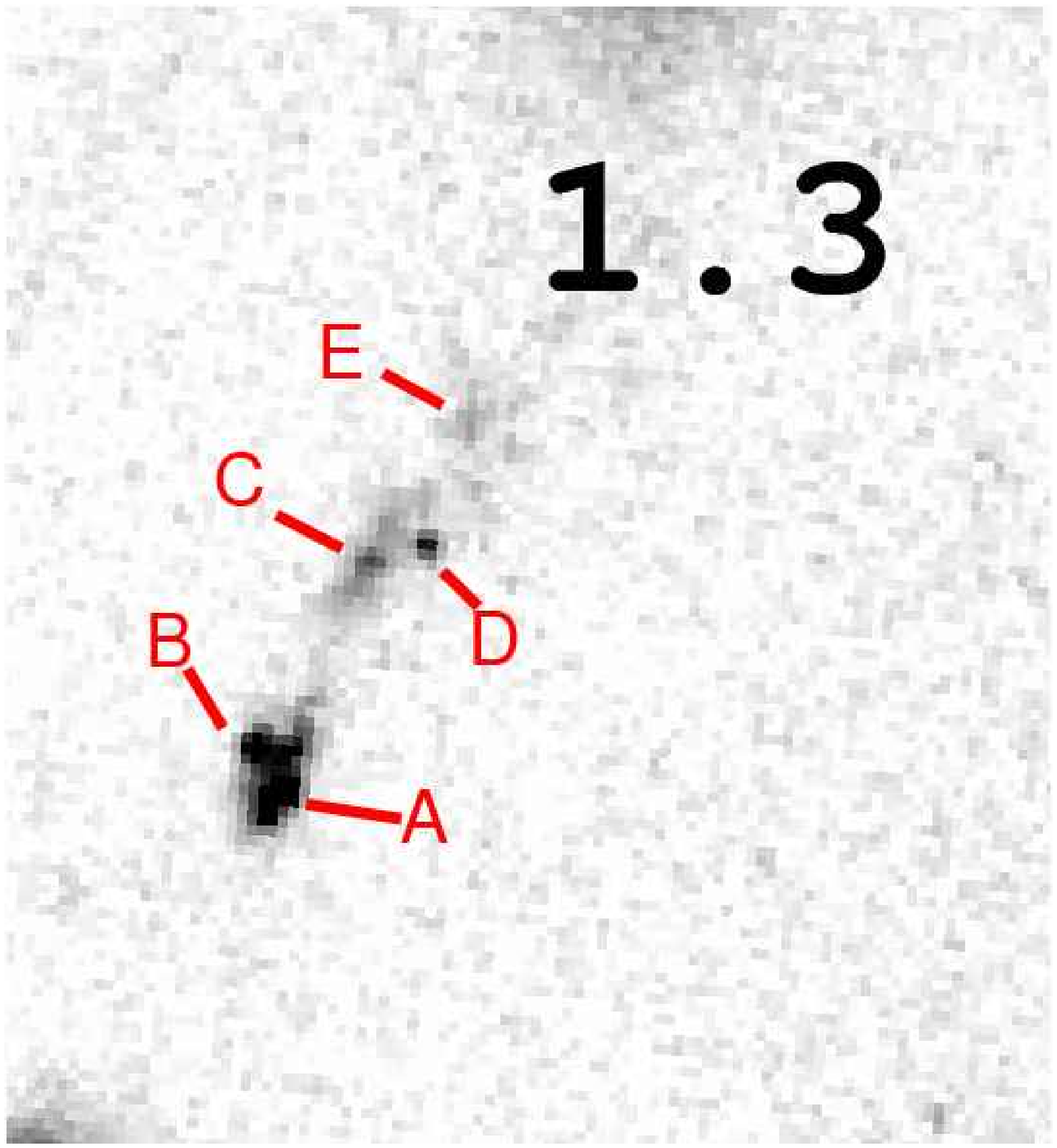}
\includegraphics[height=1.4cm,angle=0]{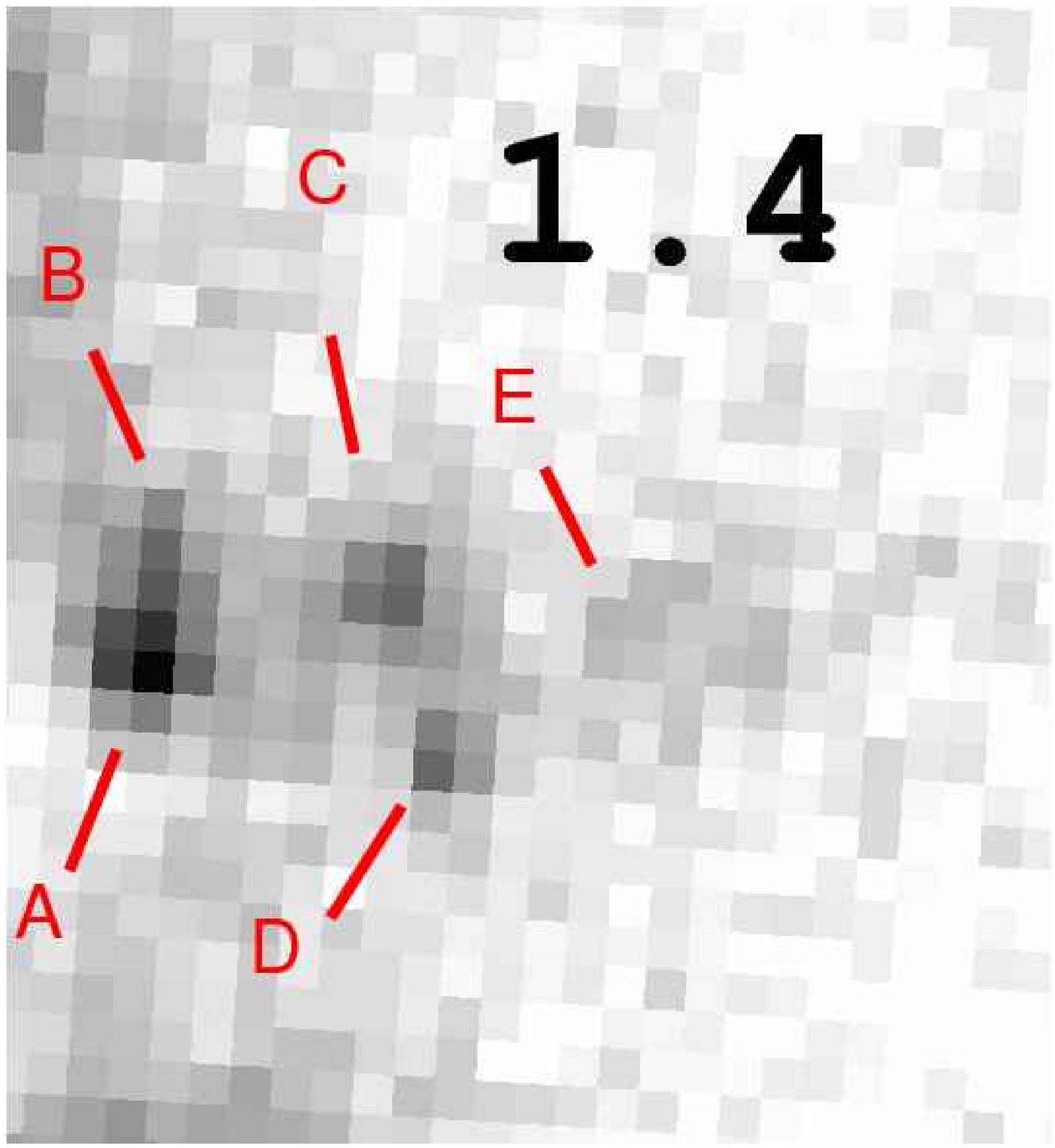}
\includegraphics[height=1.4cm,angle=0]{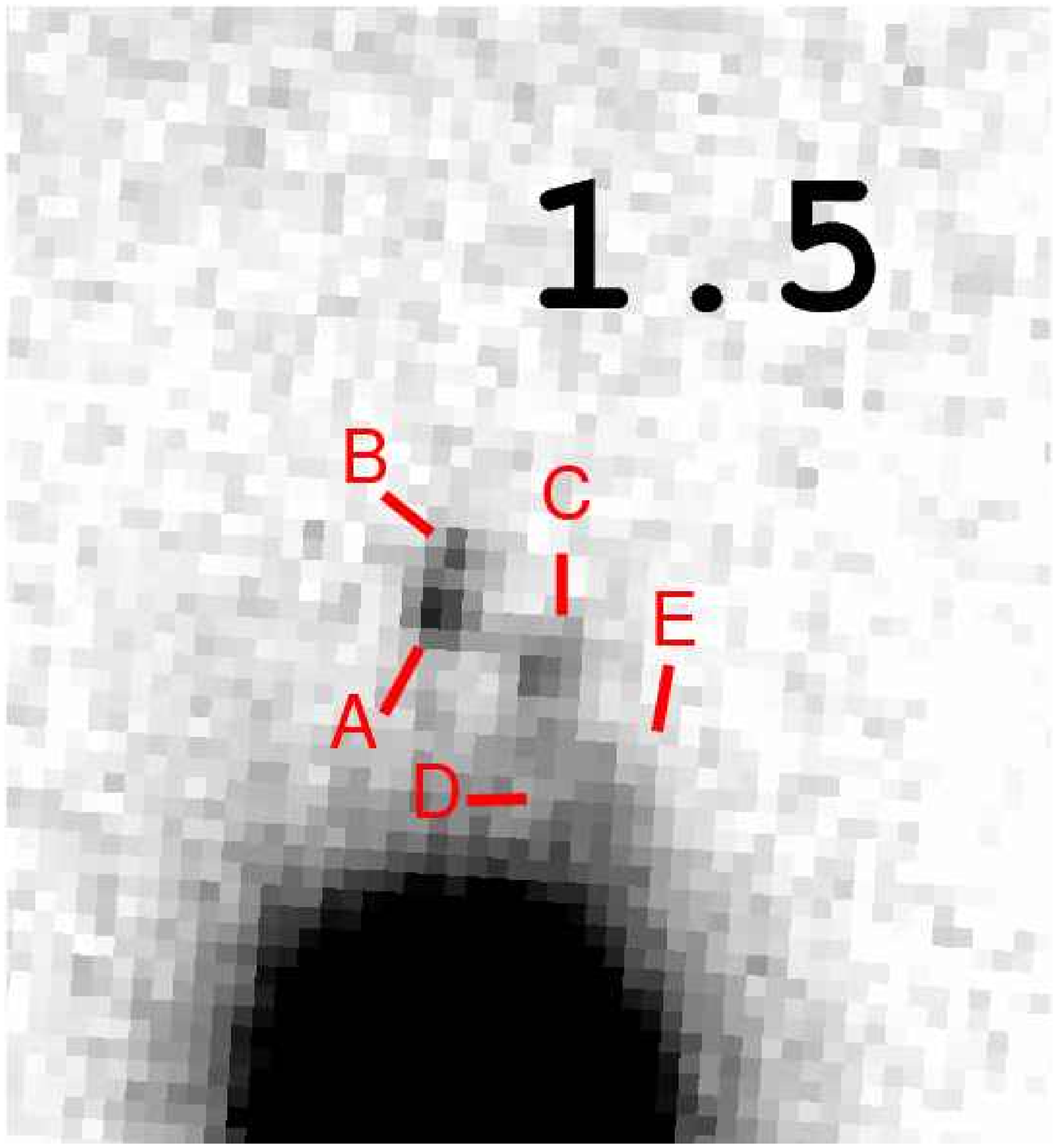}
\end{minipage}
\begin{minipage}{8cm}
\includegraphics[width=8cm,angle=0]{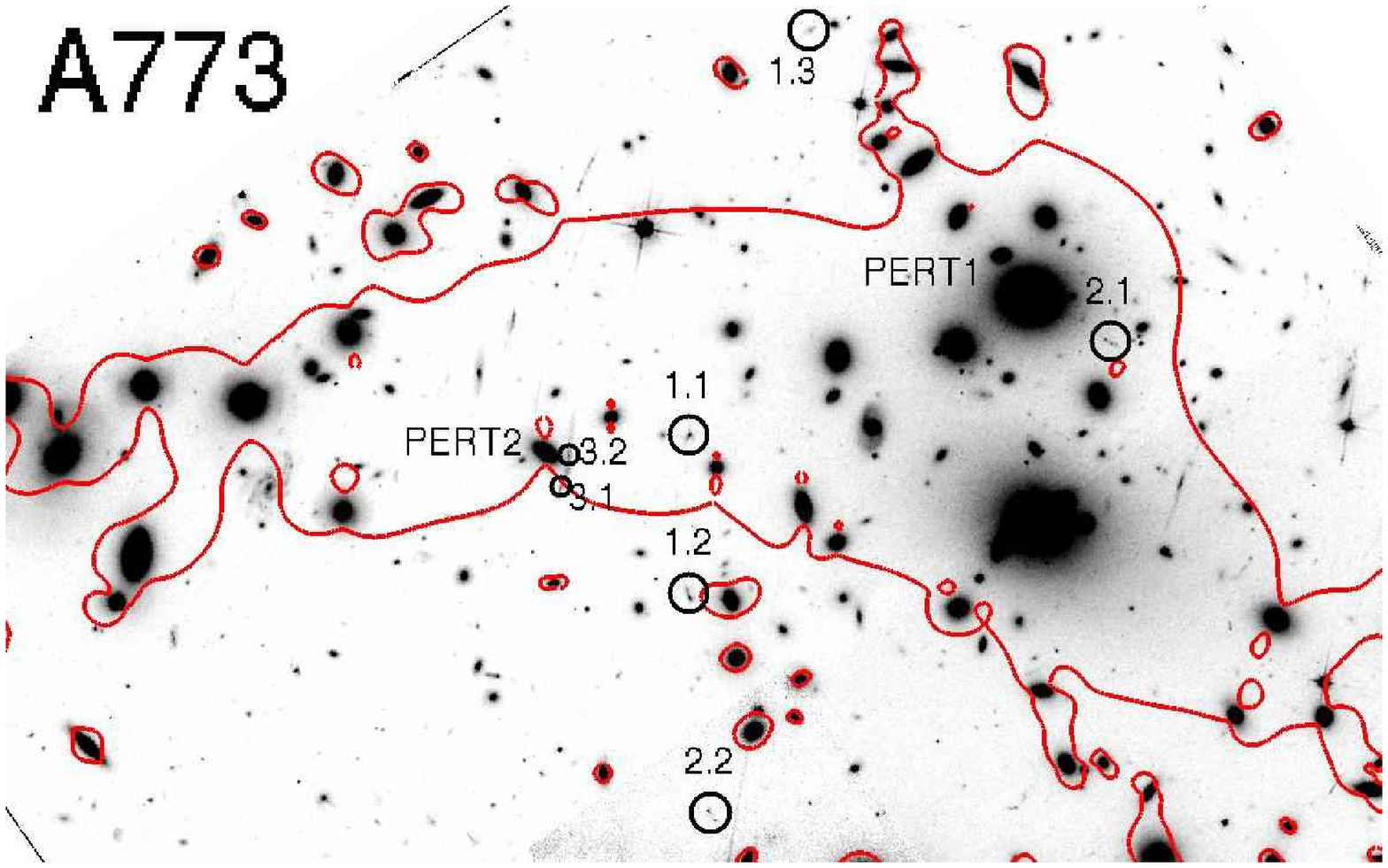}
\includegraphics[height=1.2cm,angle=0]{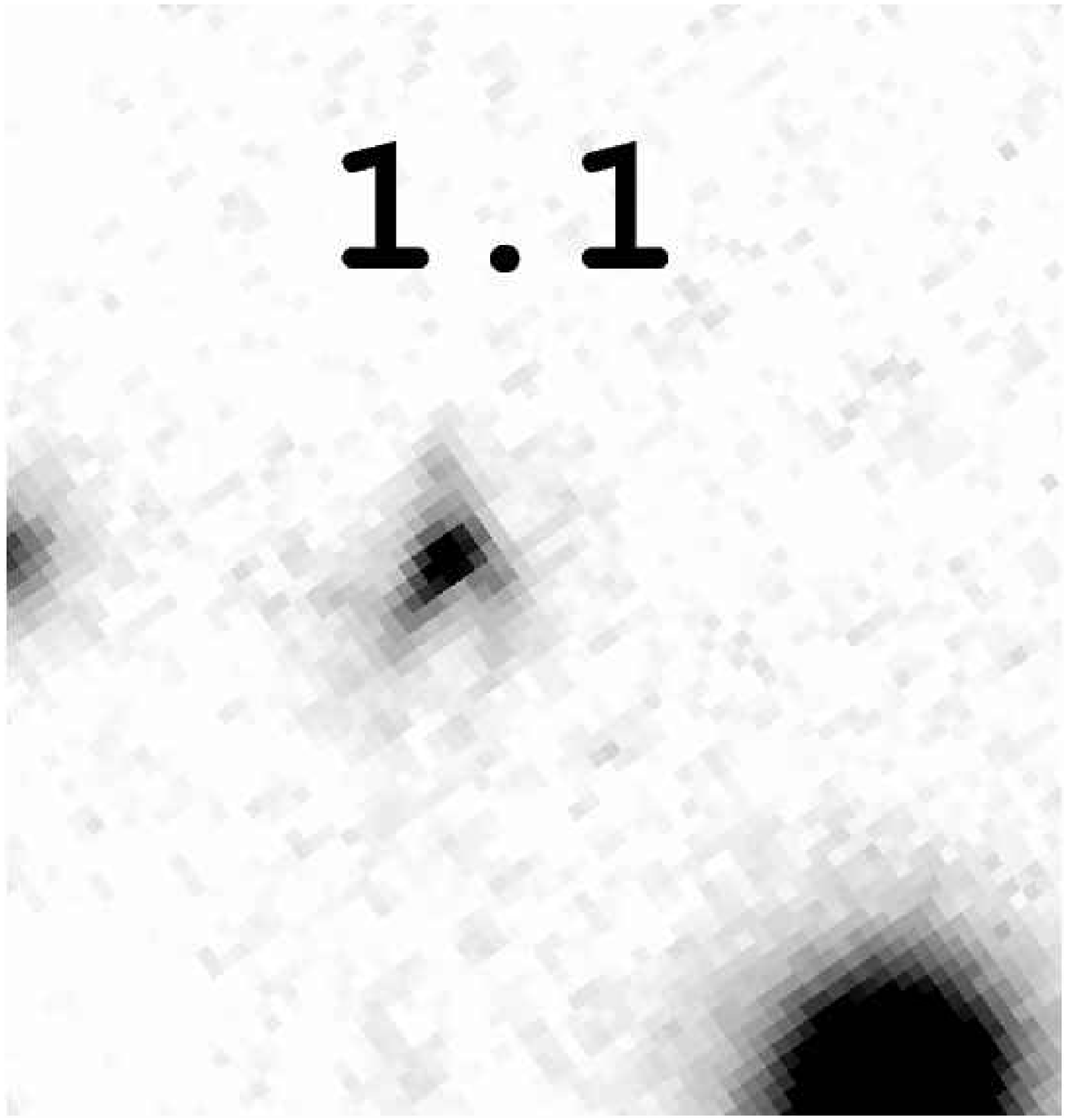}
\includegraphics[height=1.2cm,angle=0]{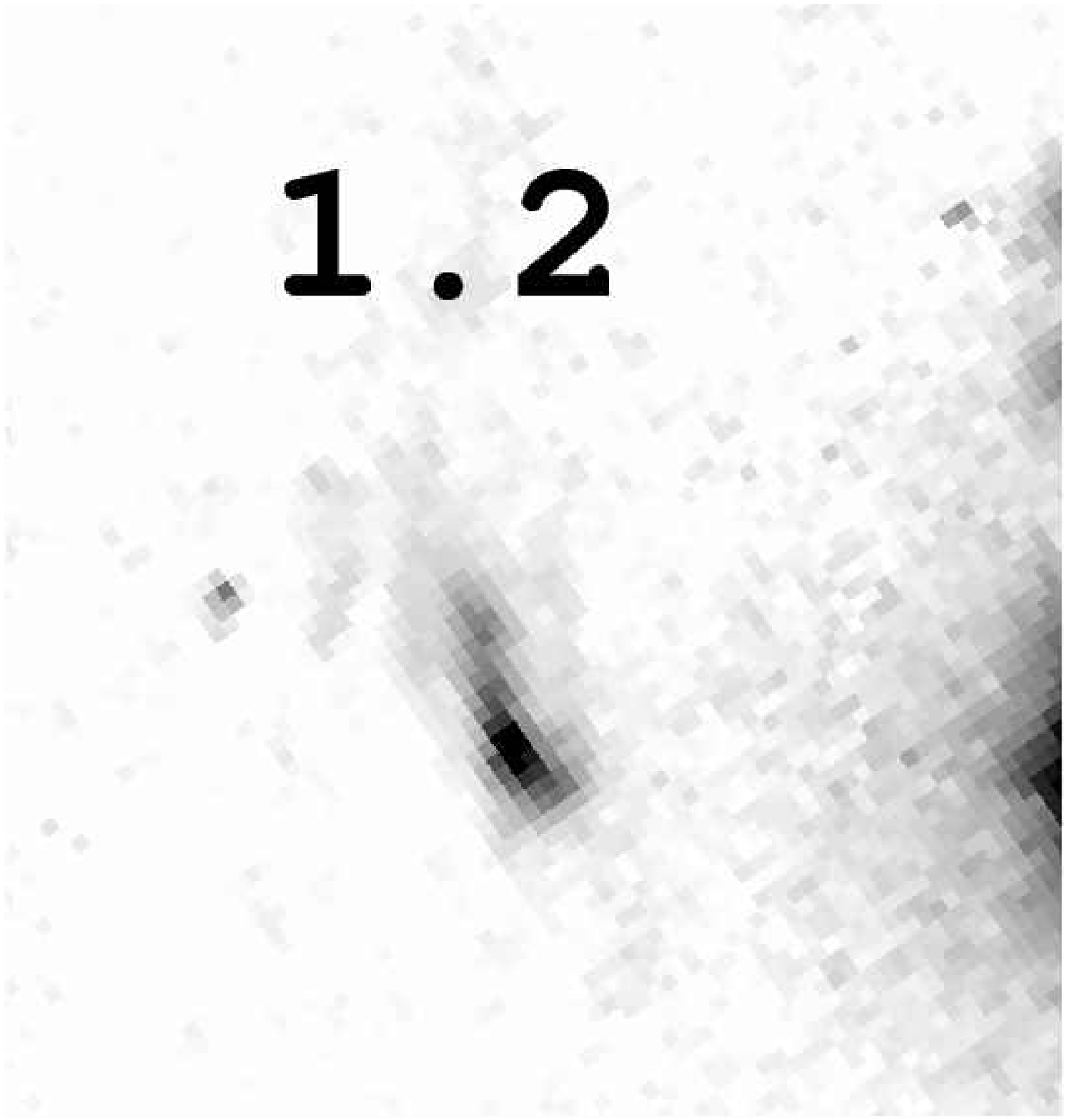}
\includegraphics[height=1.2cm,angle=0]{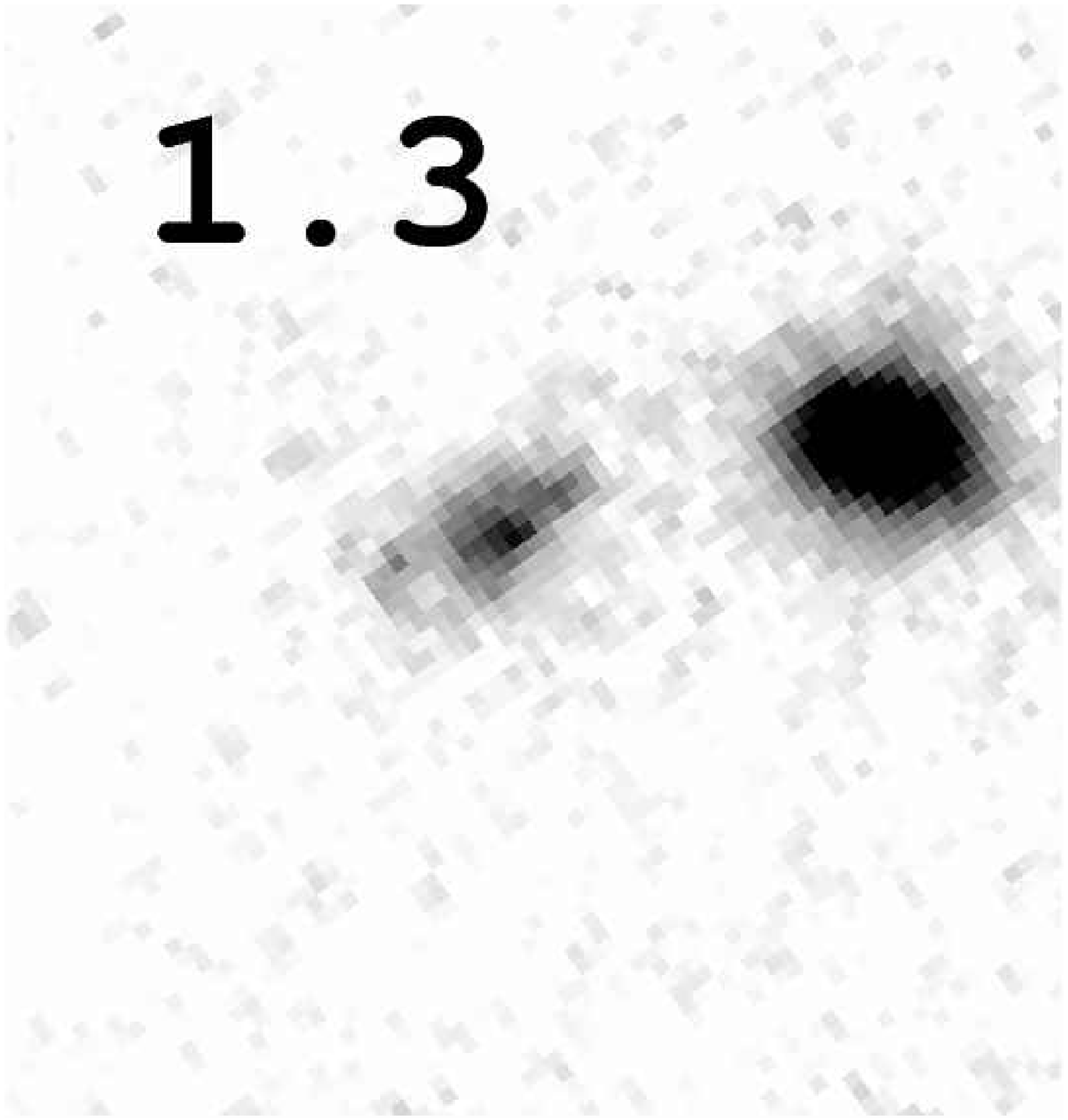}
\includegraphics[height=1.2cm,angle=0]{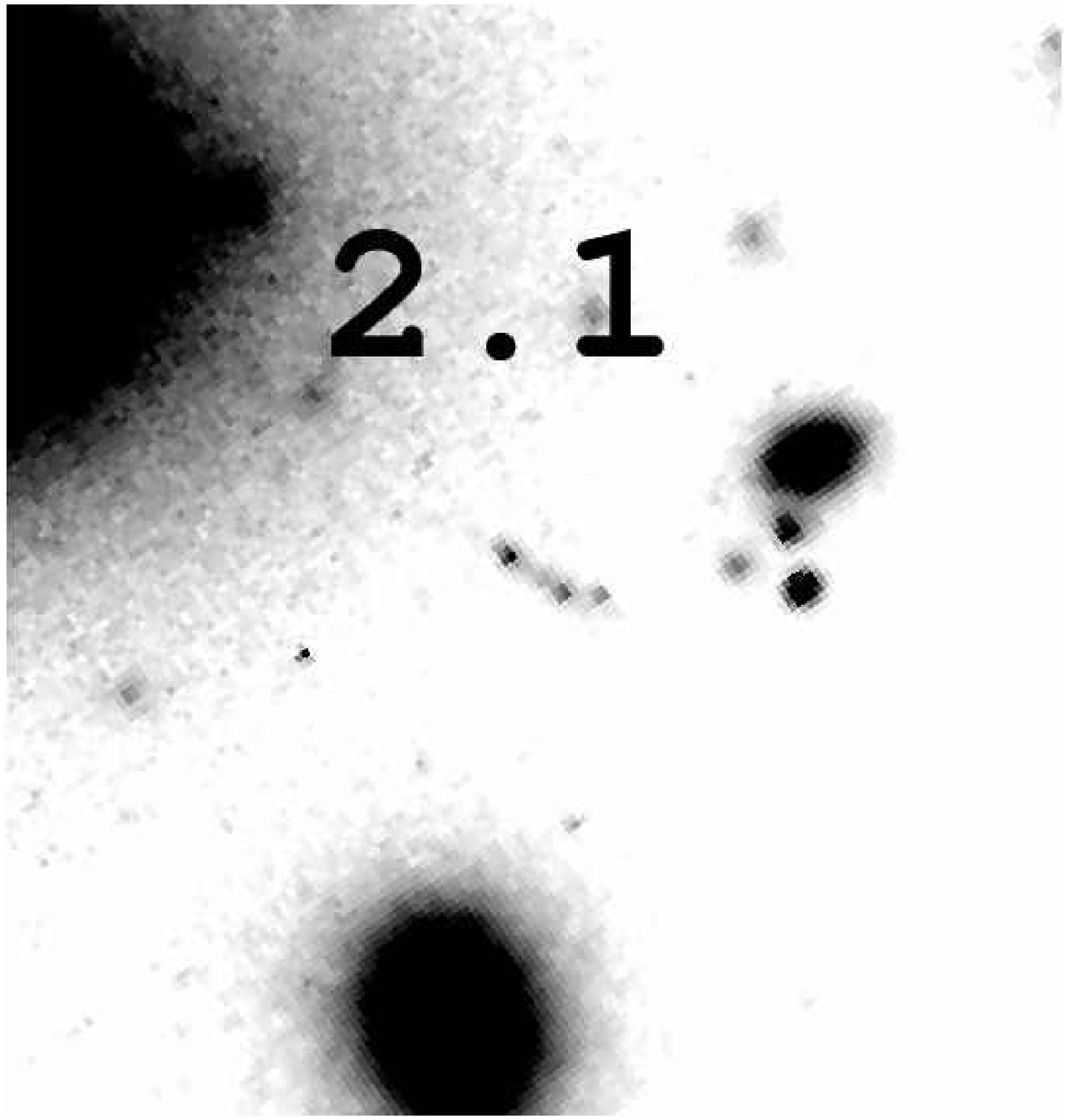}
\includegraphics[height=1.2cm,angle=0]{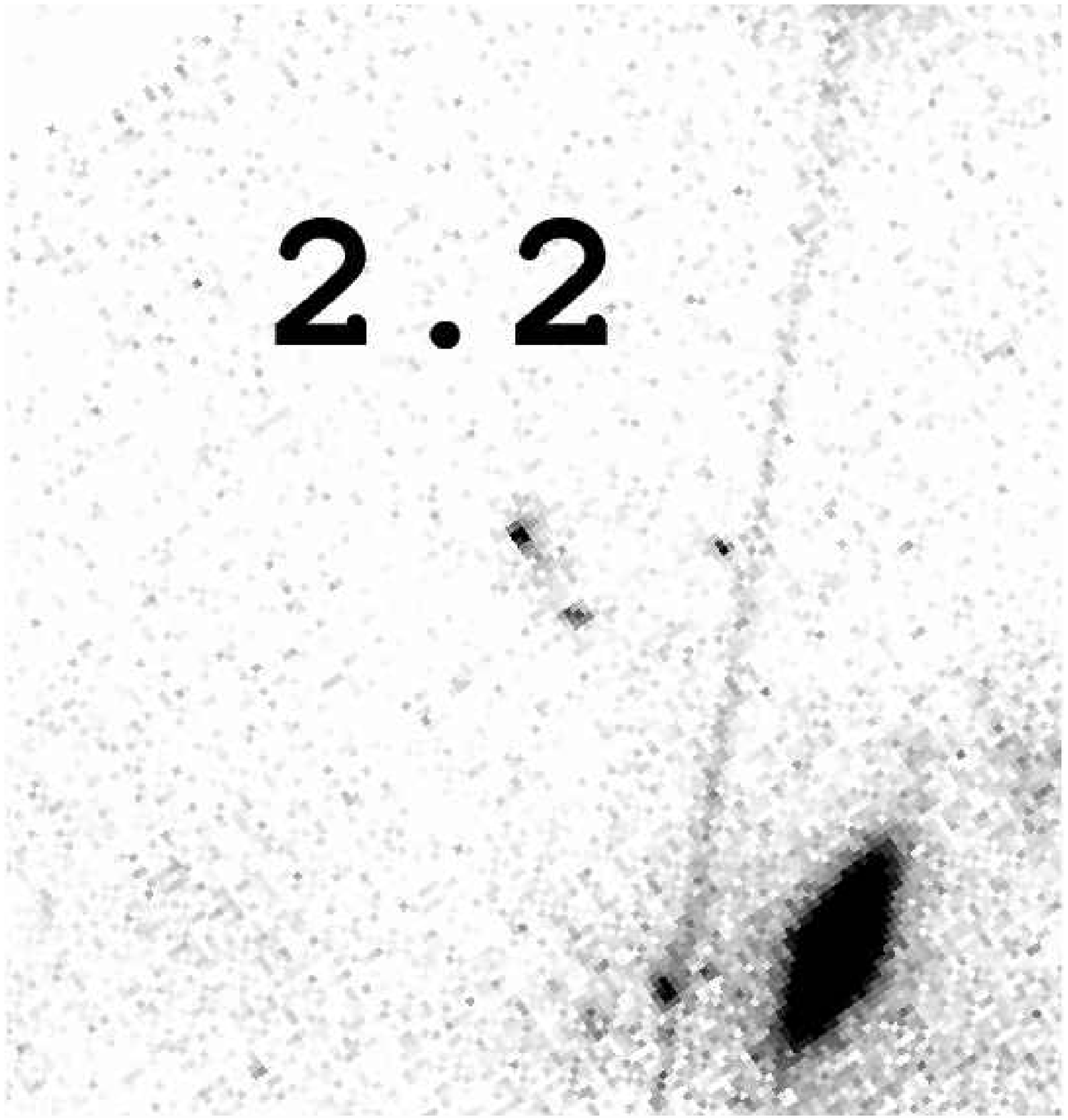}
\includegraphics[height=1.2cm,angle=0]{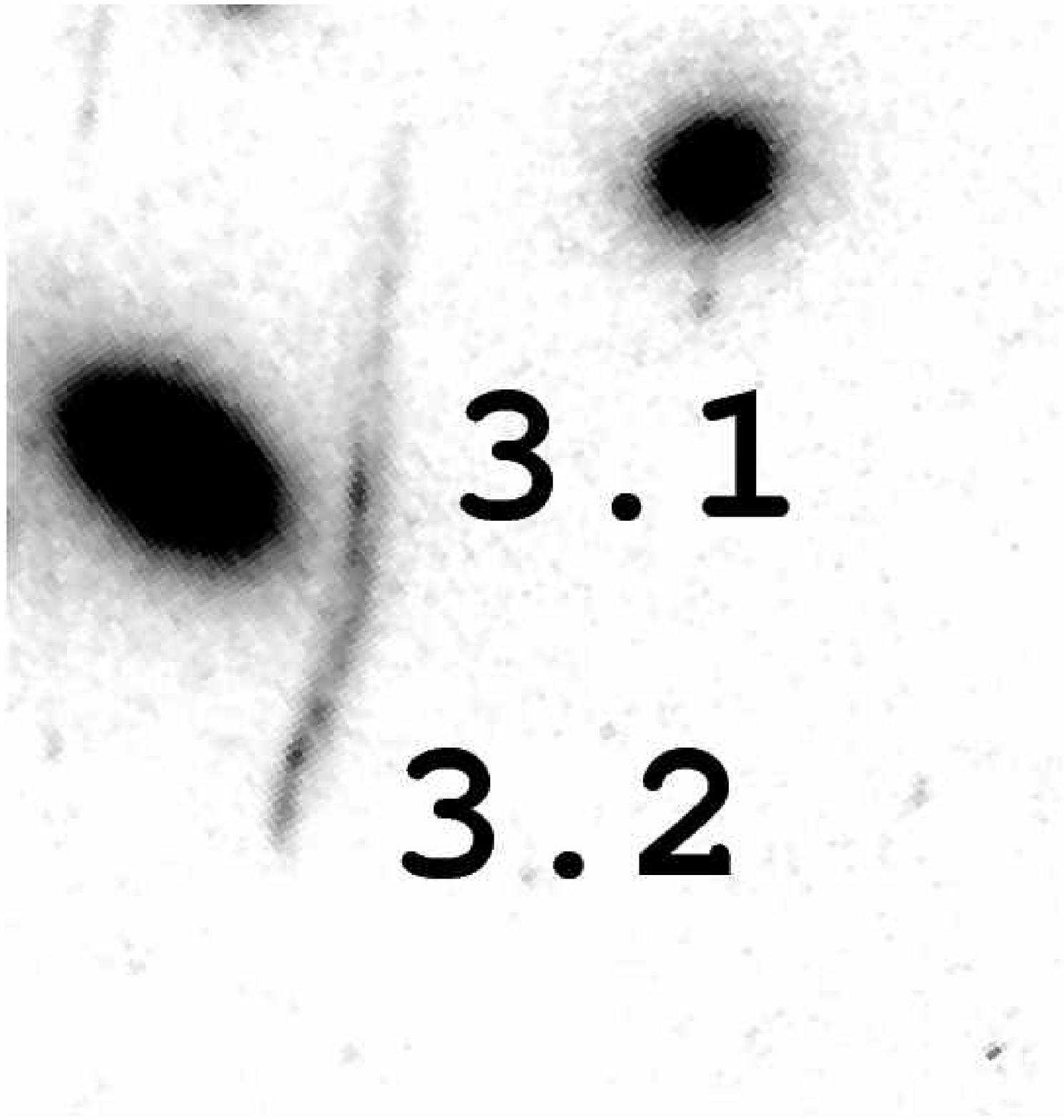}
\end{minipage}
\begin{minipage}{8cm}
\includegraphics[width=8cm,angle=0]{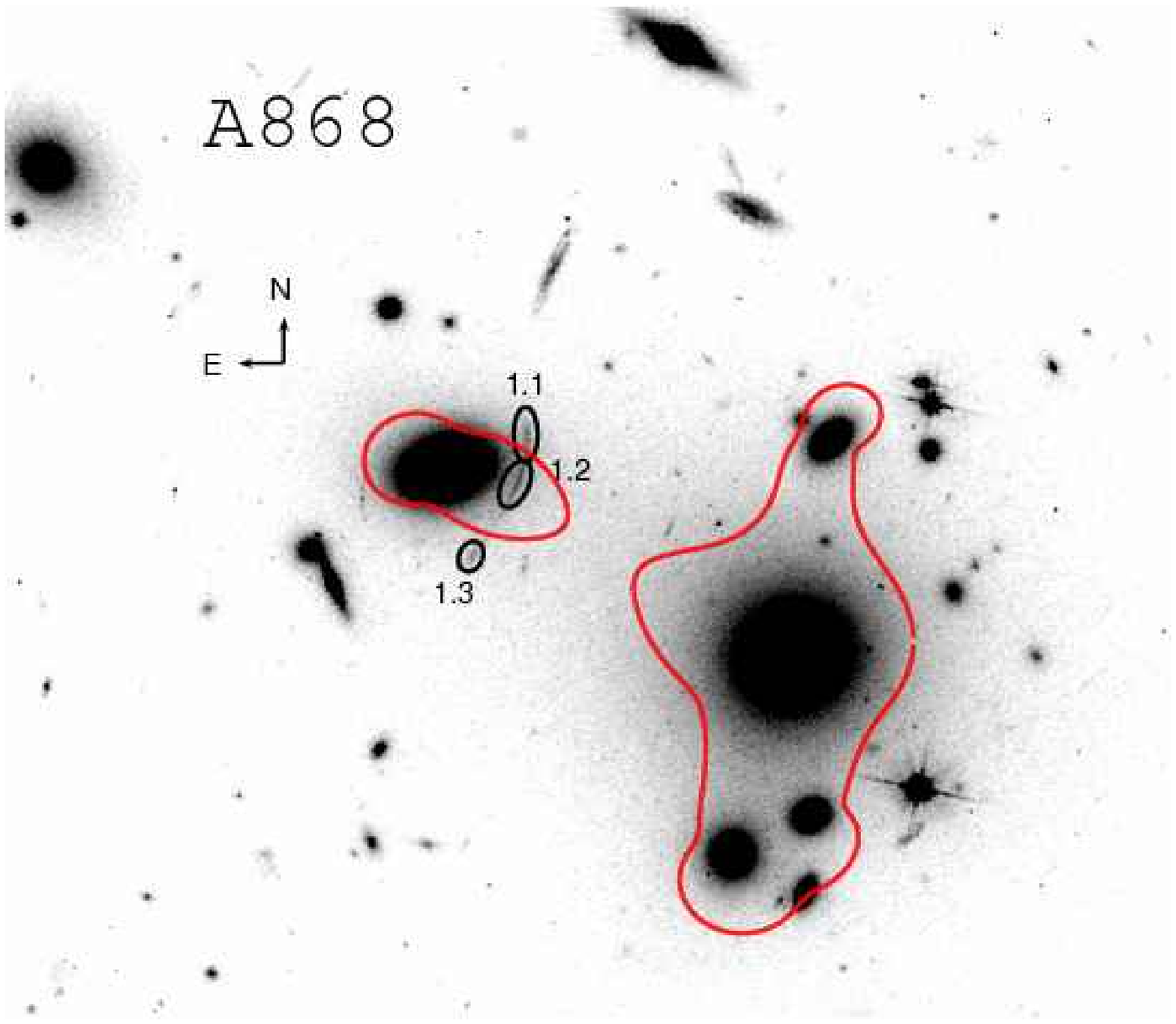}
\end{minipage}
\begin{minipage}{8cm}
\includegraphics[width=8cm,angle=0]{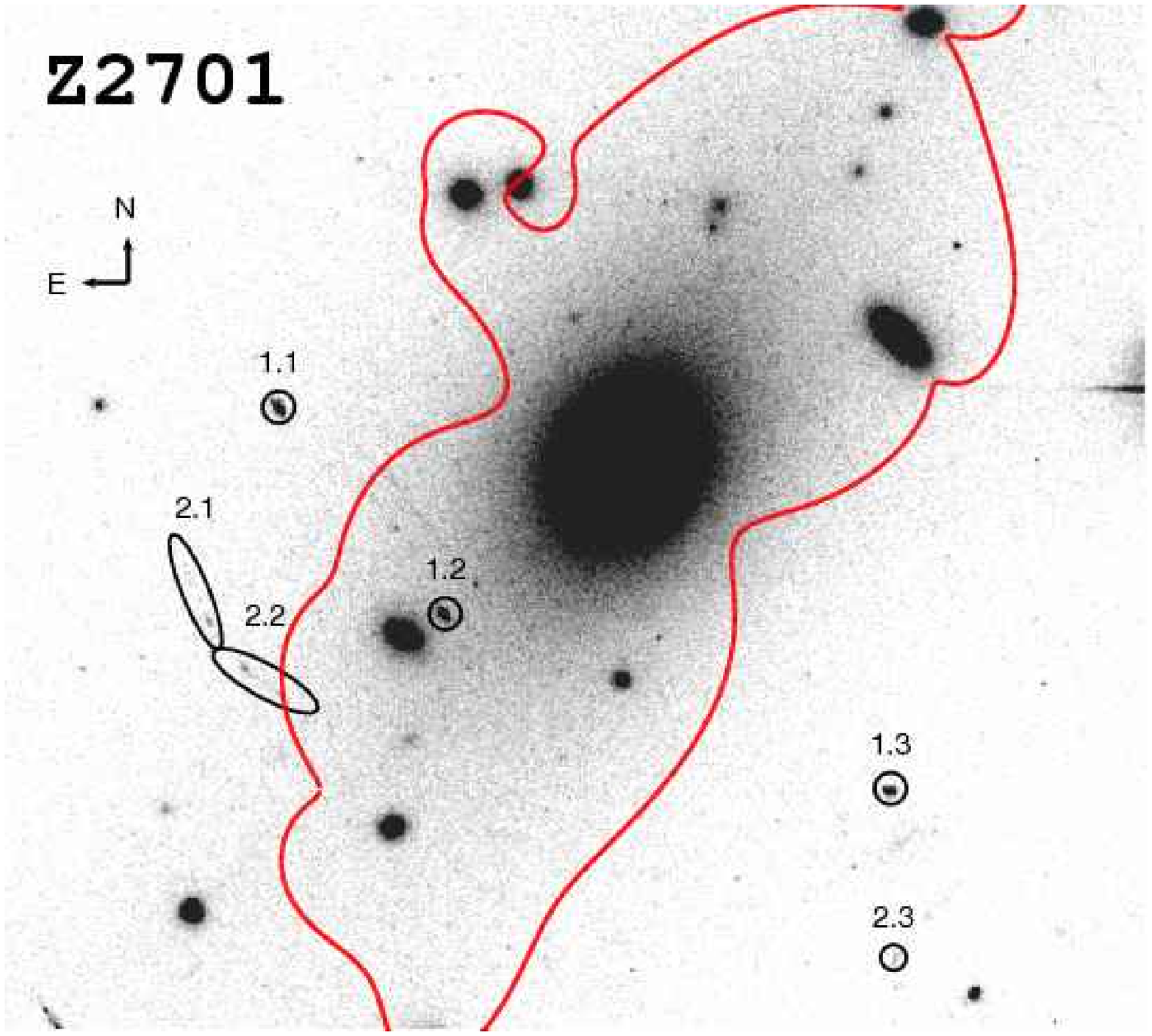}
\end{minipage}
\begin{minipage}{8cm}
\includegraphics[width=8cm,angle=0]{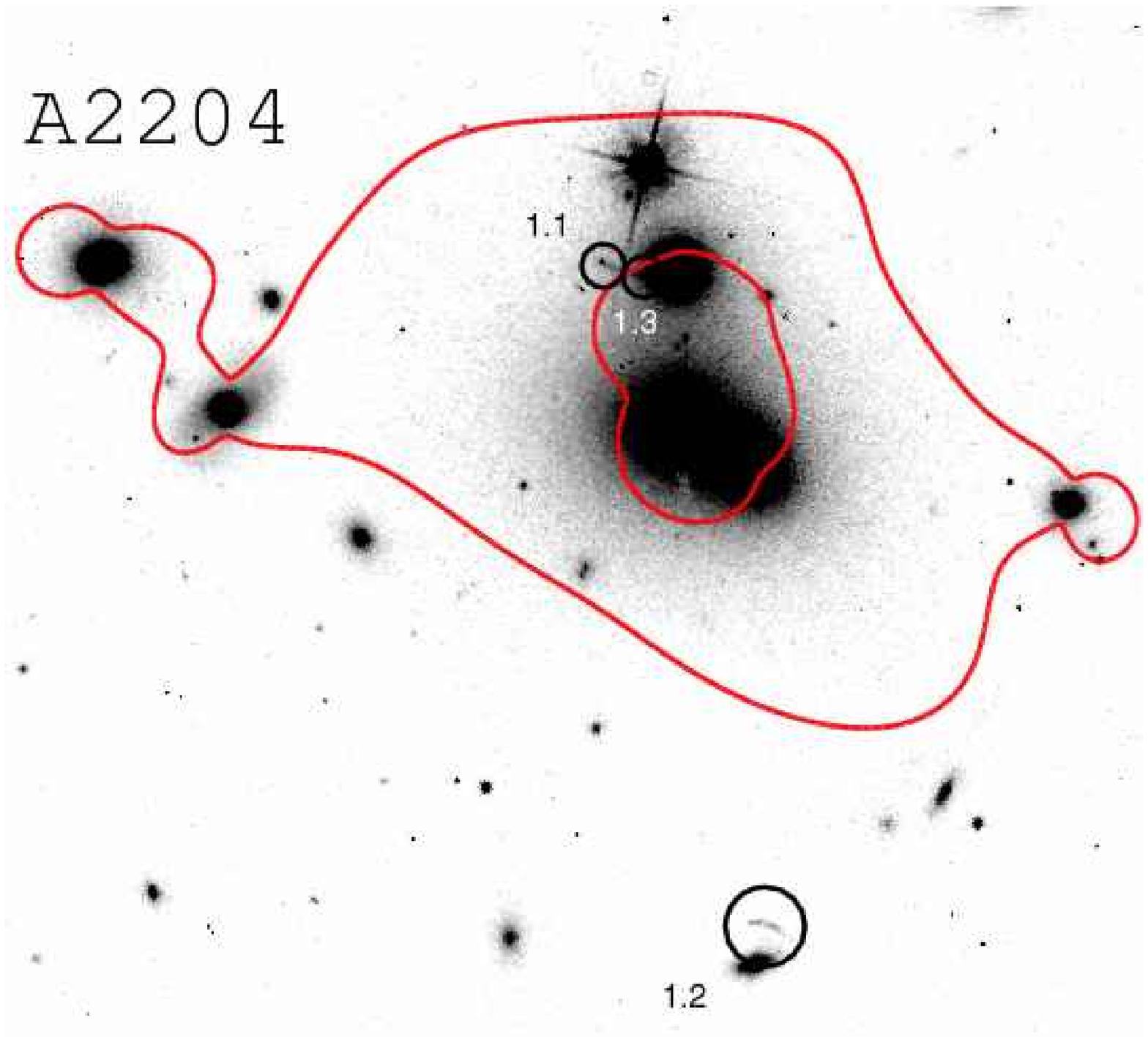}
\end{minipage}
\caption{\label{cluster}Multiple-image identifications and critical
  line at redshift 1.0 (A\,521), 0.91 and 2.1 (A\,611), 2.3 (A\,773), 0.55 (A\,868), 
  1.2 (Z2701) and 1.1 (A\,2204). Individual knots used as independent constraints are marked in
  A\,521 and A\,611.}
\end{figure*}

\begin{figure*}
\begin{minipage}{16cm}
\includegraphics[width=16cm,angle=0]{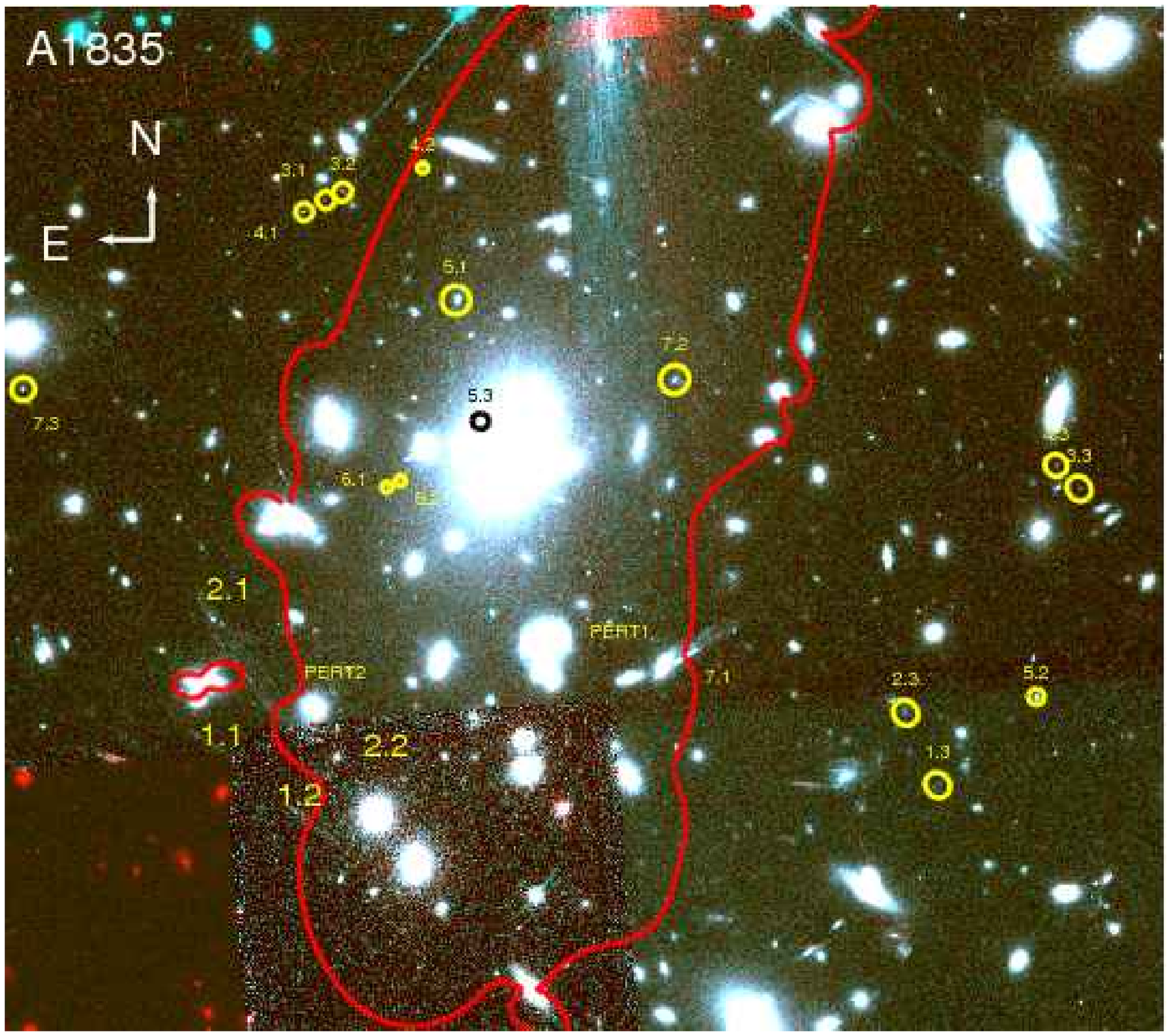}
\end{minipage}
\begin{minipage}{10cm}
\includegraphics[width=10cm,angle=0]{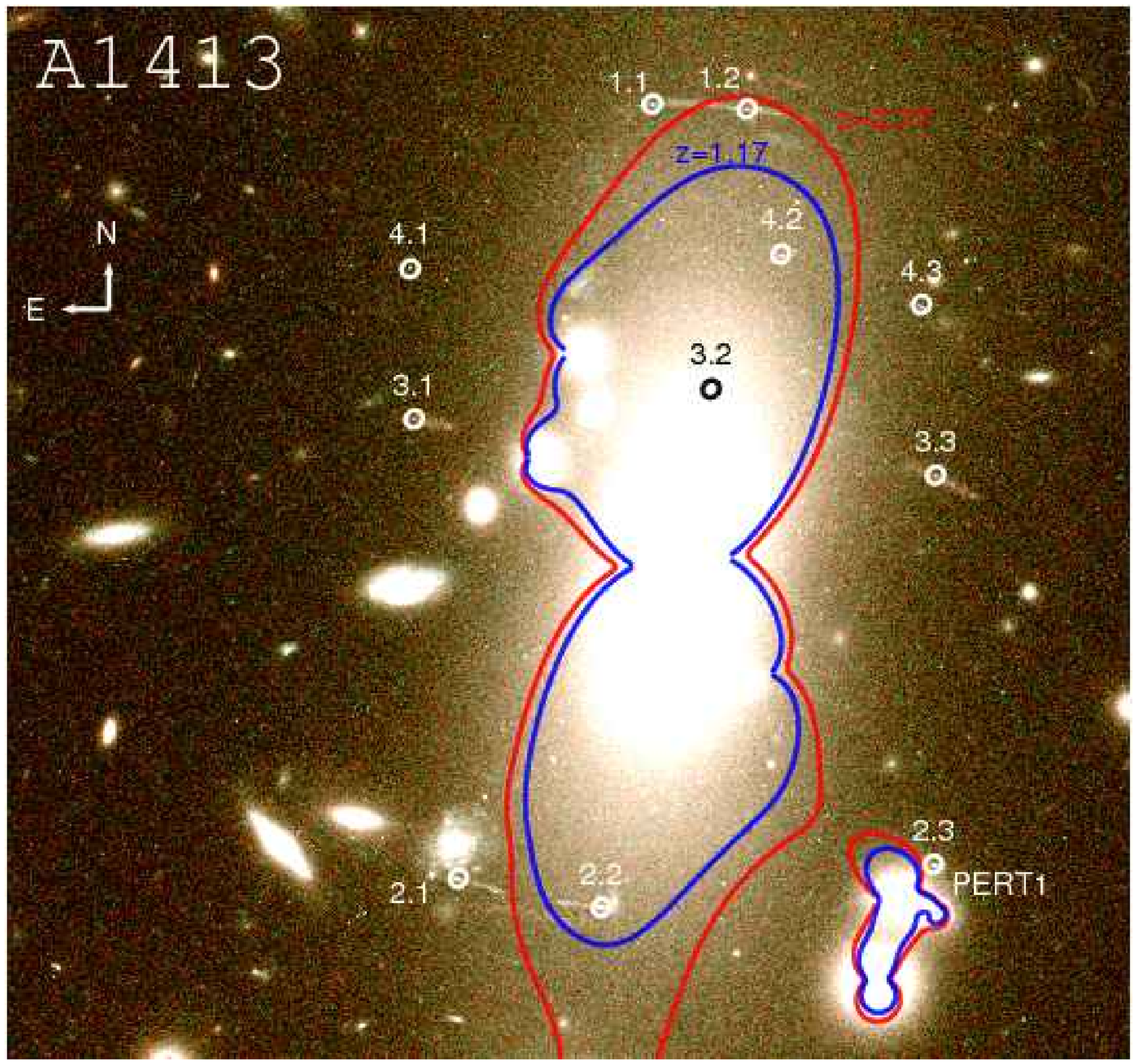}
\end{minipage}
\begin{minipage}{6cm}
\includegraphics[width=6cm,angle=0]{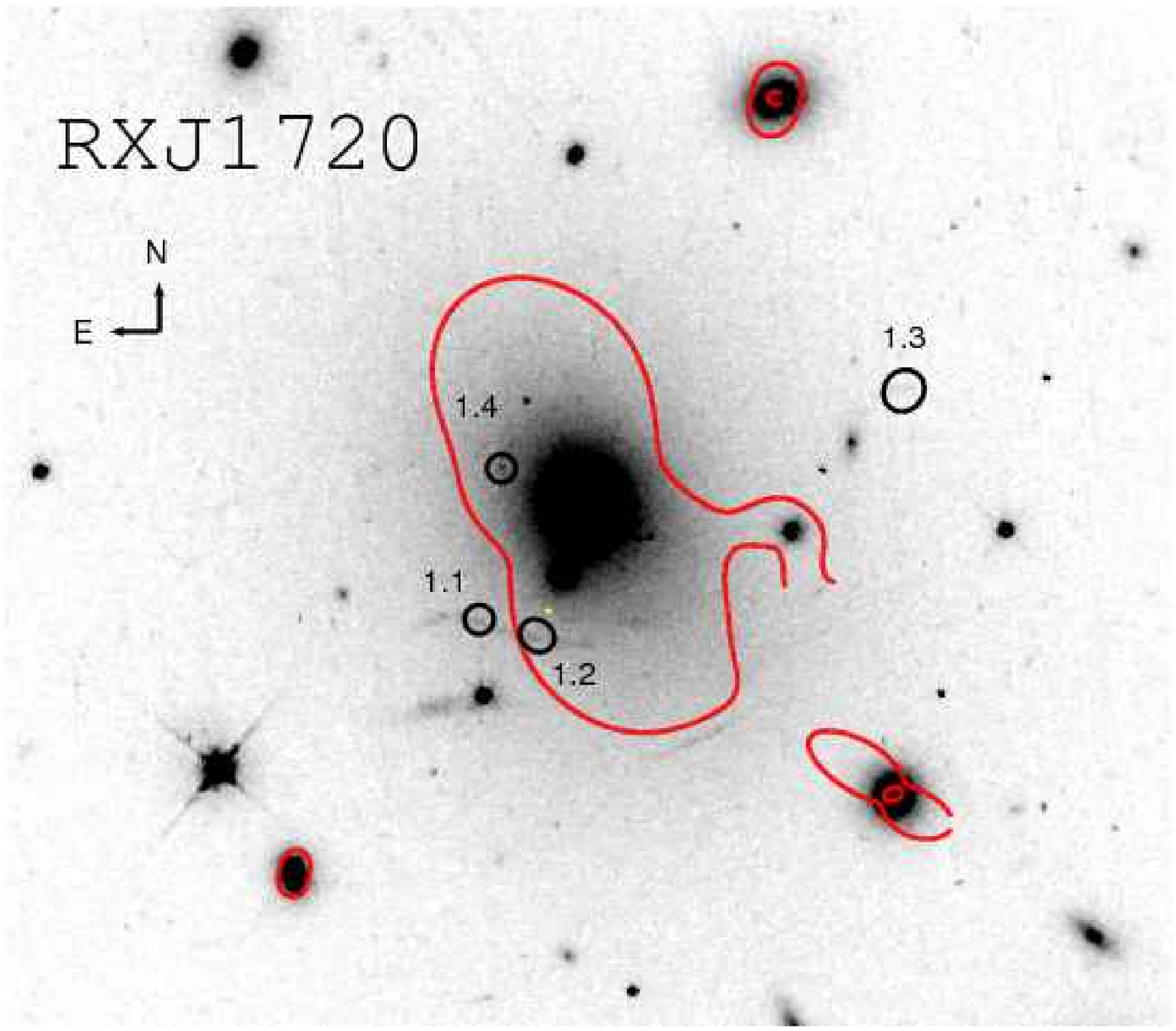}
\includegraphics[width=6cm,angle=0]{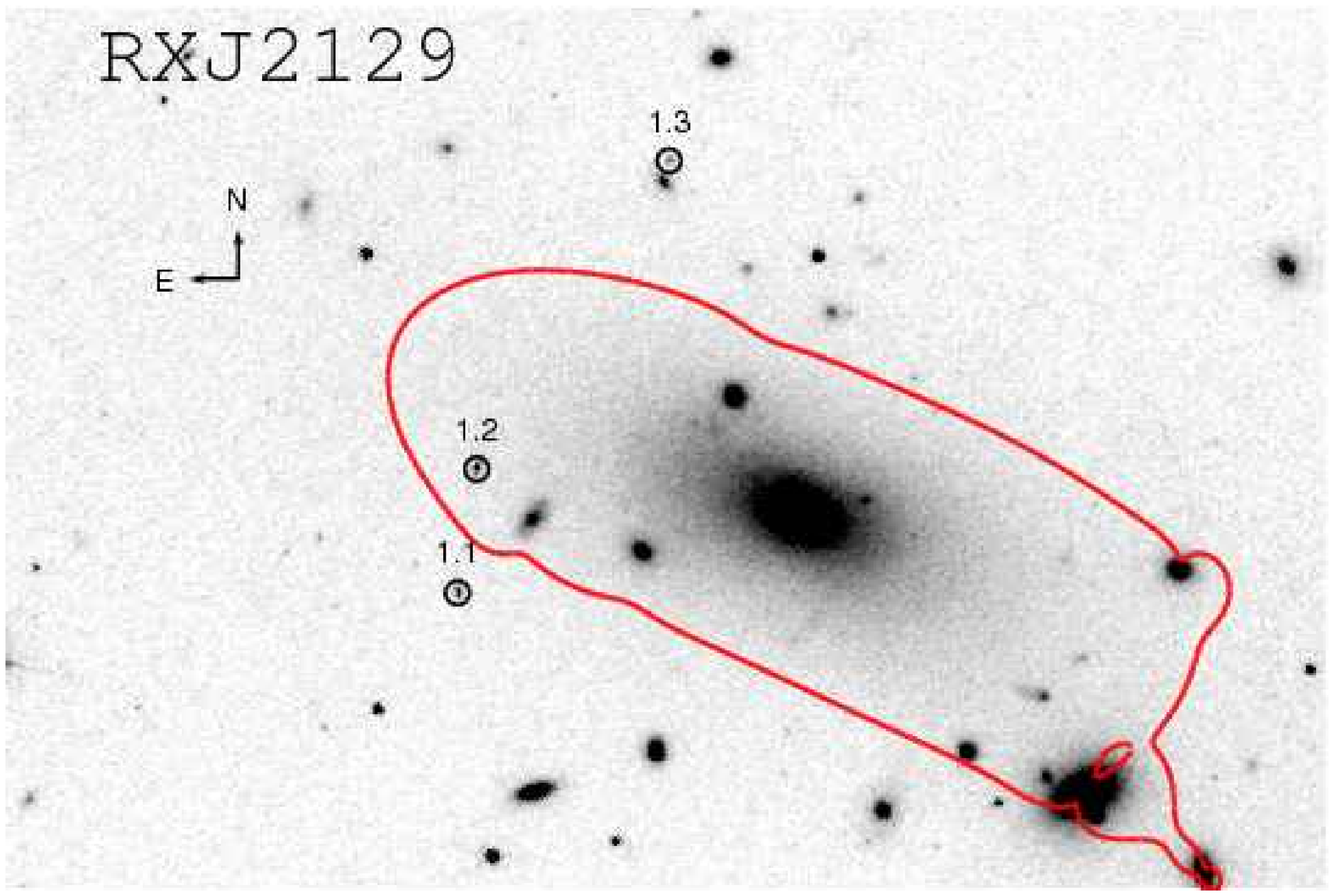}
\end{minipage}
\medskip
\noindent{\bf Fig. \ref{cluster}} Continued. The critical lines are 
presented at redshift 2.07 (A\,1835, system 7), 2.0 and 2.7 (A\,1413), 
2.1 (RXJ1720) and 2.0 (RXJ2129).
\end{figure*}

\subsection{Cluster mass models}
\label{slsub}

We have used Lenstool\footnote{Publicly available at {\tt
    http://www.oamp.fr/cosmology/lenstool}}\citep{Kneib93,Jullo} to
reconstruct the distribution of mass in the cluster cores, which is
described as a superposition of analytic mass components that account
for both cluster- and galaxy-scale mass, following Sm05.  We use all
the multiple-image systems described in the previous section and
Table~\ref{mult} as model constraints, using spectroscopic redshifts
whenever available.  The main differences between these models and
those of Sm05 are that unknown multiple-image redshifts are free
parameters in the models, and two different parameterisations are used
for the smooth cluster-scale mass components (\S\ref{clscale}), and
parameter space is explored using the Bayesian MCMC sampler that is
now available within Lenstool (\S\ref{opt}).  In common with Sm05 and
other studies of strong-lensing clusters by our group,
\citep[e.g.][]{Kneib96,Smith01a,Smith02b,Richard07,Limousin07,Limousin08,Richard09,Smith09a}),
the mass modelling is an iterative process, as we first produce a
rudimentary mass model, based on the most reliable constraints, and
then gradually incorporate additional constraints.

\subsubsection{Cluster-scale mass components}
\label{clscale}

The mass distribution of cluster cores within the strong-lensing
region is generally well-described by a smooth elliptical profile
centred near the BCG.  As a starting point in our analysis, we try to
reproduce the strong lensing constraints using a double
pseudo-isothermal elliptical mass distribution (dPIE, also known as a
smoothly truncated PIEMD, \citealt{Ardis}).  The dPIE profile, which
was also used in Sm05, is characterized by its central position,
position angle and ellipticity, the central velocity dispersion
$\sigma_0$ and two characteristic radii: a core radius $r_{\rm core}$
and a cut radius $r_{\rm cut}$.  In an ($x$, $y$) coordinate system
oriented along the angle $\theta$ of the ellipse, the surface mass
density takes the form:
\begin{equation}
\Sigma(x,y)=\frac{\sigma_0^2}{2G}\ \frac{r_{\rm cut}}{r_{\rm cut}-r_{\rm core}}\Big[\frac{1}{(r_{\rm core}^2+\rho^2)^{1/2}}-\frac{1}{(r_{\rm cut}^2+\rho^2)^{1/2}}\Big]
\label{dpie}
\end{equation}
with $\rho^2=[(2-e)\ (x-x_c)/2]^2+[(2-e)\ (y-y_c)/(2-2e)]^2$, and
$(x_c,y_c)$ is the centre of the mass distribution (see also Sm05).

The mass models generally include a single dPIE component centred at
the BCG location, but in some cases this description is unable to
reproduce the observed multiple-image systems.  When this is the case,
we add a second or third cluster-scale mass component.  A full
discussion of our capacity to identify cluster-scale clumps from the
strong lensing signal is provided in \S\ref{clumps}.

In order to estimate the model dependence of our strong lensing
results, we also independently fit the cluster-scale mass
distributions with the same number of Navarro-Frenk-White
(\citealt{NFW}, hereafter NFW) profiles, characterized by a
concentration $c_{200}$ and a scale radius $r_s$, described by usual
density profile:
\begin{equation}
\rho(r)=\frac{\rho_c\delta_c}{(r/r_s)\ (1+(r/r_s))^{2}}
\end{equation}
where $\rho_c$ is the critical density and $\delta_c$ is related to
$c_{200}$ through the relation:
\begin{equation}
\delta_c=\frac{200}{3}\frac{c_{200}^{3}}{ln(1+c_{200})-c_{200}/(1+c_{200})}.
\end{equation}
The generalisation of this spherical model into a projected elliptical
mass distribution follows a similar relation as Eq.~\ref{dpie}.  The
full analytical description of the pseudo-elliptical NFW is presented
in \citet{Golse2002}.

\subsubsection{Galaxy-scale mass component}\label{gals}

The presence of numerous cluster galaxies near the cluster centre, as
identified by our near-infrared photometric selection (\S\ref{nir})
affects the shape of the mass distribution locally, and therefore the
associated lensing signal.  Following \citet{Kneib96} and Sm05, we
account for galaxy-scale mass in the models by adding individual dPIE
mass components at the location of each galaxy in the cluster catalogs
discussed in \S\ref{nir}, within $250\kpc$ of the respective BCGs.
The geometrical parameters of these components ($x_c$, $y_c$, $e$,
$\theta$) are matched to the values measured by SExtractor.  The
parameters $\sigma_0$, $r_{\rm core}$ and $r_{\rm cut}$, upon which
the total mass of each galaxy-scale dark matter halo depend, are
assumed to scale with their $K$-band luminosity $L_K$, relative to an
$L_K^\star$ galaxy, following the \citet{FJ} relation and a constant
mass-to-light ratio:
\begin{eqnarray}
  r_{\rm core}=r_{\rm core}^\star\ (L_K/L_K^\star)^{1/2},\nonumber\\
  r_{\rm cut}=r_{\rm cut}^\star\ (L_K/L_K^\star)^{1/2},\nonumber\\
  \sigma_0=\sigma_0^\star\ (L_K/L_K^\star)^{1/4}
\end{eqnarray}
Similar scaling relations have been used in the past
\citep{Covone,Limousin07,Ardis}, showing that $r_{\rm core}^\star$ is
small and has little effect on the modelling results.  We therefore
adopt $r_{\rm core}^\star=0.15\kpc$, following \citet{Limousin07}; see
also \citet{Brainerd}.

The two remaining parameters ($\sigma_0^\star$, $r_{\rm cut}^\star$)
are degenerate, because the perturbation of the cluster lensing signal
caused by an individual galaxy depends on the total galaxy mass, given
by $\sigma_0^2\,r_{\rm cut}$.  After performing various tests with
Lenstool, we concluded that fixing one of these parameters does not
affect the ability of the models to reproduce the observed
multiple-image systems, provided that galaxies located within
$2\arcsec$ of a highly magnified arc are excluded from the scaling
relations (see \S\ref{pert}).  We therefore adopted a fixed $r_{\rm
  cut}^\star=45\kpc$, which matches recent galaxy-galaxy weak-lensing
results \citet{Natarajan09}.  For our adopted value of $r_{\rm
  core}^\star=0.15\kpc$ the measured galaxy velocity dispersions
($\sigma$) and dPIE velocity dispersions ($\sigma_0$) are in good
agreement -- $0.94<\sigma/\sigma_0<1.04$ -- the range of values
arising from the range of radii at which $\sigma$ is measured
\citep{Ardis}.  We therefore employ a Gaussian prior of
$\sigma_0^\star=(158\pm27)\kms$, making use of \citeauthor{Bernardi}
(\citeyear{Bernardi})'s observational results on $\sigma^\star$.

\subsubsection{BCG and individual galaxy perturbers}
\label{pert}

We model separately the brightest cluster galaxy (BCG) in each cluster
using a similar dPIE profile as other cluster members, but optimizing
the values of $\sigma_{0,\rm BCG}$ and $r_{\rm cut, BCG}$
independently. The same approach was used for 6 local galaxy-scale
perturbers in A\,611, A\,773, A\,1413, A\,1835, identified
in Fig. \ref{cluster}.

\subsubsection{Optimization}
\label{opt}

The models are fitted to the multiple-image constraints using the new
Bayesian Markov chain Monte-Carlo (hereafter MCMC) sampler, described
in detail in \citet{Jullo}.  This process uses the observational
constraints (positions of the multiply imaged systems) to optimize the
parameters describing the mass distribution by matching the location
of each image of a given system in the source plane.  The quality of
the models can be estimated using the root mean square (RMS) deviation
in the image-plane from the observed positions of the multiple-image
positions predicted by the model, defined as:
\begin{equation}
\sigma_i=\sqrt{\sum_{j,k}({\rm xobs}_{j,k}-{\rm xpred}_{j,k})^2+({\rm yobs}_{j,k}-{\rm ypred}_{j,k})^2}
\end{equation}
where (${\rm xobs}_{j,k},{\rm yobs}_{j,k}$) and (${\rm
  xpred}_{j,k},{\rm ypred}_{j,k}$) are the observed and predicted
locations of image $j$ in system $k$, respectively. Following Sm05,
\citet{Limousin07, Richard09} we have used $0.2\arcsec$ as the
positional uncertainty of the multiple image identifications.

\subsection{Best fit parameters}

\begin{table*}
  \caption{\label{best} Best-fit parameters of the mass models. For
    each mass component, we give the centre, ellipticity, orientation,
    core and cut radii, as well as central velocity dispersion of the
    dPIE profile. The following column gives the image plane RMS of
    this model, and the 2 rightmost columns  present the best-fit
    concentration and scale radius of the corresponding NFW profile,
    used to derive the systematic errors on the model.}
\begin{tabular}{lrrrrrrrllllllll}
Comp. & $\Delta\alpha$ & $\Delta\delta$ & $e$ & $\theta$ & r$_{\rm core}$ & r$_{cut}$ & $\sigma_0$ & rms & $c_{\rm NFW}$ & r$_{s,NFW}$ & rms$_{\rm NFW}$\\
 & [\arcsec] & [\arcsec] & & [deg] & [kpc] & [kpc] & [km\ s$^{-1}$] & [\arcsec] & & [kpc] & [\arcsec]\\
\hline 
A521 \\
DM1 & [0.0] & [0.0] & 0.67$\pm$0.03 & 49.5$\pm$0.7 & 18.5$\pm$3.2 & [1000.0] & 553$\pm$23 & 0.23 & 10.4$\pm$3.2  & 64.4$\pm$44.8  & 0.14\\
BCG & [0.0] & [0.0] & [0.238] & [47.6] & [0.0] & 14.1$\pm$15.2 & 20$\pm$71 &  &   &   & \\
L$^*$ gal & & & & & [0.15] & [45] & 124$\pm$12 & & & & \\
\hline
A611 \\
DM1 & [0.0] & [0.0] & 0.37$\pm$0.01 & -47.3$\pm$0.4 & 42.3$\pm$2.8 & [1000.0] & 854$\pm$9 & 0.21 & 8.7$\pm$0.4  & 161.7$\pm$11.1  & 0.30\\
PERT1 & [-10.8] & [10.3] & [0.346] & [88.10] & [0.109] & [32.60] & 133$\pm$8 &  &   &   & \\
BCG & [0.0] & [0.0] & [0.346] & [-61.5] & [0.0] & 99.6$\pm$26.5 & 304$\pm$32 &  &   &   & \\
L$^*$ gal & & & & & [0.15] & [45] & 124$\pm$5 & & & & \\
\hline
A773\\
DM1 & [0.0] & [0.0] & 0.62$\pm$0.15 & -37.3$\pm$6.3 & 42.1$\pm$27.5 & [1000.0] & 501$\pm$79 & 0.45 & 6.1$\pm$1.9  & 187.8$\pm$43.4  & 0.34 \\
DM2 & [0.0] & [24.0] & 0.47$\pm$0.08 & -20.2$\pm$7.7 & 128.2$\pm$20.3 & [1000.0] & 836$\pm$99 &  & 4.1$\pm$1.0  & 267.4$\pm$86.1  & \\
DM3 & -119$\pm$11 & 6$\pm$7 & 0.42$\pm$0.17 & -54.4$\pm$54.8 & [75.0] & [1000.0] & 996$\pm$94 &  & 9.2$\pm$2.1  & 196.2$\pm$148.3  & \\
BCG & [0.0] & [0.0] & [0.297] & [-41.0] & [0.396] & [79.272] & 353$\pm$103 &  &   &   & \\
PERT1 & [-0.6] & [24.0] & [0.208] & [10.0] & [0.421] & [84.163] & 411$\pm$111 &  &   &   & \\
PERT2 & [-52.2] & [7.5] & [0.373] & [-43.80] & [0.138] & [41.42] & 169$\pm$13 &  &   &   & \\
L$^*$ gal & & & & & [0.15] & [45] & 177$\pm$10 & & & & \\
\hline
A868\\
DM1 & [0.0] & [0.0] & [0.0] & [-66.5] & 71.4$\pm$26.4 & [1000.0] & 1078$\pm$257 & 0.05 & 7.9$\pm$3.5  & 33.4$\pm$193.1  & 0.04\\
DM2 & [-21.5] & [11.7] & 0.42$\pm$0.12 & 26.2$\pm$17.0 & 62.8$\pm$18.3 & [1000.0] & 426$\pm$93 &  & 2.3$\pm$2.6  & 748.7$\pm$209.5  & \\
L$^*$ gal & & & & & [0.15] & [45] & 161$\pm$26 & & & & \\
\hline
Z2701\\
DM1 & [0.0] & [0.0] & 0.28$\pm$0.05 & 55.4$\pm$1.9 & 64.6$\pm$16.5 & [1000.0] & 1008$\pm$70 & 0.11 & 3.3$\pm$1.2  & 711.8$\pm$151.5  & 0.19\\
BCG & [0.0] & [0.0] & [0.18] & [60.9] & [0.0] & 9.3$\pm$27.7 & 292$\pm$55 &  &   &   & \\
L$^*$ gal & & & & & [0.15] & [45] & 79$\pm$26 & & & & \\
\hline
A1413\\
DM1 & [0.0] & [0.0] & 0.67$\pm$0.02 & 85.1$\pm$0.5 & 67.1$\pm$5.9 & [1000.0] & 941$\pm$23 & 0.53 & 2.9$\pm$0.5  & 691.4$\pm$108.8  & 0.57\\
BCG & [0.0] & [0.0] & [0.710] & [65.0] & [0.06] & 125.7$\pm$35.9 & 334$\pm$16 &  &   &   & \\
PERT1 & [13.2] & [-19.9] & [0.116] & [36.60] & [0.104] & [31.247] & 168$\pm$8 &  &   &   & \\
L$^*$ gal & & & & & [0.15] & [45] & 107$\pm$4 & & & & \\
\hline
A1835\\
DM1 & 4.8$\pm$0.0 & 0.7$\pm$0.3 & 0.57$\pm$0.01 & 77.7$\pm$0.1 & 99.1$\pm$1.3 & [1000.0] & 1219$\pm$2 & 3.15 & 5.7$\pm$0.1  & 341.0$\pm$8.3  & 2.96\\
BCG & 1.4$\pm$0.1 & -0.3$\pm$0.1 & [0.142] & [70.0] & [4.384] & 24.7$\pm$1.5 & 880$\pm$13 &  &   &   & \\
PERT1 & [15.0] & [-19.8] & [0.720] & [60.6] & [0.098] & 94.1$\pm$34.6 & 111$\pm$30 &  &   &   & \\
PERT2 & [-17.6] & [-23.7] & [0.229] & [-36.7] & [0.124] & 2.5$\pm$3.5 & 363$\pm$43 &  &   &   & \\
L$^*$ gal & & & & & [0.15] & [45] & [158] & & & & \\
\hline
A2204\\
DM1 & [0.0] & [0.0] & 0.54$\pm$0.15 & 134.6$\pm$5.3 & 13.2$\pm$18.5 & [1000.0] & 556$\pm$158 & 0.29 & 3.5$\pm$3.2  & 687.5$\pm$197.8  & 0.20\\
L$^*$ gal & & & & & [0.15] & [45] & 238$\pm$25 & & & & \\
\hline
RXJ1720\\
DM1 & [0.0] & [0.0] & 0.59$\pm$0.19 & [-66.9] & 9.9$\pm$17.3 & [1000.0] & 539$\pm$143 & 0.15 & 13.9$\pm$3.2  & 61.4$\pm$168.8  & 0.24\\
L$^*$ gal & & & & & [0.15] & [45] & 127$\pm$22 & & & & \\
\hline
RXJ2129\\
DM1 & [0.0] & [0.0] & 0.46$\pm$0.15 & -16.4$\pm$2.5 & 45.2$\pm$13.9 & [1000.0] & 755$\pm$98 & 0.11 & 5.9$\pm$2.0  & 198.6$\pm$79.8  & 0.23\\
BCG & [0.0] & [0.0] & [0.490] & [-35.4] & [0.172] & [1.988] & 335$\pm$129 &  &   &   & \\
L$^*$ gal & & & & & [0.15] & [45] & 229$\pm$35 & & & & \\
\hline
\end{tabular}

\end{table*}

The best fit parameters of the mass models are listed in Table
\ref{best}, for both models that parameterize the cluster-scale mass
components as dPIE and NFW profiles.  The parameter uncertainties are
based on the MCMC chains generated by Lenstool \citep[see][for
details]{Jullo}.  We obtain $\sigma_i\ls0.5\arcsec$ for 9/10 clusters,
with 7/10 having $\sigma_i\ls0.2\arcsec$.  These results are typical
of strong lensing studies using a similar number of multiples images
\citep{Richard07,Limousin07,Richard09}.  However in the case of
A\,1835, where we have the largest number of multiple-image
constraints (7 systems), we obtained $\sigma_i\sim3.15\arcsec$, which
is comparable with recent results on A\,1689 ($2.87\arcsec$,
\citealt{Limousin07}) using 32 systems.  We discuss the origin of this
large value of $\sigma_i$ for A\,1835 in \S\ref{a1835}.  

The best fit ellipticity $e$ and position angle $\theta$ of the
cluster-scale dark matter halos are in good agreement between the dPIE
and NFW models, with $0\ls e\ls0.7$, and $\theta$ of the main
cluster-scale halo agreeing with the orientation of the BCG.  A few
clusters have $e>0.5$; \citet{Golse2002} have shown that the
pseudo-elliptical NFW profile presents a boxy/peanut shape at such
high ellipticities.  We therefore adopt the dPIE models as the
fiducial models for the rest of the paper, and use the NFW models to
quantify systematic errors in \S\ref{syst}.

For 8/10 clusters a single dPIE or NFW cluster-scale dark matter halo
is needed in addition to the cluster galaxies to reproduce accurately
the observed multiply-imaged systems. However, for two clusters, the
strong lensing constraints reveal the influence of additional
cluster-scale dark matter halos:

\begin{itemize}
\item{A\,868: The overall shape of the giant arc providing the
    constraints in this cluster is quite straight, and slightly curved
    towards the second brightest galaxy instead of the cluster
    centre. This shows that it is mostly influence by the presence of
    a secondary component, which we parametrize as centred on this
    second galaxy.}
\item{A\,773: The large number of highly sheared arcs and multiple
    images show the presence of a dual component at the centre, as
    well as a third dark matter clump towards the east. Using the new
    spectroscopic redshifts of multiple images, we confirm the result
    of Sm05, who included 3 components in the lens model to explain the
    overall weak-lensing signal.}
\end{itemize}

We compare the magnification factors $\mu$ computed from the the lens
models with the photometry of each image in a given multiple system
(last columns of Table~\ref{mult}).  We find only a marginal agreement
between the measured photometry and these magnifications, which is
certainly due to the fact that the magnification factors are derived
only at a single position, whereas many of the arcs are fairly
extended, and also due to the surface brightness limits when measuring
the photometry of the faintest images. The majority of the multiple
images are strongly magnified, with typically $\mu\sim2$ mags (or
$5-10$ on linear scale).  Errors on the magnification reach very large
values ($\Delta\mu>1$ mag) for multiple images in the vicinity of the
critical line ($\mu\gtrsim4$ mags, or $\gtrsim30$ on linear scale).


\subsection{Systematic Errors}\label{syst}

In this section we discuss four systematic uncertainties in our mass
models: inability of parameterized models to fit the multiple-image
constraints in A\,1835 (\S\ref{a1835}), the impact of fixing the size
of $L^\star$ galaxies on the modeling results (\S\ref{lstar}),
reliability of identification of substructures (\S\ref{clumps}), and
choice of parameterization for cluster-scale dark matter halos
(\S\ref{pchoice}).

\subsubsection{Quality of Fit for A\,1835}\label{a1835}

We tried to improve the quality of the fit to the multiple-image
constraints in A\,1835 ($\sigma_i\simeq3\arcsec$) by adding a
secondary dPIE cluster-scale halo on the next brightest peaks seen on
the $K$-band light map (see \S\ref{clumps}), as well as allowing the
positions of the main cluster-scale halo (previously with its centre
fixed on that of the BCG) to be free parameters.  This did not yield
any significant improvements in the quality of the fit.  The high
value of $\sigma_i$ is due to the inability of the parameterized model
to reproduce simultaneously the multiple-image systems located to the
North and to the South of the BCG.  It therefore appears that the main
limitation of the current model for this cluster is the assumption of
elliptical symmetry in the mass components.

To explore this, we used the new, more flexible method presented by
\citet{Jullo2} that employs a multi-scale adaptive grid to refine a
parametric model that is constrained by a large number of strong
lensing constraints -- thus dropping the assumption of elliptical
symmetry.  The resulting mass map shows a mass extension to the North
of the BCG and less extended to the South, compared to the original
dPIE model.  The quality of the fit is improved to
$\sigma_i=1.7\arcsec$ for the same multiply-imaged systems.  The lack
of spectroscopic redshifts in the Northern multiple images precludes
strong conclusions about this asymmetry.  Securing
spectroscopic redshifts for these multiple-image systems is therefore
a necessary step to further improve this mass model.  Overall, the
absence of a secondary cluster-scale component in the simple dPIE
model and the wide distribution of multiply-imaged galaxies across the
cluster core reassure us that the absolute calibration of the total
mass, Einstein radius and cluster substructure are reasonably
accurate.

\subsubsection{Fixed Size of $L^\star$ Cluster Galaxies}\label{lstar}

We note that other galaxy-galaxy weak-lensing works have obtained
different values for $r_{\rm cut}^*$, but within the range $10<r_{\rm
  cut}^\star<100\kpc$ \citep{Natarajan02,Limousin07b,Halkola07}.  To quantify
how much our choice of $r_{\rm cut}^\star=45\kpc$ (\S\ref{gals})
affects mass model results, we therefore ran the following test: we
included $r_{\rm cut}^\star$ as a free parameter for the 5 models
having the largest number of constraints (i.e. A\,611, A\,773,
A\,1413, A\,1835 and Z\,2701), allowing it to vary in the range
$[10-100]\kpc$.  The best fit parameters and quality of fit (judged by
$\sigma_i$) of these 5 models are all consistent with the fiducial
models in Table~\ref{best}, specifically, the best-fit values of
$r_{\rm cut}^\star$ are all consistent with $r_{\rm cut}^\star=45\kpc$
within their respective statistical uncertainties.  The degeneracy
between $r_{\rm cut}^\star$ and $\sigma^\star$ also obey the expected
$r_{\rm cut}^\star\ \sigma^{\star 2}$=Const. relation -- i.e.\ a constant
amount of mass is assigned to galaxy-size haloes (within 10\%)
regardless of the specific values of these two parameters.  This folds
through to an uncertainty of just 15\% on the cluster substructure
fractions, which is negligible in comparison with the statistical
errors.

\subsubsection{Identification of Massive Substructures}
\label{clumps}

\begin{figure*}
\includegraphics[width=17cm,angle=0]{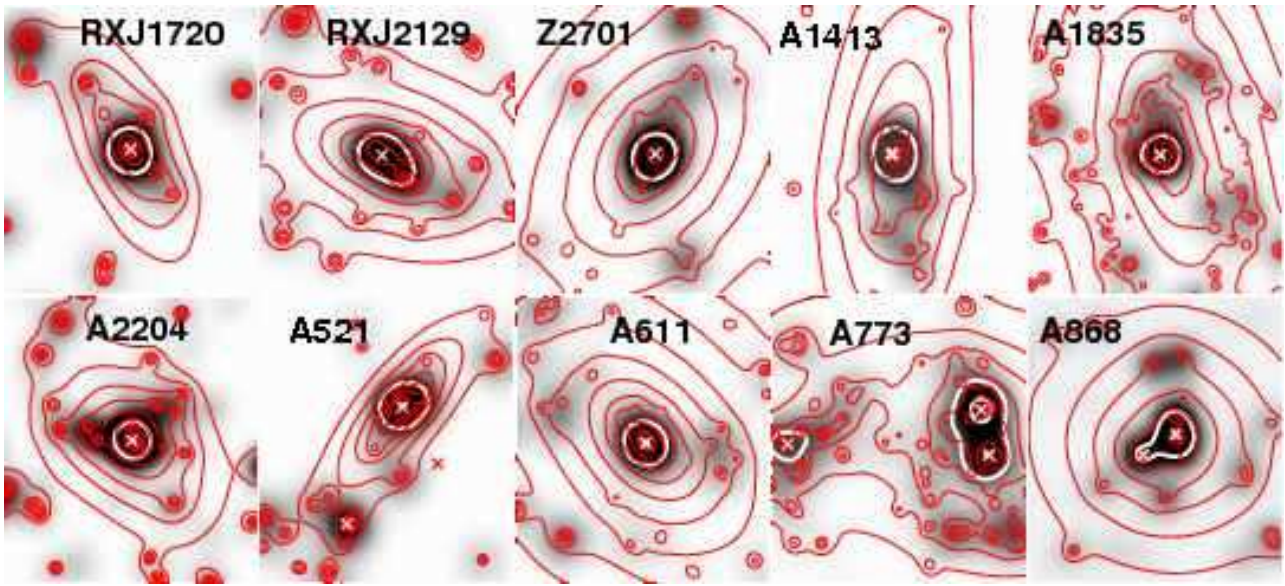}
\caption{\label{klight} $K$-band cluster luminosity density maps (gray
  scale) smoothed with a FWHM=20\arcsec gaussian. Red contours show
  the reconstructed dark matter distribution with a constant
  logarithmic scale. The white contours show the selection of $K$-band
  light peaks as local maxima (crosses, see \S\ref{clumps} for
  details). }
\end{figure*}

The necessity or not to add a second or third cluster-scale mass
component to reproduce the observed multiple-image systems has been
determined by the number and location of the strong-lensing
constraints.  The presence of massive substructures has direct
implications on the substructure measurements discussed in
\S\ref{results}.  We therefore test the fiducial mass models by
comparing quantitatively the distribution of $K$-band light in the
cluster cores with the mass modeling results.

We constructed $K$-band luminosity density maps from the cluster
galaxy catalogues discussed in \S\ref{nir}, with a Gaussian smoothing
scale of ${\rm FWHM}=20\arcsec$ (corresponding to $\sim70\kpc$ at
$z=0.2$), which is the smallest scale over which we measured the
influence of a secondary clump on the strong lensing images.  The
location of the highest peak ($L_{\rm K,max}$) in the $K$-band light
maps always coincides (within 1\arcsec) with the centre of the
cluster-scale mass distribution.  We searched for secondary peaks in
the $K$-band light by adopting a threshold $L>0.5L_{\rm K,max}$ and
look for the local maxima within these regions (Figure \ref{klight},
white crosses).  When comparing their location with the cluster-scale
dark matter clumps, we find that every local maximum is coincident
with a cluster-scale mass component in our models within the modelling
errors (1\arcsec), except for the secondary peak in A\,521 (located
$\sim1\arcmin$ South of the BCG).  It therefore appears that we are
unable to detect this structure because the strong-lensing constraints
lie exclusively North of the BCG.  The presence of substructure in
A\,521 is consistent with the dynamical study of \citet{Maurogordato},
who identified many large-scale components in A\,521, and suggested
that it is a highly disturbed merging cluster.  Our inability to
detect any of these structures may explain why A\,521 is an outlier in
the $\fsub-\dm$ relation discussed in \S\ref{gap}.

\subsubsection{Parameterization of Dark Halos}\label{pchoice}

\begin{figure*}
\begin{minipage}{5.0cm}
\includegraphics[height=5cm,angle=270]{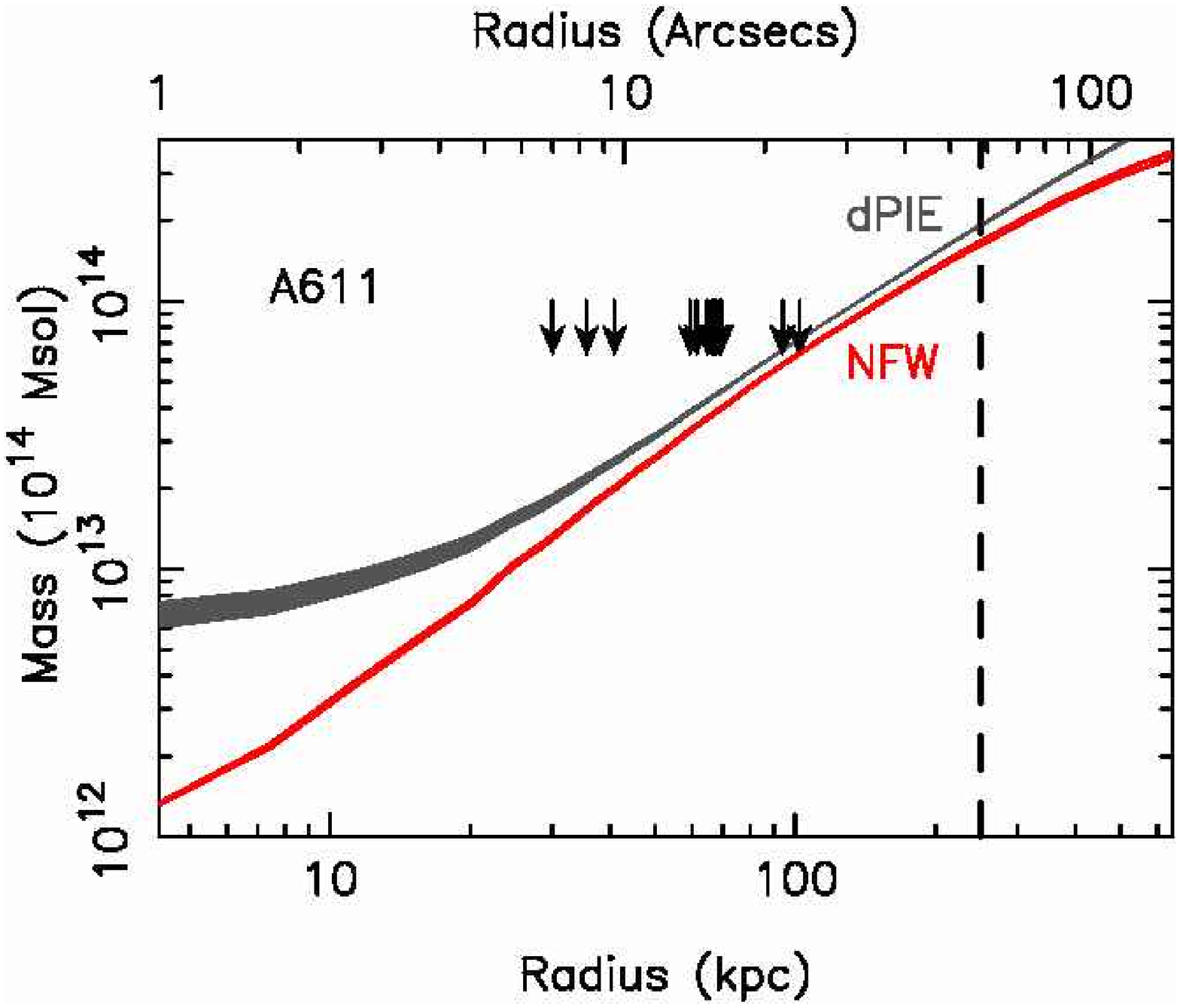}
\end{minipage}
\begin{minipage}{5.0cm}
\includegraphics[height=5cm,angle=270]{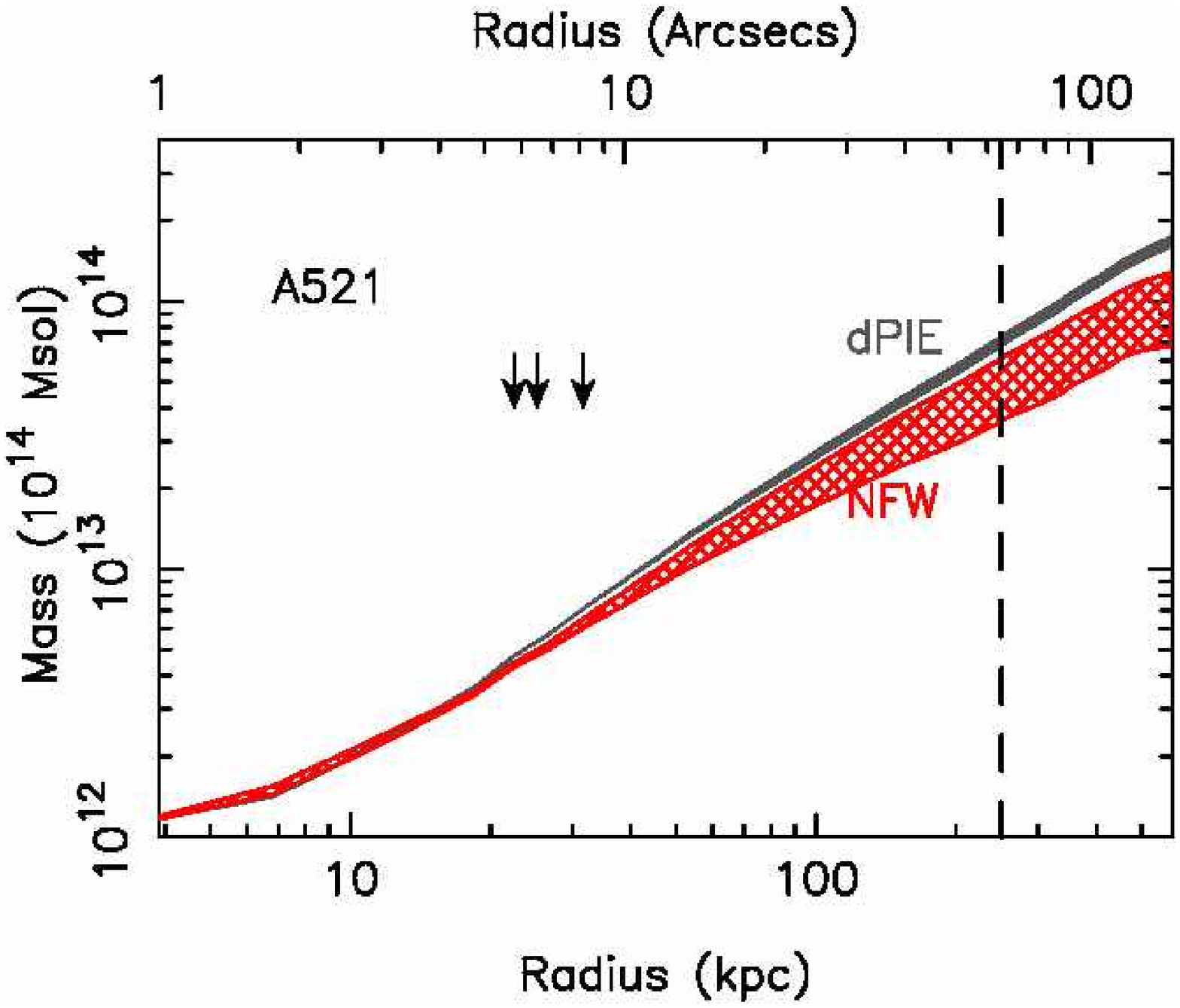}
\end{minipage}
\begin{minipage}{5.0cm}
\includegraphics[height=5cm,angle=270]{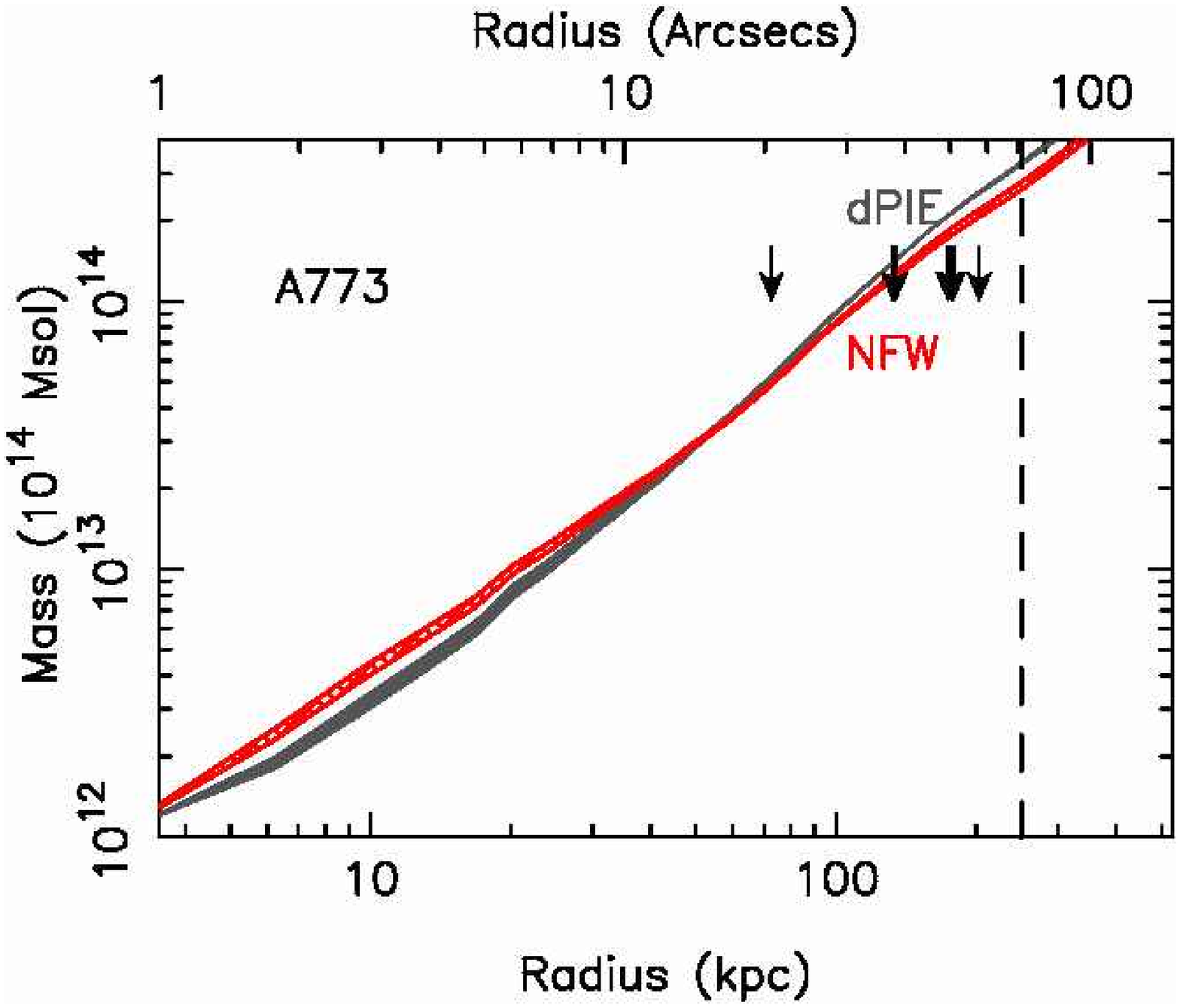}
\end{minipage}
\begin{minipage}{5.0cm}
\includegraphics[height=5cm,angle=270]{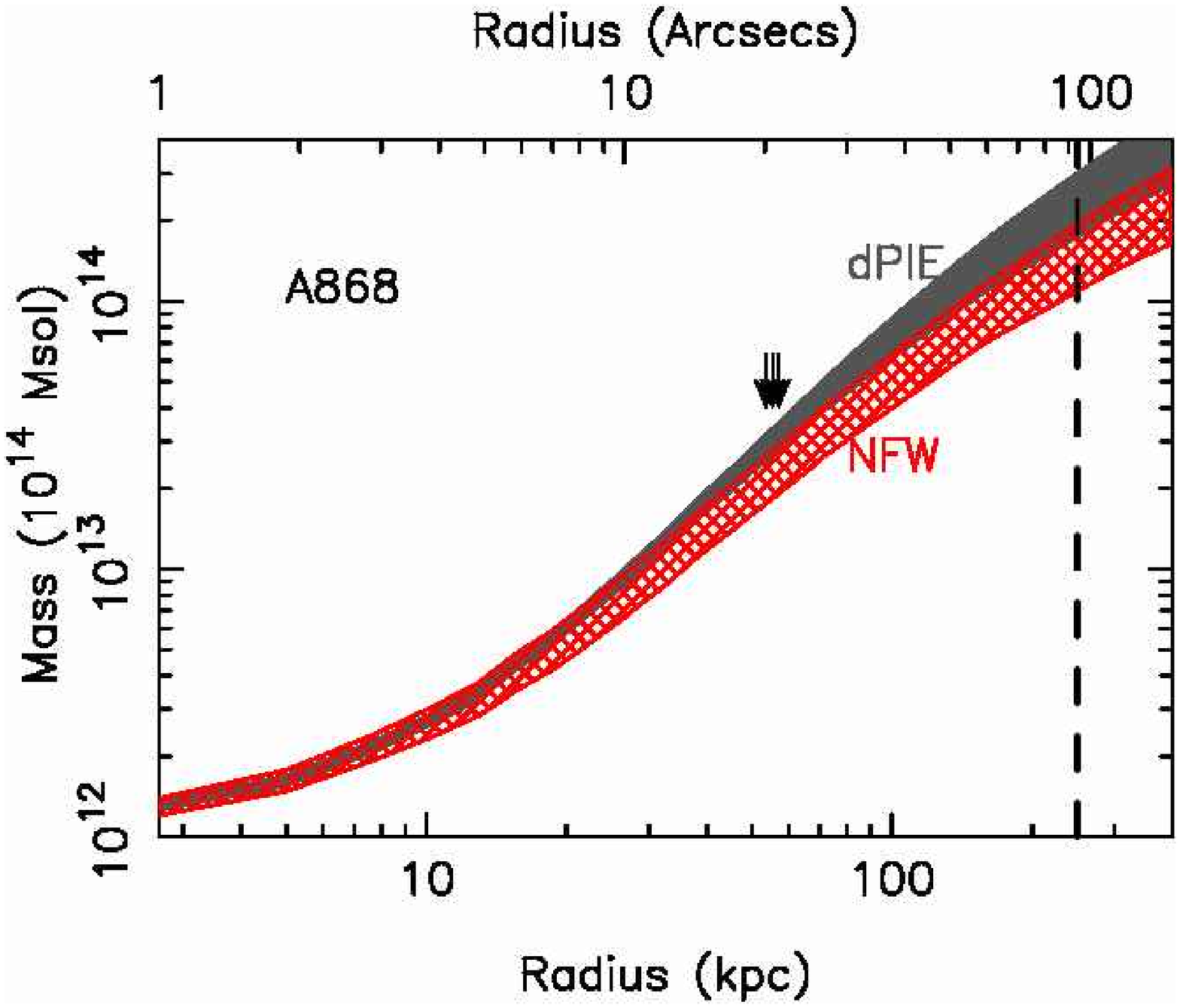}
\end{minipage}
\begin{minipage}{5.0cm}
\includegraphics[height=5cm,angle=270]{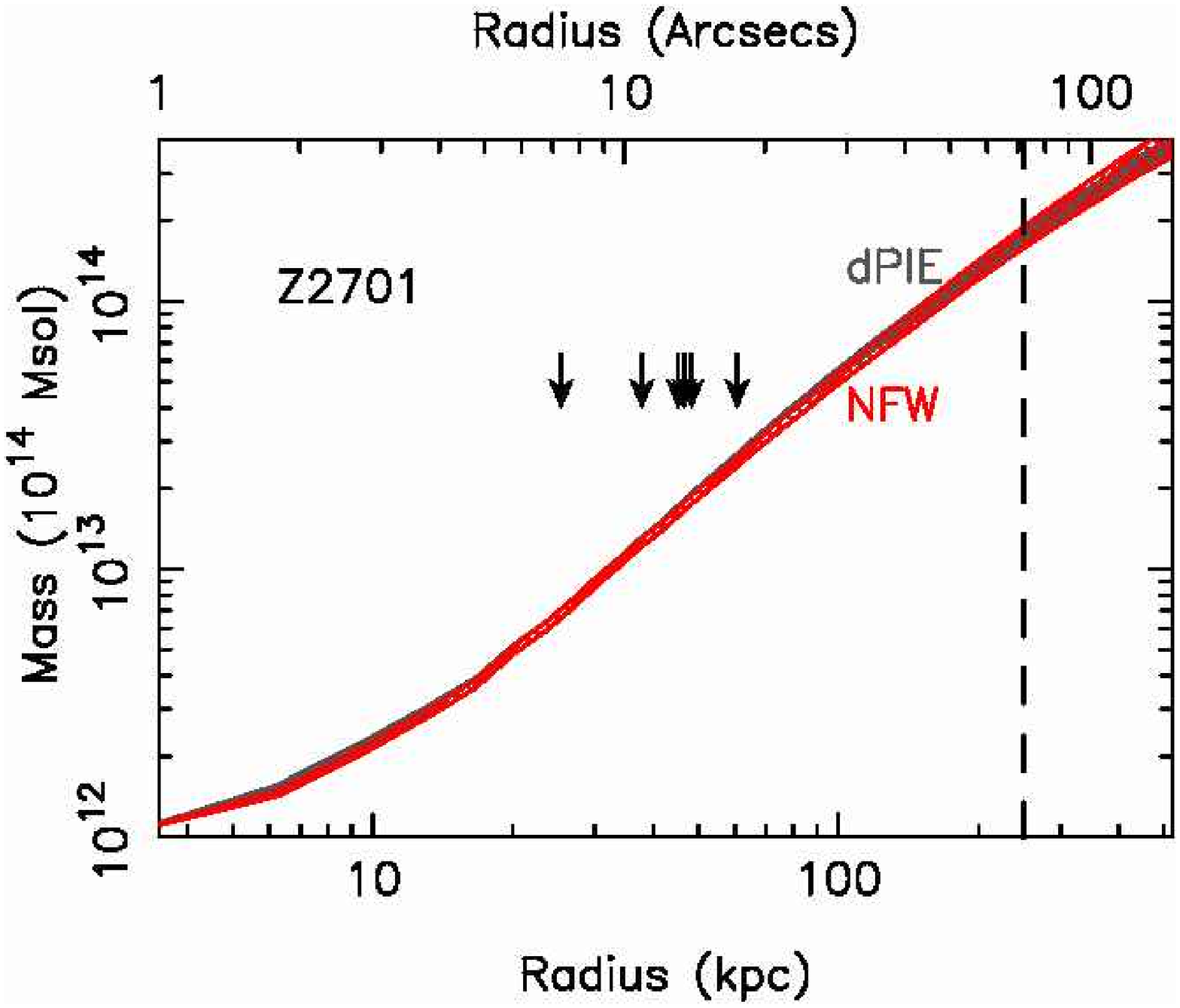}
\end{minipage}
\begin{minipage}{5.0cm}
\includegraphics[height=5cm,angle=270]{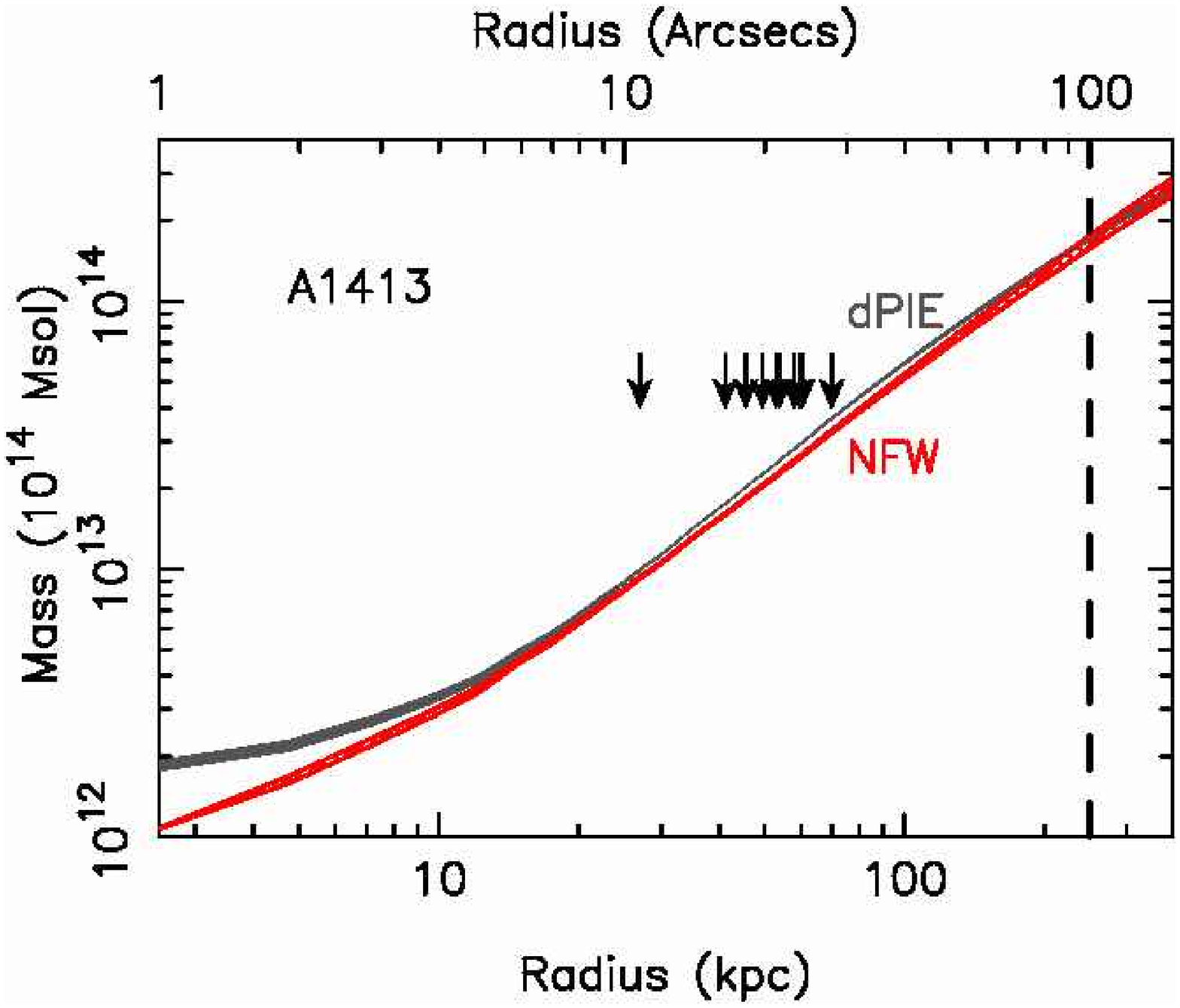}
\end{minipage}
\begin{minipage}{5.0cm}
\includegraphics[height=5cm,angle=270]{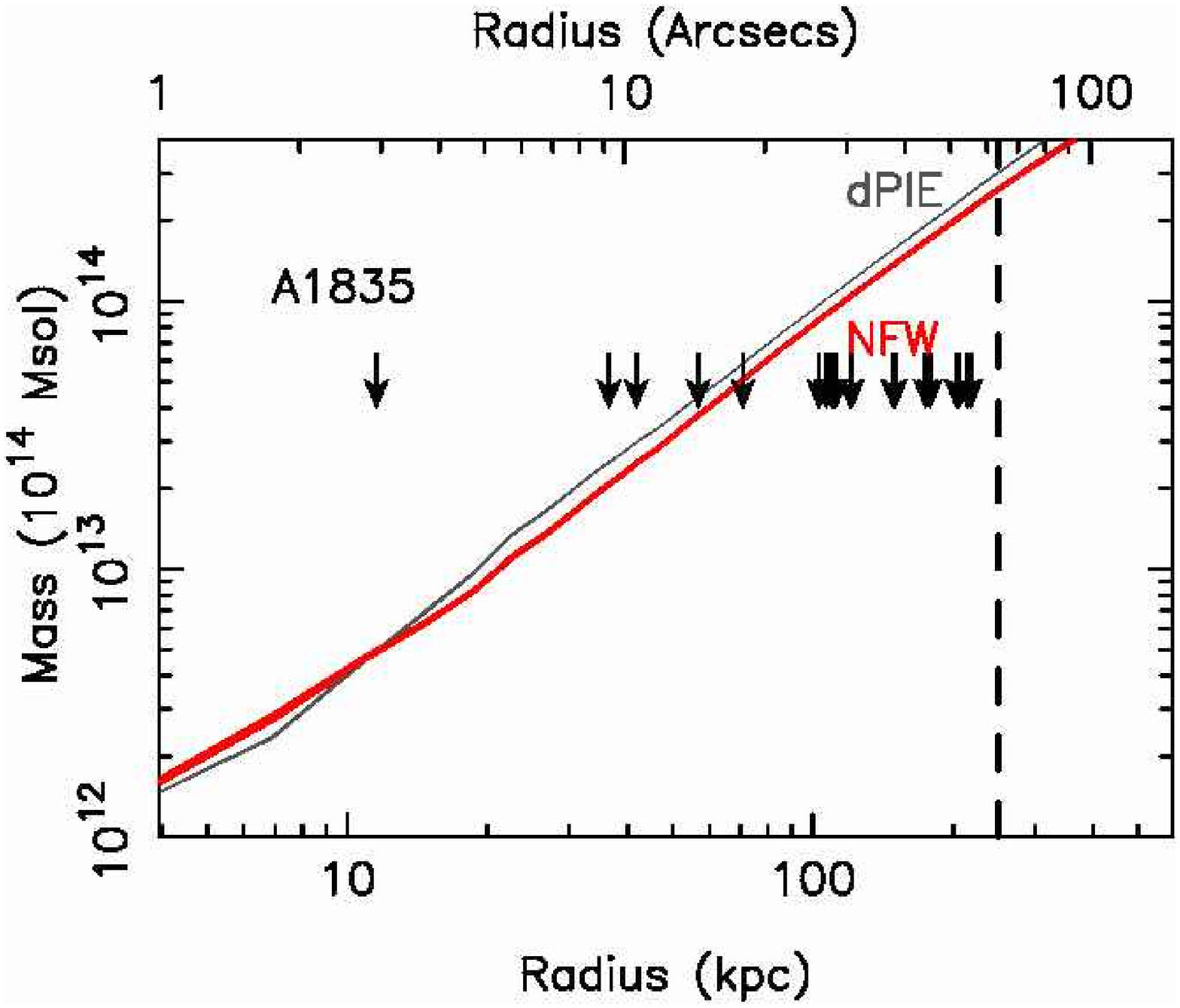}
\end{minipage}
\begin{minipage}{5.0cm}
\includegraphics[height=5cm,angle=270]{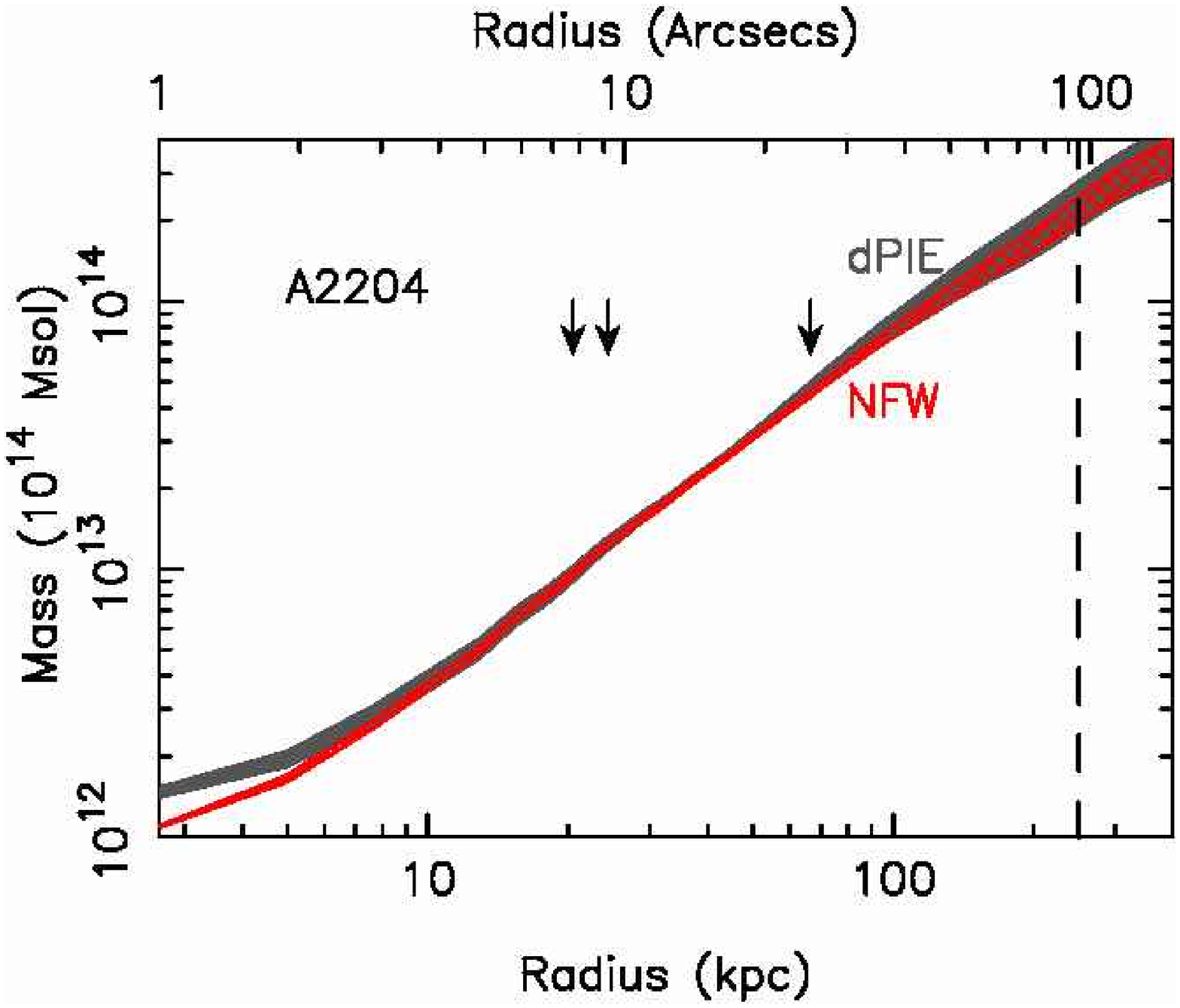}
\end{minipage}
\begin{minipage}{5.0cm}
\includegraphics[height=5cm,angle=270]{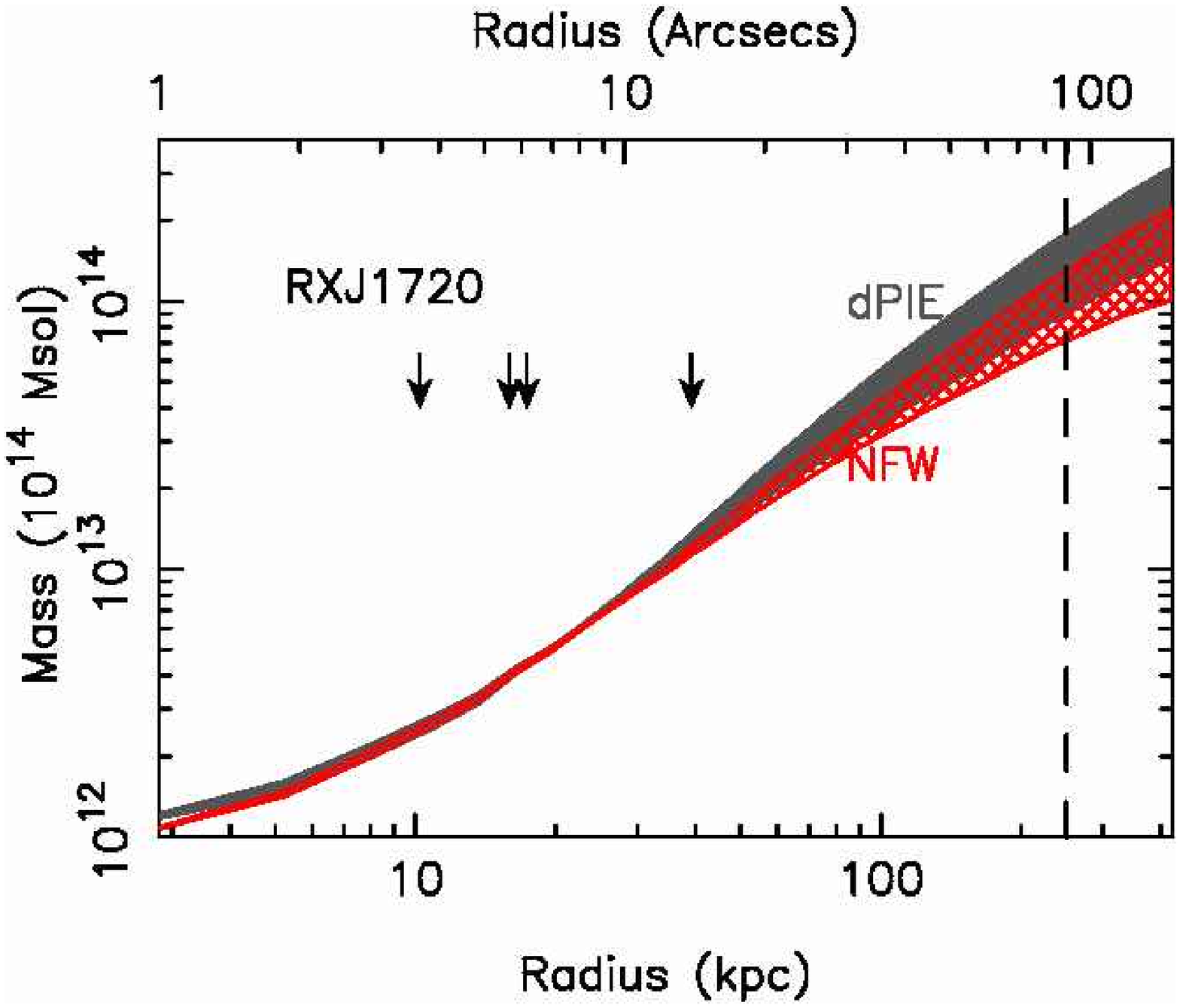}
\end{minipage}
\begin{minipage}{5.0cm}
\includegraphics[height=5cm,angle=270]{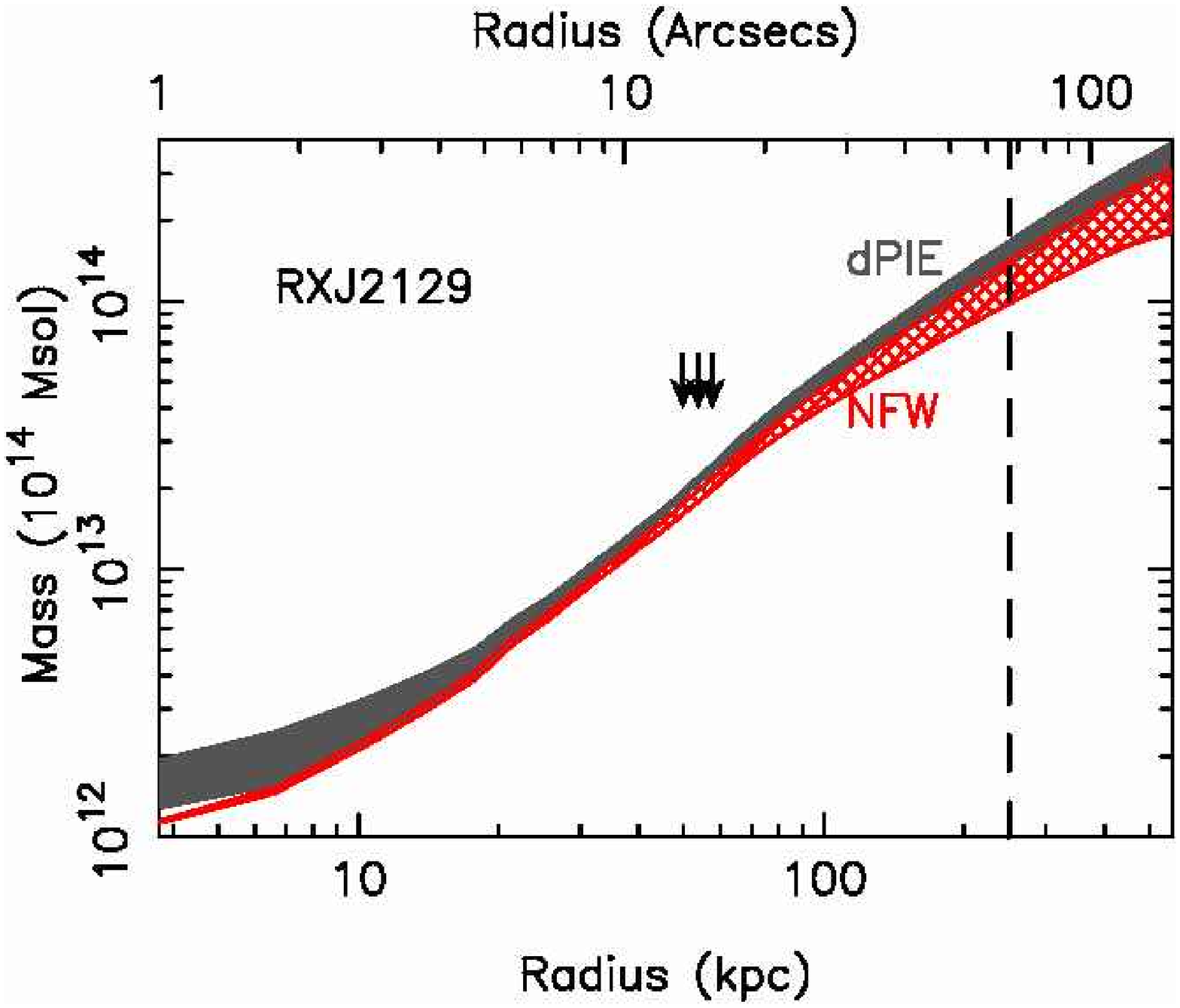}
\end{minipage}
\caption{\label{massprof} Integrated mass profiles assuming a dPIE
  (solid grey) or a NFW (hatched red) profile for the dark-matter
  distribution. The arrows mark the distance of the multiple images
  used as constraints in each cluster, and the vertical dashed
  corresponds to the radius of $250\kpc$ used to derive the enclosed
  masses.}
\end{figure*}

To examine the amplitude of any systematic error in cluster mass
measurement arising from choice of dPIE or NFW mass profiles, we show
the projected mass profiles $M_{\rm SL}(R)$ in Fig.~\ref{massprof}
from both dPIE and NFW models.  The statistical error on $M_{\rm
  SL}(R)$ is typically $0.05$ to $0.2\dex$, depending on the number of
lensing constraints.  We find a general agreement between the $M_{\rm
  SL}(R)$ profiles obtained from the best-fitting dPIE and NFW models,
and use the average difference between the two as an estimate of the
systematic error made by assuming a specific cluster-scale profile --
this systematic error is comparable with the statistical errors.

\section{The Mass and Structure of Cluster Cores}
\label{results}

In this section we use the strong lens models of the full sample of 20
clusters to construct a statistical sample of measurements of the mass
and structure of cluster cores, for comparison with theoretical
predictions.  Despite the modest statistical significance achieved in
these comparisons, due to the sample size, this is so far the largest
sample of strong lensing clusters analyzed in a uniform manner, and
represents a four-fold increase on the strong-lensing clusters studied
by Sm05.  Most importantly, these results provide a statistical
context within which to view results from detailed analysis of
spectacular individual cluster lenses.

We begin by describing integrated cluster masses and related
quantities such as Einstein radii in \S\ref{mass}, and then
concentrate on structural quantities such as substructure fractions
and cluster ellipticities in \S\ref{struct}.

\subsection{Cluster Masses and Einstein Radii}\label{mass}

\subsubsection{Einstein radii}\label{einstein}

\begin{figure}
\centerline{\includegraphics[width=5.5cm,angle=270]{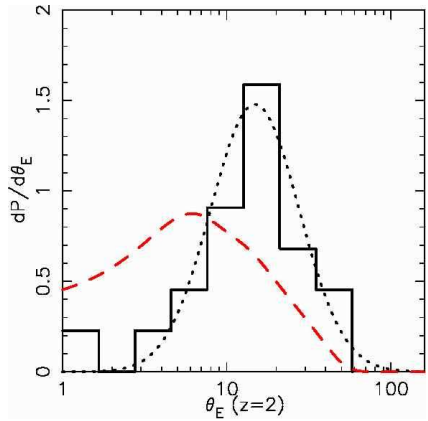}}
\caption{\label{redist} The observed distribution of effective
  Einstein radii for a source at $z_s=2.0$ (solid black histogram),
  and best-fit lognormal distribution (black dotted line).  We also
  plot the predicted distribution based on the NFW model parameters 
  (Table~\ref{best}), and the predicted distributions of halo properties
  in strong-lensing selected simulated clusters -- see
  \S\ref{einstein} for more details.}
\end{figure}

We first measure the effective Einstein radius $\theta_E$ of each
cluster at $z_S=2$.  $\theta_E$ is defined as the angular radius
$\theta$ from the centre of the cluster at which the average
convergence $\bar{\kappa}(\theta<\theta_E)=1$ \citep{Broadhurst08b}.
We choose $z_s=2$ because this is the typical redshift of the
spectroscopically confirmed multiply-imaged galaxies
(Table~\ref{mult}).  Values of $\theta_E$ range between $\sim4\arcsec$
and $47\arcsec$, are listed in Table~\ref{physics}, and the
distribution is plotted in Fig.~\ref{redist}.  The $\theta_E$
distribution is best-fitted by a log-normal distribution with
$\langle\log_{10}\theta_E\rangle=1.16\pm0.28$, where $\theta_E$ is
measured in arcseconds.  Two clusters (A\,1689 and Abell 1703) are in
common between this study and the \citet{Broadhurst08b} sample of
clusters with large Einstein radii.  These two clusters are located at
$2\sigma$ and $1.5\sigma$ above $\langle\log_{10}\theta_E\rangle$,
suggesting that $\sim2.3\%$ and $\sim6.7\%$ of clusters found in
larger sample will have $\theta_E$ at least as large as those of
A\,1689 and A\,1703 respectively.

Comparison of observed Einstein radii with theoretical predictions
from numerical simulations is problematic because the simulations
require both sufficiently large volume to contain a large sample of
clusters as massive as observed systems, and sufficient numerical
resolution to allow $\theta_E$ to be measured reliably from the
simulated data.  Even modern simulations such as the Millennium
Simulation are unable to satisfy both requirements, the main
shortcoming being the simulation volume.  Nevertheless, it is
interesting to make a comparison.  We therefore take the best-fit mass
and concentration of the ten NFW models in Table~\ref{best} and
convolve the measured concentrations with the predicted concentration
distributions in \citet{Broadhurst08b}.  We then calculate the
predicted $\theta_E$ distributions and sum them to produce the red
dashed curve in Fig.~\ref{redist}.  This distribution is much broader
than the observed distribution, mainly because the virial masses of
the NFW models are poorly constrained by strong-lensing constraints
alone.  In spite of this, we find that the predicted distribution
peaks at $\theta_E\simeq5\arcsec$, a factor of 2 lower than the
observed distribution.  This difference may be caused by an important
difference between the simulations and observations, namely the
presence of baryons in observed universe and the absence of them from
the simulations.  However it is interesting to note that in a recent
weak-lensing study of similarly sized sample, \citet{Okabe} found that
the normalization of the mass-concentration of observed clusters is
roughly a factor of 2 higher than predicted from numerical
simulations.  The physical origin of both differences, if confirmed by
larger samples, may be similar.

\begin{table*}
  \caption{\label{physics}The mass and structure of the cluster cores
    inferred from the strong-lensing mass models, plus related
    quantities from the near-infrared cluster galaxy catalogs. } 
\begin{tabular}{lccllrlll}
Cluster &  $N_{\rm mult}^a$ & Radial?$^b$ & $M_{\rm SL}^c$ & $e_{\rm 2D}^d$ & $\theta_{E}$ ($z=2$) & $\fsub$ & $L_{\rm K,BCG}/L_{\rm K,tot}$ & $\dm$ \\
        &                    &         &               &             & (arcsec)            &         &                            & (mags) \\
\hline 
A\,521  &1 & No & 0.61$\pm$0.33 & 0.66 & $3.6\pm0.8$ &$0.13\pm0.04$ & $0.22\pm0.03$ & 0.05 \\
A\,611  & 4 & Yes & 1.76$\pm$0.33 & 0.36 & $21.0\pm1.3$ & $0.10\pm0.01$ & $0.44\pm0.06$ & 2.16\\
A\,773  & 3 & Yes & 3.01$\pm$0.58 & 0.385& $30.1\pm1.2$ & $0.78\pm0.03$ & $0.13\pm0.02$ & 0.13\\
A\,868  & 1 & No & 1.97$\pm$1.11 & 0.06 & $14.2\pm5.6$ & $0.26\pm0.12$ & $0.25\pm0.04$ & 0.81\\
Z2701   & 2 & No & 1.74$\pm$0.14 & 0.34 & $9.0\pm0.5$ & $0.04\pm0.02$ & $0.45\pm0.06$ & 2.33\\
A\,1413 & 4 & No & 1.71$\pm$0.20 & 0.64 & $11.9\pm0.5$ & $0.07\pm0.01$ & $0.32\pm0.05$ & 1.80\\
A\,1835 & 7 & Yes&  2.83$\pm$0.41 & 0.49 & $30.5\pm0.5$ & $0.13\pm0.01$ & $0.26\pm0.09$ & 1.57\\
A\,2204 &  1 & Yes & 2.29$\pm$0.50 & 0.27& $23.9\pm2.2$ & $0.25\pm0.10$ & $0.13\pm0.02$ & 0.14\\
RXJ1720 & 1& Yes &  1.18$\pm$0.59 & 0.59 & $7.0\pm0.5$ & $0.10\pm0.05$& $0.47\pm0.07$ & 1.60\\
RXJ2129 & 1 & No & 1.37$\pm$0.37 & 0.56 & $9.0\pm1.4$ &  $0.15\pm0.06$& $0.40\pm0.06$ & 1.26\\
\hline
\multicolumn{9}{c}{Extended sample -- see Table~\ref{extsample} for mass model references} \\
A68     & 6 & No & 2.16$\pm$0.23 & 0.23 & 7.5$\pm$0.5    & 0.33$\pm$0.04   & 0.32$\pm$0.04   & 1.40 \\
A383    & 3 & Yes & 1.87$\pm$0.26 & 0.22  & 10.4$\pm$2.6  & 0.02$\pm$0.01   & 0.46$\pm$0.06  & 1.90 \\
A963    & 2 & No & 1.74$\pm$0.44  & 0.355 & 7.5$\pm$0.5    & 0.13$\pm$0.07 &   0.38$\pm$0.05  & 1.26 \\
A1201   & 1 & No & 0.80$\pm$0.33 & 0.66 &1.5$\pm$0.15  & 0.02$\pm$0.01 & 0.64$\pm$0.09 & 2.54 \\
A2218   & 6 & Yes & 3.00$\pm$0.24 & 0.23& 18.3$\pm$0.5   & 0.54$\pm$0.01  &  0.18$\pm$0.03  & 0.46 \\
A2219   &6 & Yes & 2.33$\pm$0.23 & 0.41 &15.6$\pm$0.6    & 0.57$\pm$ 0.04  &  0.22$\pm$0.03  & 0.75 \\
A2390   & 4  & No  & 1.99$\pm$0.07 &0.14 & 17.5$\pm$0.5& 0.03$\pm$0.03& 0.30$\pm$0.04  & 1.53 \\
A2667   & 3 & No   & 2.41$\pm$0.07 & 0.39 & 13.0$\pm$0.6 & 0.14$\pm$0.03 &0.17$\pm$0.02 & 0.87\\
A1689   & 34& Yes & 4.53$\pm$0.13 & 0.22 & 47.1$\pm$0.5& 0.22$\pm$0.03 & 0.24$\pm$0.02 & 0.68 \\
A1703   & 16 & Yes             & 2.98$\pm$0.09& 0.39 & 36.8$\pm$1.5 &0.15$\pm$0.02 & 0.52$\pm$0.07 & 1.57 \\
\hline
\end{tabular}

$^a$~Number of  strongly-lensed galaxies used to constrain the mass model.
$^b$~Whether the strong-lensing constraints include a radial image pair.
$^c$~Projected mass within $R<250\kpc$ in units of $10^{14}\Msol$.
$^d$~Ellipticity of mass distribution in the cluster core inferred from the projected mass maps.
\end{table*}

\subsubsection{Projected Mass Measurements}

\begin{figure}
\centerline{\includegraphics[width=6cm,angle=270]{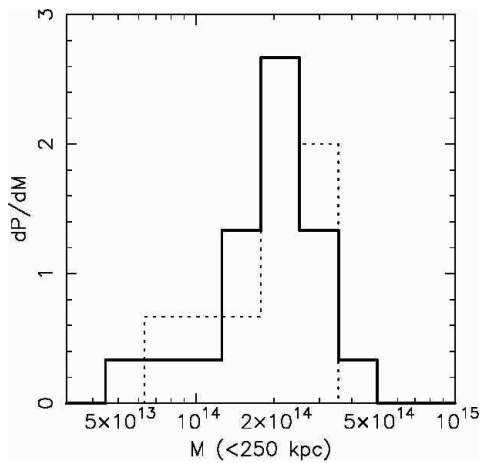}}
\caption{\label{mdist} Probability distribution of lensing masses
  (measured in a projected radius of 250 kpc) for the Sm05 sample
  (dashed histogram) and the current sample (solid histogram). }
\end{figure}

We show projected mass maps of the 10 new strong-lensing clusters in
Fig.~\ref{klight}, and integrate these maps and their equivalents for
the extended sample to measure the projected mass of the cluster cores
$M_{\rm SL}(R<250\kpc)$ -- see Table~\ref{physics}.  The mass
measurements for the extended sample were based on the published dPIE
models of these clusters listed in
Tables~\ref{extsample}~\&~\ref{best2}, adjusted to the cosmology used
in this paper where relevant.  The aperture of $R<250\kpc$ is chosen
because this ensures that the region within we measure mass lies
within the \emph{HST} field of view for all clusters; it is also
$\sim2\times$ the largest observed Einstein radius within this sample.

We compare the $M_{\rm SL}(R<250\kpc)$ distribution of the current
sample with the 10 clusters studied by Sm05 in Fig.\ref{mdist}, and
fit a log-normal distribution to both samples, obtaining:
$\langle\log_{10}(M_{\rm SL})\rangle=14.29\pm0.19$ and $14.27\pm0.17$
for our sample and Sm05 respectively.  This confirms that both studies
are probing clusters of the same mass, which is consistent with close
match in the range of X-ray luminosity probed: $4.3\times10^{44}\le
L_X\le22.8\times10^{44}\ergs$ in the case of Sm05, and
$3.5\times10^{44}\le L_X\le22.8\times10^{44}\ergs$ here
(Table~\ref{physics}). 

\begin{figure}
\includegraphics[width=8cm,angle=270]{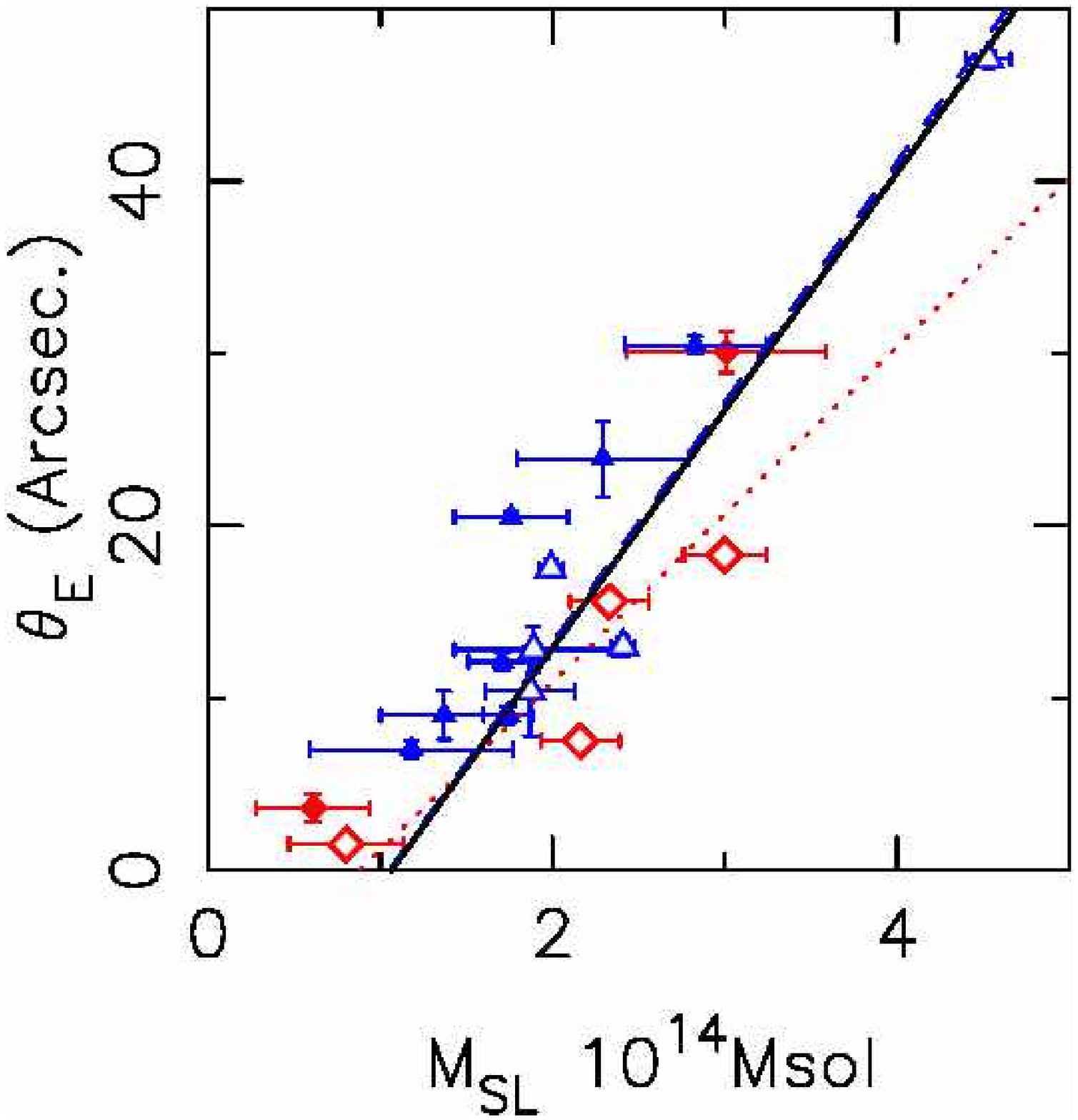}
\caption{\label{re}Effective Einstein radius (assuming a $z=2$ source)
  vs strong-lensing mass (measured within 250 kpc). The lines show the
  general trends when least-squared fitting a linear relation for all
  clusters (black solid line), undisturbed clusters (blue dashed line)
  and disturbed clusters (red dotted lines).  Red diamonds denote
  disturbed clusters and blue triangles undisturbed clusters.  Filled
  symbols are the new strong lensing models, and open symbols are
  taken from the extended sample.}
\end{figure}

 We also plot in Fig.~\ref{re} the relationship between
  $\theta_E$ and $M(<250\kpc)$, revealing a clear positive linear
  correlation between the two quantities, as is expected from the
  properties of analytic descriptions of dark matter density profiles
  from simulations.  The best-fit relation for the whole sample is
  $\theta_E=(-14.7\pm4.7)+(13.8\pm2.5)\,M_{\rm SL}$; the best-fit
  relations for cool-core and non-cool core sub-samples are both
  statistically indistinguishable from this relation.  However, we
  find that disturbed clusters tend to lie below the best-fit relation
  for the full sample -- indeed, we obtain best-fit relations of:
  $\theta_E=(-8.6\pm4.5)+(9.75\pm2.3)\,M_{\rm SL}$ and
  $\theta_E=(-14.7\pm5.0)+(14.0\pm2.2)\,M_{\rm SL}$ for disturbed and
  undisturbed clusters, respectively.  We interpret this as implying
  that disturbed clusters have flatter density profiles than
  undisturbed clusters of comparable mass.  This is likely due to a
  combination of (i) the cluster-cluster mergers that are likely to
  both soften the density profile, and also to cause the observed
  disturbance (i.e.\ cause the offset between the X-ray and optical
  centers of the clusters), and (ii) disturbed clusters tending to
  have their merger axis preferentially aligned in the plane of the
  sky, thus making the disturbance possible to measure.  This latter
  point is of particular importance with respect to A\,1689, because
  this cluster is classified as undisturbed despite there being strong
  evidence for it being a line-of-sight merger \citep{Limousin07}.  If
  this cluster were viewed side-on, then it would probably lie below
  the relation for undisturbed clusters, and be classified as
  disturbed. 

\subsubsection{Comparison of Lensing and X-ray Mass Measurements}\label{mlmx}

We further investigate the integrated mass of the cluster cores within
$R<250\kpc$ by comparing the $M_{\rm SL}$ with masses measured within
the same aperture from the X-ray models discussed in \S\ref{xprop}.
We used least-squared minimization, taking account of errors in both
quantities to find the best fit parameters of the relation: $M_{\rm
  SL}=B\,M_X$.  For the full sample, we obtain a best-fit of
$log_{10}(B)=0.13\pm0.04$, i.e.\ a mean X-ray/lensing mass discrepancy
of a factor of $M_{\rm SL}/M_X=1.3$ at $\sim3\sigma$ significance.

\begin{figure}
\includegraphics[width=8.5cm,angle=270]{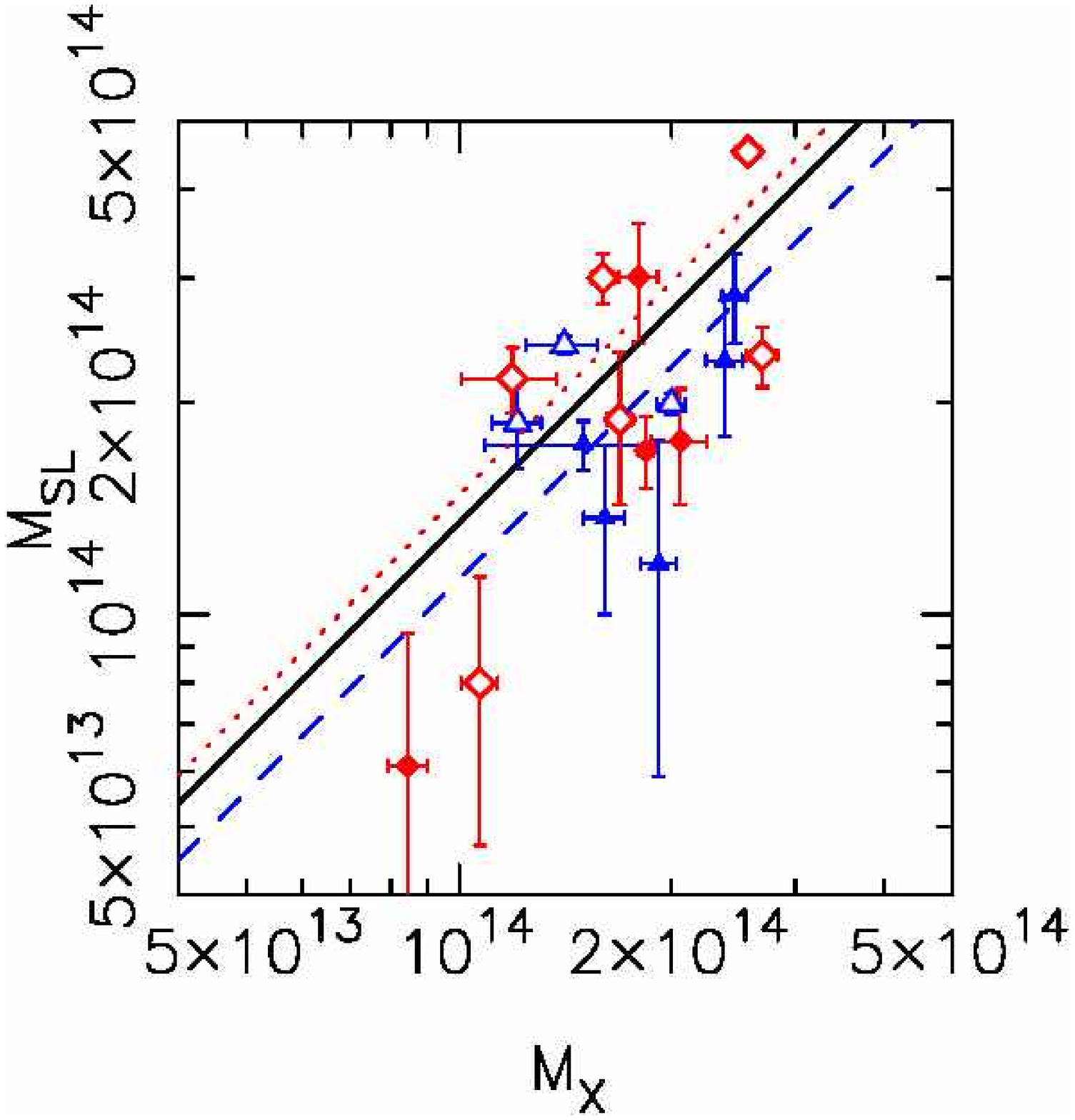}
\caption{\label{xfig1}Relationship between $M_X$ and $M_{\rm SL}$,
  both measured within $R<250\kpc$.  
  The lines show the general trends when least-squared fitting a linear relation for all
  clusters (black solid line), cool-core clusters (blue dashed line)
  and non-cool core clusters (red dotted line). Blue triangles denote cool-core clusters 
  and red diamonds non-cool core clusters.  Filled symbols are the new strong lensing 
  models, and open symbols are taken from the extended sample.
  }
\end{figure}

 We then fit the same relations for the cool core/non-cool core
  as well as disturbed/undisturbed cluster subsamples, and report the
  best fit values in Table \ref{fitx}.  We find that the
  disturbed/undisturbed sub-samples are statistically
  undistinguishable from both each other and from the entire
  sample. However, we find that the normalization, $log_{10}(B)$, of
  the relation for non-cool-core clusters is $2\sigma$ higher than for
  cool-core clusters. These differences are illustrated in
  Fig. \ref{xfig1}.  

\begin{figure*}
\begin{minipage}{5.5cm}
\centerline{\mbox{\includegraphics[width=5.5cm,angle=270]{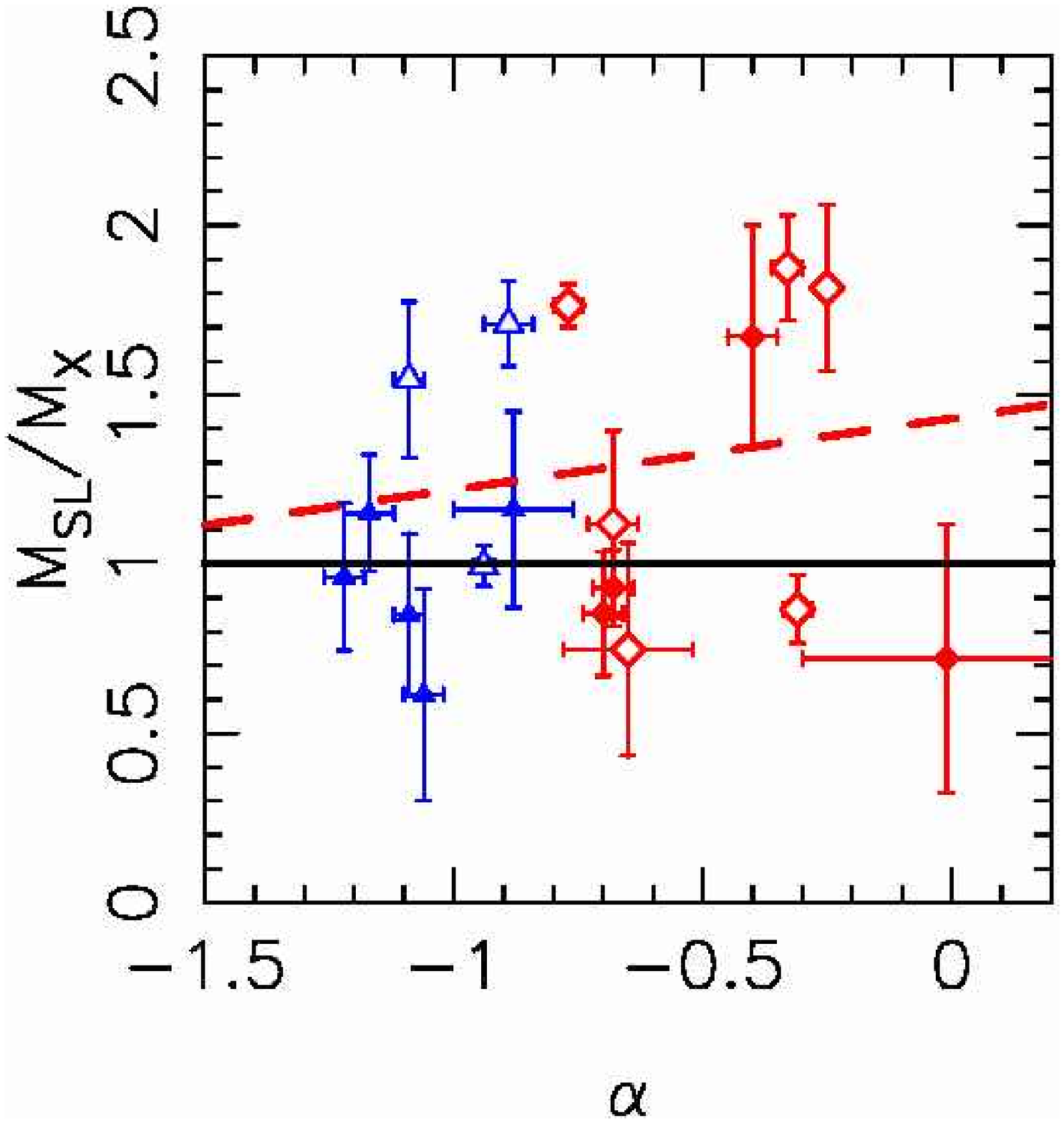}}}
\end{minipage}
\begin{minipage}{5.5cm}
\centerline{\mbox{\includegraphics[width=5.5cm,angle=270]{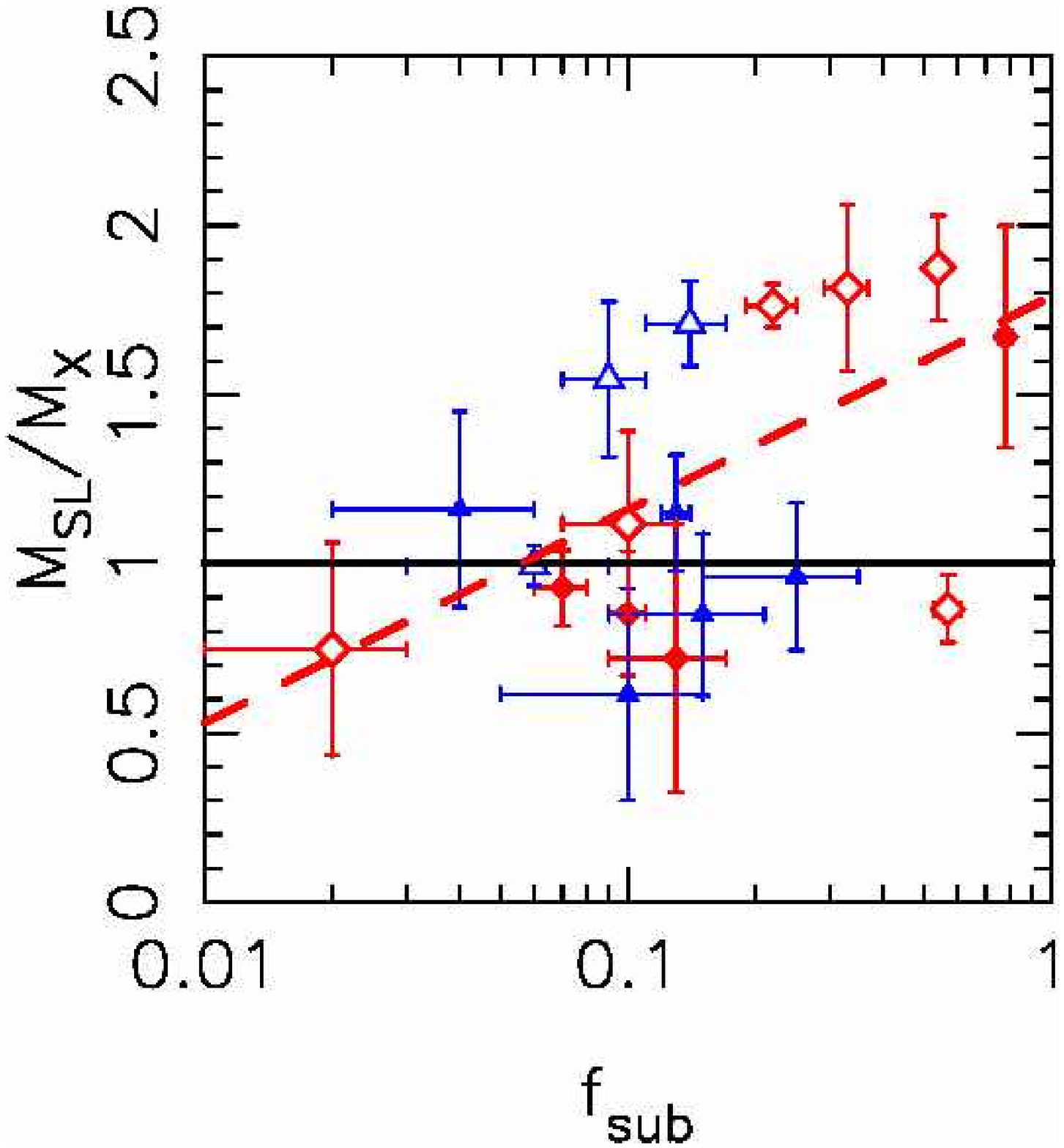}}}
\end{minipage}
\begin{minipage}{5.5cm}
\centerline{\mbox{\includegraphics[width=5.5cm,angle=270]{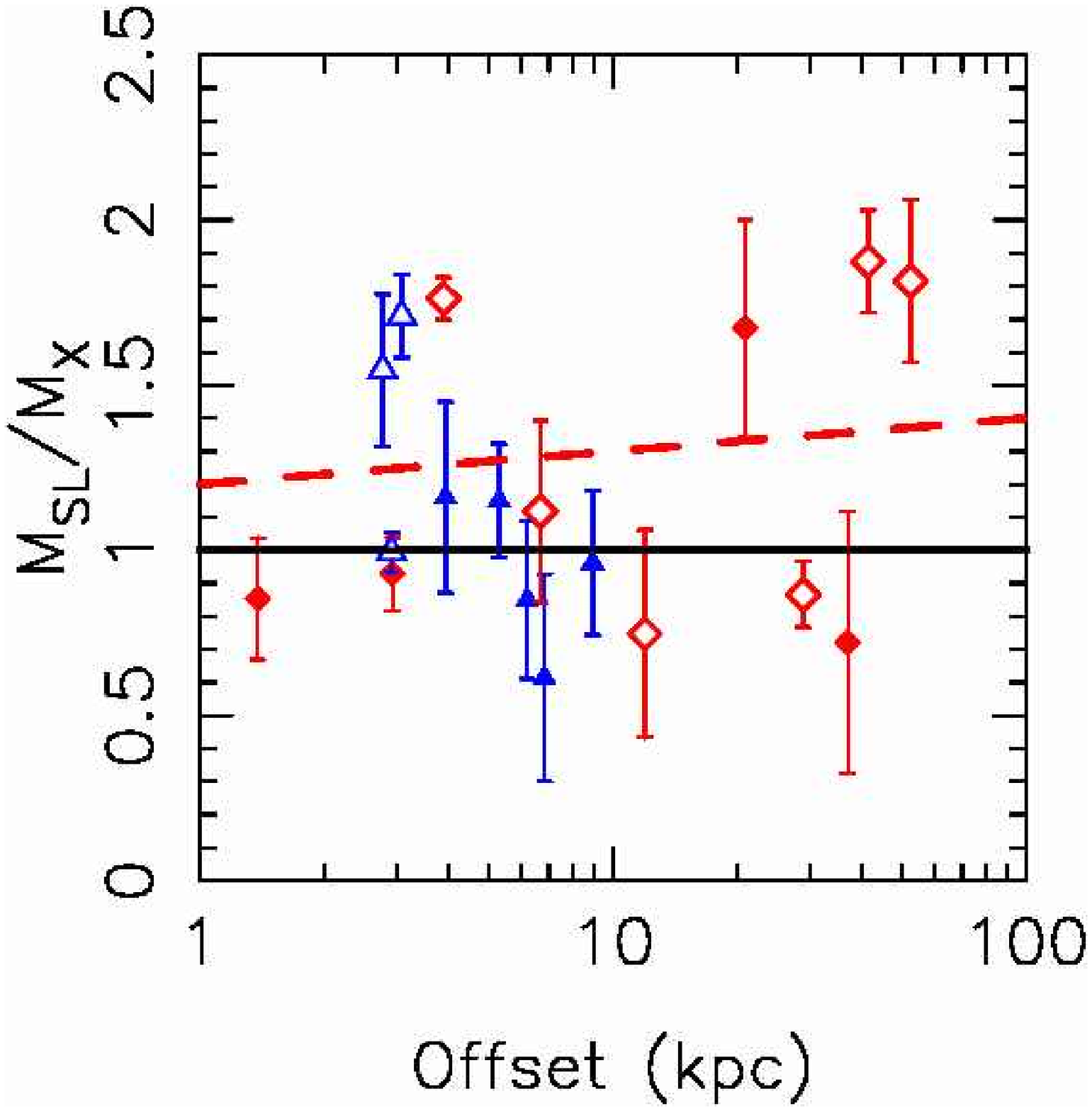}}}
\end{minipage}
\caption{\label{xtrend}Residuals of the $M_{\rm SL}$/$M_X$ relation
  against the central X-ray slope $\alpha$ (left panel), the
  strong-lensing derived substructure fraction (middle panel) or the
  offset between the X-ray centre and the BCG (right panel). The red
  dashed line shows the best fitting linear relation for each
  diagram. Symbols are identical to Fig.~\ref{xfig1}.}
\end{figure*}

To test the link between the agreement between lensing and X-ray mass
measurements and the mass estimates and the structure of the cluster
cores in more detail, we plot $M_{\rm SL}/M_X$ against $\alpha$,
$\fsub$ (\S\ref{fsub}) and X-ray/BCG offset in Fig.~\ref{xtrend}.  There is
considerable scatter in all three panels, however we try to fit a
straight-line relation to the data in each case.  We obtain a
  positive correlation between $M_{\rm SL}/M_X$ and all of $\alpha$,
  $\fsub$ and offset, as shown by the dashed lines in
  Fig.~\ref{xtrend}, however the $M_{\rm SL}/M_X$ vs $f_{\rm sub}$ is
  the only one of the three with a slope that is statistically
  distinguishable from flat, with a best-fit of $M_{\rm
    SL}/M_X=(1.8\pm0.3)+(0.63\pm0.35)log_{10}(f_{\rm sub})$ -- i.e.\ a
  $\sim2\sigma$ detection of a dependence of $M_{\rm SL}/M_X$ on
  $\fsub$.

We interpret this as evidence that the assumption of hydrostatic
equilibrium required by the X-ray mass estimates is less reliable in
clusters with larger substructure fractions.  Specifically, the
merging activity signaled by larger substructure fractions may be
adding non-hydrostatic pressure support to the ICM through bulk
motions of gas in a manner similar to that identified by
\citet{Rasia06} and \citet{Nagai07} at larger radii in simulations.

\begin{table}
\caption{\label{fitx} Best fit parameters for the relationship between
  $M_{\rm SL}(R<250\kpc)$ and $M_X(R<250\kpc)$ discussed in \S\ref{mlmx}.}
\begin{tabular}{llll}
\hline
Sample    & $N_{\rm clus}^a$ & Normalization    & Scatter \\
          &               & $\log_{10}(B)$    & $\sigma_M$ (dex) \\
\hline
All$^b$   & 18            & $0.13\pm0.04$    & $0.17$ \\
Cool core & 8            & $0.05\pm0.05$    &  $0.19$ \\
Non cool core & 10  &  $0.17\pm0.04$   &  $0.13$ \\
Undisturbed & 12     & $0.15\pm0.05$    &  $0.20$\\  
Disturbed & 6            & $0.12\pm0.04$    &  $0.15$\\
\hline
\end{tabular}

$^a$~The number of clusters in each sub-sample are taken from
Table~\ref{xray} 
$^b$~Chandra data are not available for two of the clusters -- A\,868
and A\,1703.
\end{table}

\subsection{Cluster Substructure, BCG Dominance and
  Ellipticity}\label{struct}

\subsubsection{Substructure Fractions}\label{fsub}

\begin{figure*}
\includegraphics[width=5cm,angle=270]{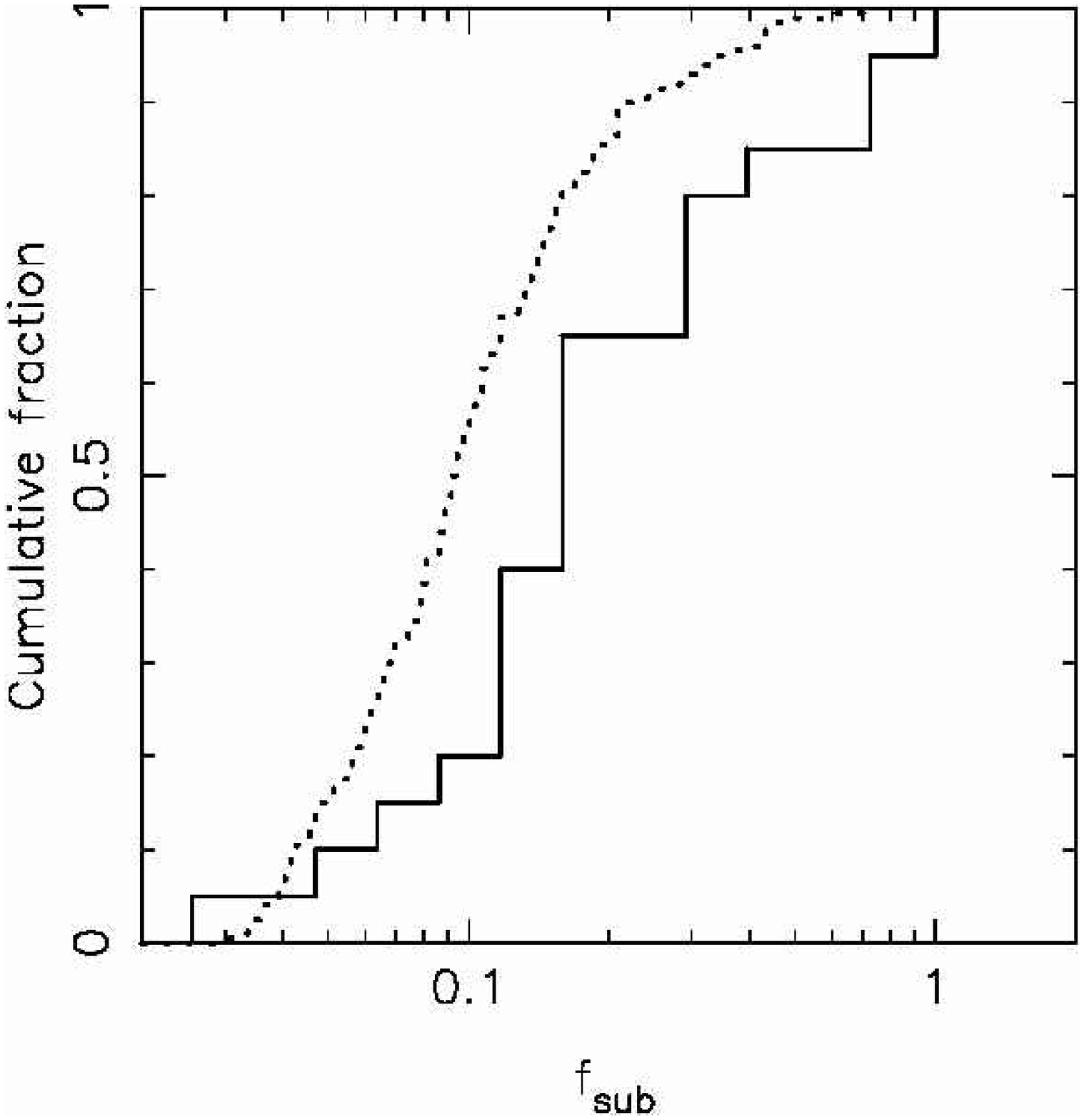}
\hspace{2cm}
\includegraphics[width=5cm,angle=270]{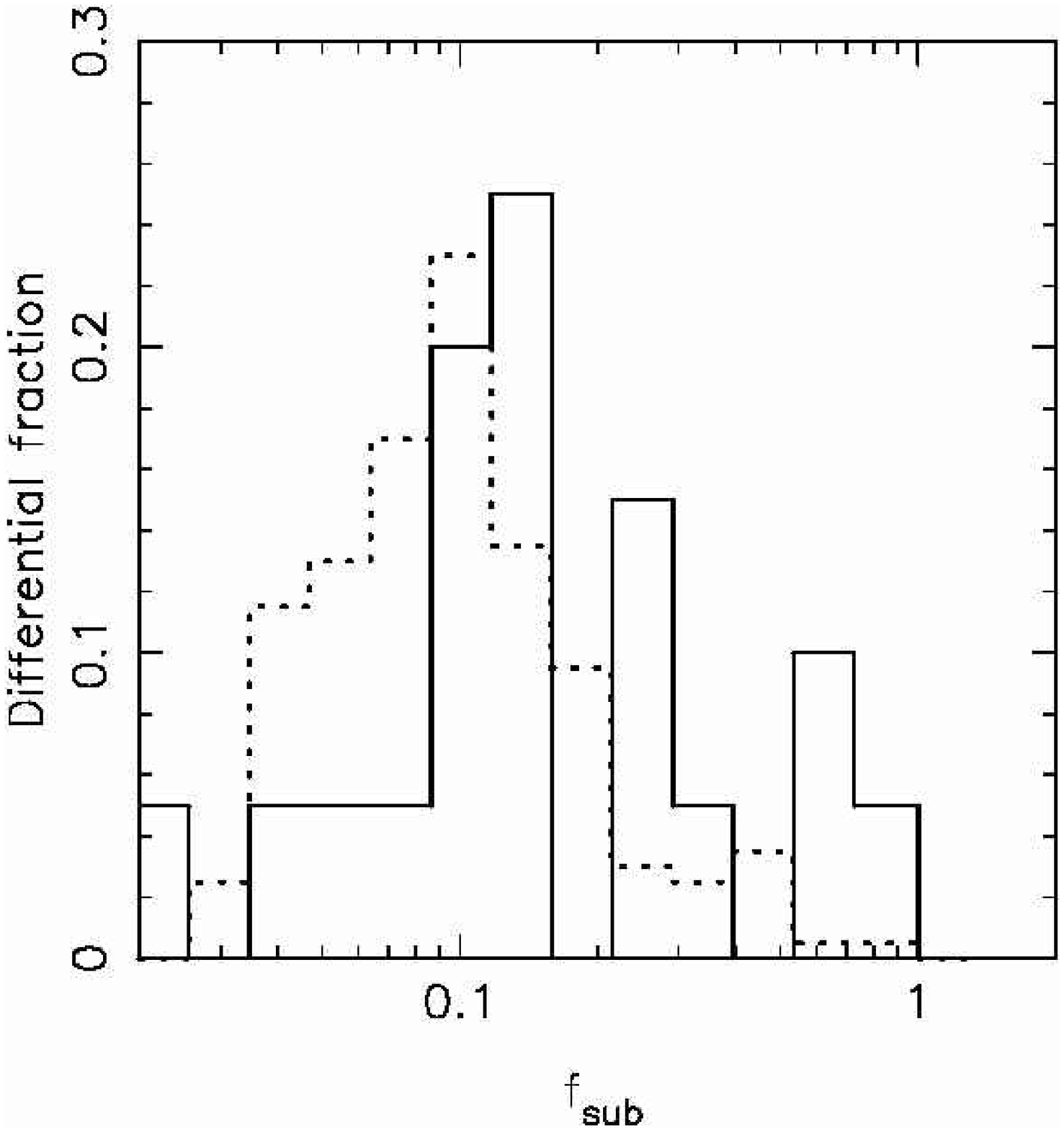}
\caption{\label{subhist} Observed substructure fraction (solid curve),
  as determined from the 20 strong-lensing models in the extended
  sample, compared with the expected distribution (dashed curve)
  presented in \citet{Taylor}. Left diagram: cumulative
  distribution. Right diagram: differential distribution.}
\end{figure*}

Following Sm05 and \citet{Taylor}, we define the substructure fraction
$\fsub$ as the fraction of the mass within a radius R that is not
contained in the BCG and the cluster-scale dark matter halo centred on
the BCG.  This allows straightforward calculation of $\fsub$ from the
mass models, and provides an estimate of the fraction of mass that
resides in galaxy- and group-scale halos within the parent cluster.
We measure $\fsub$ within the same aperture as for $M_{\rm SL}$, i.e.\
$R<250\kpc$.  At these scales $\fsub$ is stable to small changes in
the aperture choice; this stability extends down to $R\gs100\kpc$ in
most cases.  The measurements of $\fsub$ are listed in
Table~\ref{physics}.

\citet{Taylor} compared the predicted \citep[from][]{Taylor05b} and
observed (from Sm05) $\fsub$ distributions, finding a possible excess
of clusters with the highest values of $\fsub$.  They speculated that,
if confirmed, these excesses may be due to X-ray selected clusters
comprising an excess of cool core and merging clusters with respect a
purely mass-selected sample (the synthetic clusters were selected on
mass from \citeauthor{Taylor05b}'s model).  We return to this
distribution in Fig.~\ref{subhist} with our enlarged sample of 20
strong-lensing clusters (Sm05 comprises 5 strong- and 5 weak-lensing
constrained cluster cores).  As seen in Fig.~\ref{subhist}, the
observed $\fsub$ distribution remains noisey with 20 clusters.  We
repeat the Kolmogorov Smirnov test of \citeauthor{Taylor}, and again
obtain an inconclusive result, with the null hypothesis that the
observed and sythetic samples are drawn from the same underlying
distribution disfavoured at $80\%$ (i.e.\ $1.3\sigma$) confidence.

\begin{figure}
\includegraphics[width=8.5cm,angle=270]{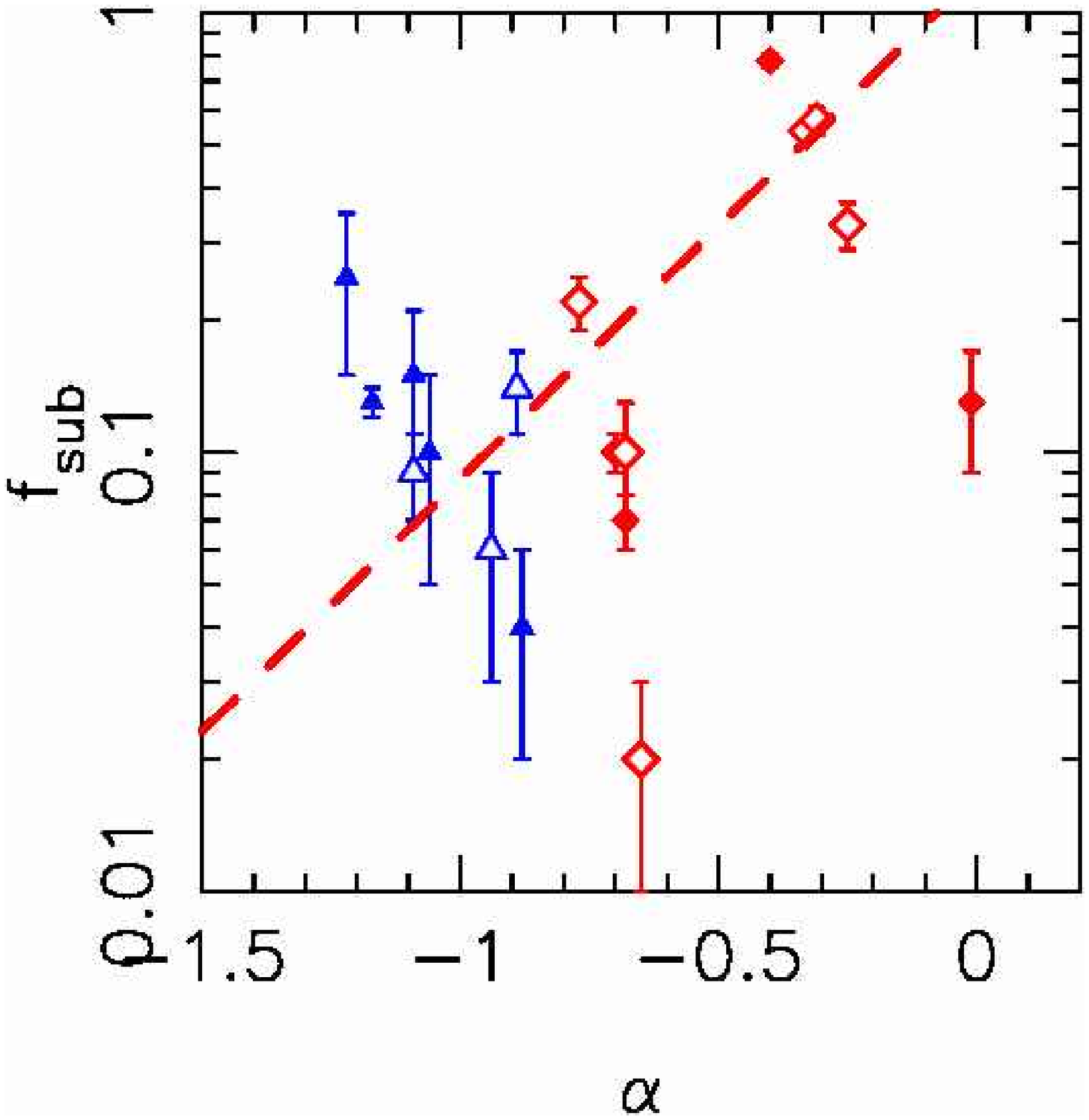}
\caption{\label{xfig2}Relationship between substructure fraction,
  $\fsub$, and $\alpha$ the slope of the gas density profile at
  $0.04,r_{500}$, as a measure of the strength of cooling in the
  cluster cores core.  Symbols are identical to Fig.~\ref{xfig1}.  The
  dashed line shows the best-fit relation to the data.}
\end{figure}

There is extensive discussion in the literature on the issue of
whether cluster-cluster mergers are capable of destroying cool cores
\cite[e.g.][]{Poole08,Burns08}.  We have previously addressed this
question indirectly through comparisons of $\alpha$ and X-ray/BCG
offsets in \citet{Sanderson09b}.  Here, we tackle it more directly, by
comparing $\alpha$ with $\fsub$ in Fig.~\ref{xfig2}. We fit a
  relation of the form $\log_{10}(\fsub)=A+B\,\alpha$ and obtain best
  fit parameters of A=$0.09\pm0.12$, and B=$1.15\pm0.22$. We
  therefore find a positive correlation at 5$\sigma$ significance
  between $\fsub$ and $\alpha$, albeit with large scatter.  Clusters
that host a cool core therefore also have less substructure in their
cluster core mass distribution, and vice versa.  This
  supports the connection found in \S\ref{mlmx} between the differing
  values of $M_{\rm SL}/M_X$ for cool core and non-cool core clusters,
  and the positive correlation between the mass ratio and $\fsub$.
  This result is also strongly suggestive that the cluster-cluster
  merger activity associated with larger values of $\fsub$ plays a
  role in destroying cool cores.  However the appreciable scatter
on this relationship also suggests that the physics is more
complicated than a simple one-to-one relationship.

\subsubsection{BCG Dominance}
\label{gap}

\begin{figure*}
\begin{minipage}{8.5cm}
\includegraphics[width=8.2cm,angle=270]{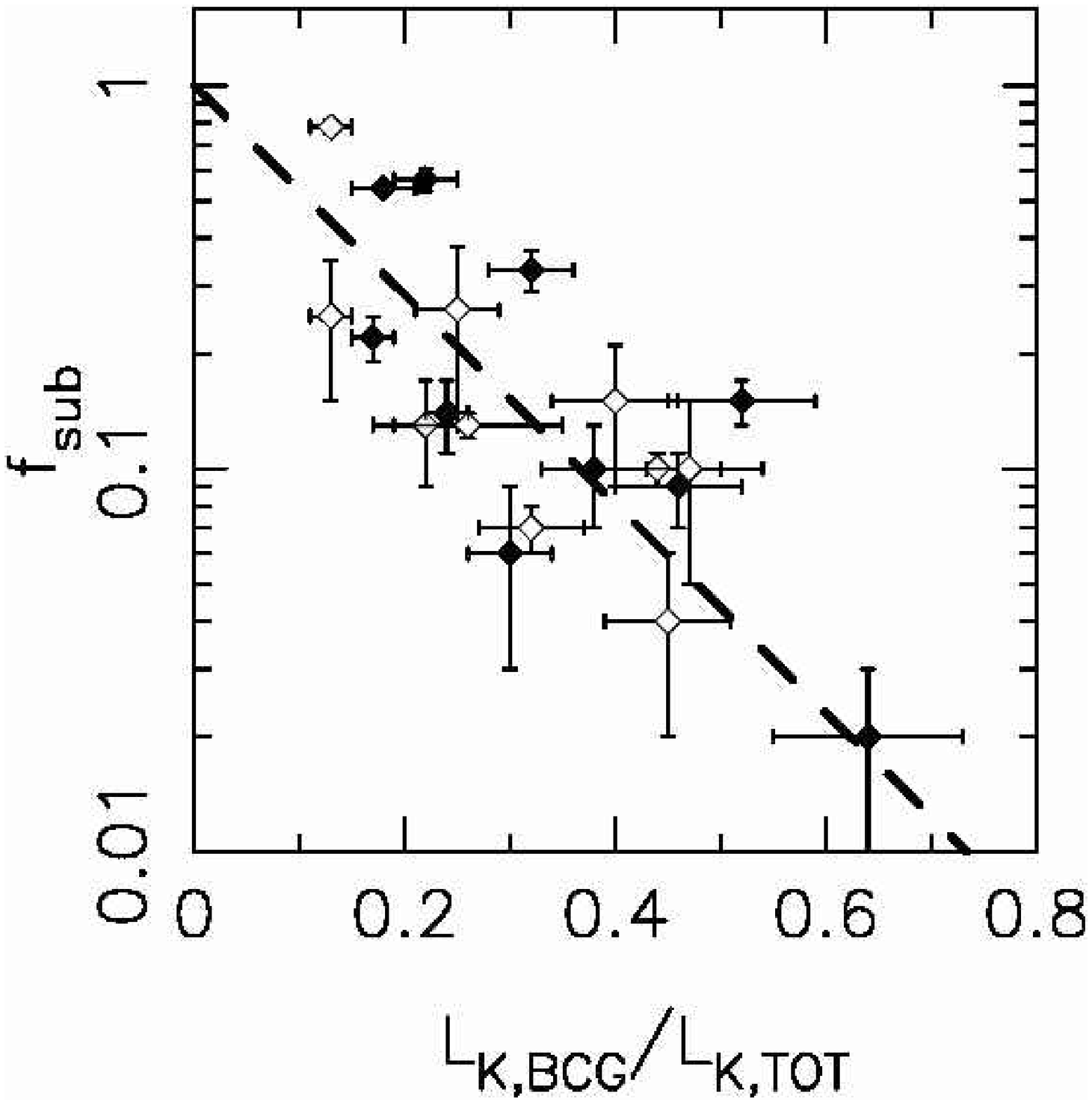}
\end{minipage}
\begin{minipage}{8.5cm}
\includegraphics[width=8.2cm,angle=270]{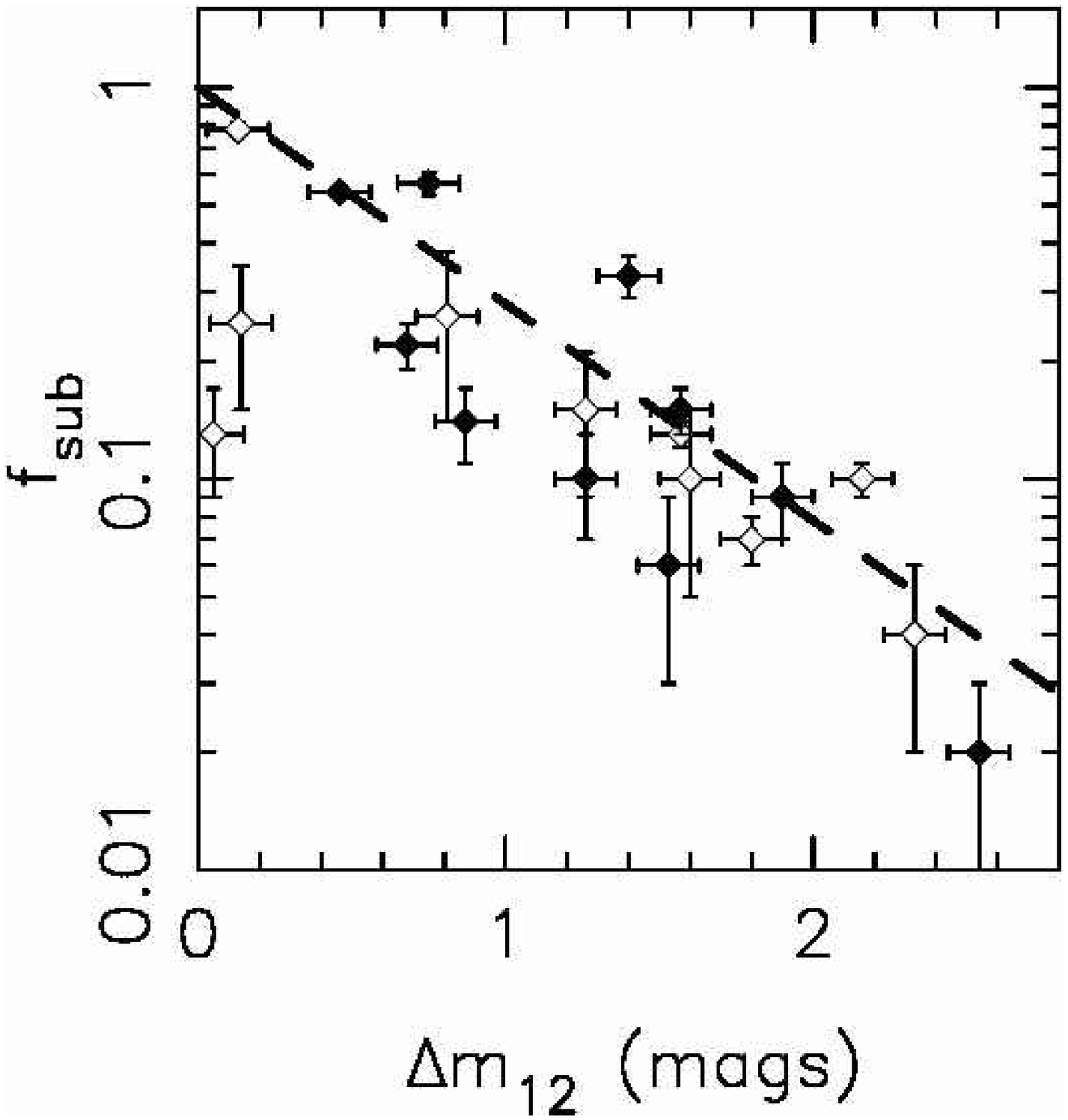}
\end{minipage}
\caption{\label{subplots}Substructure fraction versus indicators of
  BCG dominance: fraction of the cluster core $K$-band luminosity that
  emanates from the BCG (left) and the magnitude gap between the BCG
  and the second ranked cluster galaxy (right). Filled symbols are the new
    clusters, and open symbols are from the extended sample.}
\end{figure*}

The dominance of the BCG over the total $K$-band luminosity of cluster
galaxies within cluster cores was investigated by Sm05, who found a
roughly monotonic relationship between $L_{K,BCG}/L_{K,TOT}$ and
$\fsub$ in the sense that more dominant BCGs live in clusters with
lower substructure fractions.  This has been investigated in more
detail recently by \citet{Smith09b} who showed that $\sim8\%$ of
$10^{15}\Msol$ clusters contain a BCG at least two magnitudes brighter
than the second ranked cluster galaxy ($\dm>2$).  We build on these
results here by comparing substructure in the total mass distribution
in the cluster cores with the dominance of the BCGs for our enlarged
sample of 20 strong-lensing clusters.  We plot $\fsub$ versus both the
fraction of the cluster $K$-band luminosity emanating from the BCG and
the magnitude gap between the BCG and the second ranked cluster galaxy
in Fig.~\ref{subplots}, with both measured within the same $R<250\kpc$
region as $\fsub$.  Both plots show an obvious correlation, confirming
and extended the results in Sm05.  We obtain best-fit relations of:
$log_{10}(f_{\rm sub})=-2.80\ L_{\rm K,BCG}/L_{\rm K,tot}+0.22$ and
$log_{10}(f_{\rm sub})=-0.53 \Delta$m$_{12}-0.07$, both relations
having a dispersion of $\sigma\sim0.3\dex$.

\subsubsection{Cluster ellipticity}

\begin{figure}
\includegraphics[width=8cm,angle=270]{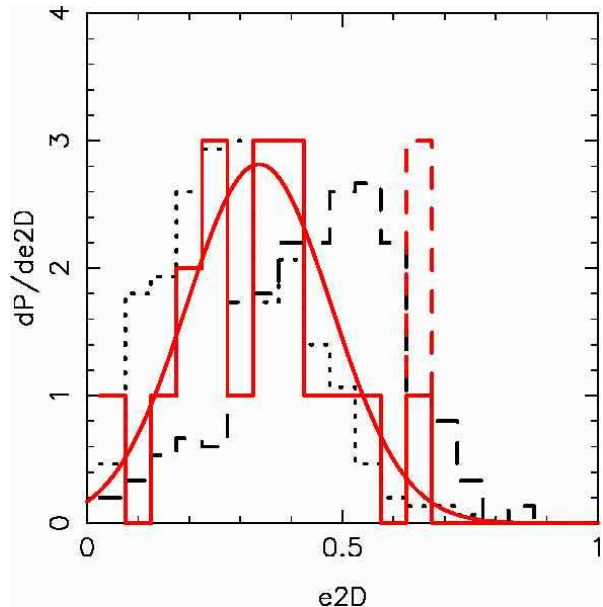}
\caption{\label{e2d} Observed distribution of 2d ellipticities (red
  solid histogram) compared with predictions from simulations by
  \citet{Oguri}, for an unbiased cluster population (black dashed
  histogram) and a cluster population producing the largest Einstein
  radii (black dotted histogram).  The observed peak at
  $e_{2D}\simeq0.65$ is dominated by A\,1201 and A\,521, the 2
  clusters with the smallest Einstein radii (red dot-dashed histogram). 
  The red solid curve shows the gaussian fit to the probability distribution.}
\end{figure}

We have measured the 2D ellipticity and orientation of each cluster
core mass distribution ($R<250\kpc$) by fitting elliptical mass
contours with the IRAF routine {\tt ellipse} to the projected mass
maps based on the mass models (see also \citealt{Richard09}).  We kept
the centre of the ellipses fixed at the peak of the mass map and let
the ellipticity $e_{\rm}(a)$ and position angle $\phi_{\rm 2D}(a)$
vary as free parameters with the semi major axis $a$.  The resulting
ellipticities are listed in Table~\ref{physics}, and the distribution
is plotted in Fig.~\ref{e2d}.  We also show in this figure the
predicted distributions from \citet{Oguri}, both for the underlying
mass-selected cluster population, and for clusters with the largest
Einstein radii.  \citeauthor{Oguri} interpreted the difference between
these two distributions, with the latter peaking at smaller
ellipticities than the former, as implying that the major axis the 3D
mass distributions of strong-lensing clusters are more likely to be
closely aligned with our line of sight through the clusters.
Strong-lensing clusters are therefore expected to be rounder on the
sky than non-strong-lensing clusters.  

Our observed distribution agrees better with the predicted
distribution of large Einstein radii clusters than with the
mass-selected population, however we detect a peak at
$e_{2D}\simeq0.65$.  This peak contains A\,521 and A\,1201, the two
clusters in our sample with the smallest Einstein radii, demonstrating
that clusters with small Einstein radii can still be strong lensing
clusters if background galaxies are fortuitously aligned.  We
therefore exclude these two clusters from the observed distribution
and fit a normal distribution to the other 18 clusters, obtaining:
$\langle e_{2D}\rangle =0.34\pm0.14$

\section{Summary and Discussion}
\label{conc}

We have presented a strong lensing analysis of 20 massive galaxy
clusters from the X-ray selected LoCuSS sample, all of which contain
at least one spectroscopically-confirmed multiply-imaged background
galaxy.  Ten of the sample are new strong-lensing clusters for which
we present the first detailed parametric lens models based on new
spectroscopic redshift measurements at the Keck I telescope.  All of
the clusters are well described by a parametric mass distribution
containing one or several cluster-scale dark-matter halos centred on
the brightest peaks of the cluster K-band light.  We used our
parametric models to compute maps of the mass distribution in the
cluster cores, and thus to measure the projected mass and fraction of
that mass associated with substructures within $R\le250\kpc$.  We have
compared these measurements with observations of the baryons within
the clusters, employing X-ray and K-band data to probe the
intracluster medium and evolved stars respectively.  Our main results
are summarized below:

\smallskip\noindent (i) The Einstein radius for a typical source
redshift of $z=2$ spans $4\arcsec\le\theta_E(z=2)\le47\arcsec$ and is
lognormally distributed, peaking at
$\langle\log_{10}\theta_E\rangle=1.16\pm0.28$, where the uncertainty
is width of the distribution.  Famous clusters with a large Einstein
Radius, A\,1689 and A\,1703, are thus ``outliers'' at $2\sigma$ and
$1.5\sigma$ above the mean respectively.  We also compare the observed
distribution with that predicted from the Millennium Simulation, and
find that the peak of the predicted distribution lies at
$\sim5\arcsec$, i.e.\ $\sim1.7\sigma$ below that of the observed
distribution.

\smallskip\noindent (ii) $\theta_E(z=2)$ is correlated with
$M_{SL}(R<250\kpc)$, the projected mass of the cluster core obtained
from the strong lens models.  We find that ``disturbed'' clusters,
i.e.\ those with an offset between the centroid of their X-ray
emission and the optical centroid of their BCG of $>0.01r_{500}$,
typically lie below the best-fit $\theta_E-M_{SL}$ relation.  We
interpret this as arising from a combination of (i) cluster-cluster
mergers (assumed to be responsible for the X-ray/BCG offset) acting to
soften the cluster density profile, and thus reduce $\theta_E$, and
(ii) an orientation effect, i.e.\ that the major axis of disturbed
clusters tends to be closer to orthogonal to the line of sight through
the cluster, whilst undisturbed clusters tend to have their major axis
parallel with the line of sight.

\smallskip\noindent (iii) The ratio of strong-lensing- and X-ray-based
projected cluster mass measurements within $R<250\kpc$ is measured to
be $M_{SL}/M_X=1.3$, discrepant with unity at $3\sigma$.  This
X-ray/lensing mass discrepancy depends on the structure of the cluster
core -- we detect a positive correlation between $M_{SL}/M_X$ and the
fraction of cluster  mass associated with substructures within
$R<250\kpc$, $\fsub$, at $2\sigma$ significance.  We interpret this as
evidence that the cluster-cluster merger activity associated with
cluster substructure is responsible for departures from hydrostatic
equilibrium.

\smallskip\noindent (iv) The substructure fraction, $\fsub$, is also
correlated with $\alpha$, the slope of the logarithmic gas density
profile at $0.04r_{500}$, in the sense that clusters with steeper
(more nagative $\alpha$) gas density profiles have smaller values of
$\fsub$, and vice versa.  The gas density profile slope is used as an
indicator of the strength of cooling in the intracluster medium, the
steepest slopes ($\alpha\ls0.85$) being identified as ``cool core
clusters''.  This direct empirical relationship between $\alpha$ and
$\fsub$, implies a connection between the cluster-cluster mergers and
the strength of cooling in cluster cores.

\smallskip\noindent (v) We also find a strong correlation between
$\fsub$ and the dominance of the BCG in the sense that clusters with
more substructure have less dominant BCGs.  This suggests that
measures of BCG dominance, including the luminosity gap statistic
(difference between the magnitudes of the first and second ranked
galaxies), may be a useful probe of cluster substructure.

\smallskip\noindent (vi) The distribution of cluster core
ellipticities, measured from the mass maps that are in turn computed
from the parametric lens models, are consistent with the predicted
ellipticity distribution of strong-lensing clusters.

\smallskip

This is the largest published sample of strong-lensing clusters to
date -- a $4\times$ increase in sample size from \cite{Smith05}.
Overall our empirical results are consistent with those of
\citeauthor{Smith05} in that a straightforward interpretation is that
cluster-cluster mergers play a prominent role in shaping the observed
properties of cluster cores.  This is most striking in the
X-ray/lensing comparisons, i.e.\ significant detection of a $30\%$
X-ray/lensing mass discrepancy, and a dependence of this discrepancy
on structure of the cluster core.  This is underlined by the
correlation between $\fsub$ and $\alpha$ -- implying that
cluster-cluster mergers both cause departures from hydrostatic
equilibrium and play a role in moderating the cooling of gas in
cluster cores.  This latter point remains controversial
\cite[e.g.][]{Poole08,Burns08}, and we caution that the distribution
of formation epochs, and other physical processes such as pre-heating
may ultimately modify our simplistic interpretation.

We also note that the clusters with the largest Einstein radii in our
sample (A\,1689 and A\,1703) are $\ls2\sigma$ above the mean of the
best-fit log-normal distribution, implying that $\sim2-7\%$ of larger
samples will contain comparable clusters.  We also speculate that the
$\sim1.7\sigma$ discrepancy between the peak of the observed and
predicted Einstein radius distributions may be in part attributable to
the presence of baryons in the observed universe, in contrast to the
dark matter only simulations on which the prediction is based.
Certainly, the discrepancy is likely over-estimated because of the
current computational limitations on numerical simulations, namely
that even the most advanced simulations such as the Millennium
simulation do not embrace sufficient volume at sufficient numerical
resolution to contain a representative sample of the most massive
strong-lensing clusters.  In summary, we find no compelling evidence
from our statistical analysis of a sample of strong lensing clusters
to support recent claims that clusters with large Einstein radii
present a challenge to the CDM paradigm \citep{Broadhurst08b}.

This article has concentrated on clusters observable from Mauna Kea
with the Keck-I telescope.  In the future we will expand this sample
to include Southern clusters that we have also observed with
\emph{HST} and followed up spectroscopically with VLT and Gemini-S
(P.\ May et al., in preparation).  Similar studies of clusters at
$z>0.3$ from the MAssive Cluster Survey \citep{Ebeling01,Ebeling07}
will allow evolution in the properties of cluster cores to be probed
\cite[e.g.][Richard et al. in preparation]{Smith09a}.  Within LoCuSS, future work will include
combining these strong lens models with our recently published
weak-lensing analysis of Subaru observations \citep{Okabe} -- this
will allow a detailed investigation of the structure of the cluster
mass distributions across a wide range of physical scales.  The
well-constrained lens models presented here are also well-suited to 
be utilized in gravitational telescope searches for very high redshift galaxies 
\citep{Maizy}, following, for example, \citet{Kneib04,Richard06,Richard08,
  Bouwens09}.  Finally, we have presented optical spectra of lensed
background galaxies at $z\sim$ $1-4$, which are magnified by $1-4$
magnitudes from strong lensing. These sources are well suited for
further high resolution spectroscopic follow-up, such as near-infrared
IFU observations \citep{Swinbank09}.

\section*{Acknowledgments}

We thank our colleagues in the LoCuSS collaborations for much support,
encouragement and help. We acknowledge useful discussions with Andrew
Newman, Masmune Oguri, Arif Babul, Giovanni Covone, Timothy James,
Alastair Edge and Ian Smail. We are grateful for Matt Lehnert,
Christian Tapken and Nicole Nesvadba for their spectroscopic
measurement of one of the systems. Mark Sullivan kindly observed
long-slit data on A\,611 for us. JR acknowledges support from an EU
Marie-Curie fellowship. GPS acknowledges support from the Royal
Society and STFC. JPK acknowledges support from the CNRS, from the
Agence Nationale de la Recherche grantÊANR-06-BLAN-0067, and from the
French-Israelian collaboration projectÊ07-AST-F9. AJRS acknowledges
support from STFC. ML acknowledges the Centre National d' Etudes
Spatiales (CNES) for their support. The Dark Cosmology Centre is
funded by the Danish National Research Foundation.  Results are
partially based on observations made with the NASA/ESA Hubble Space
Telescope and the Keck telescope.The authors recognize and acknowledge
the very significant cultural role and reverence that the summit of
Mauna Kea has always had within the indigenous Hawaiian community. We
are most fortunate to have the opportunity to conduct observations
from this mountain.

\bibliography{references}

\label{lastpage}

\appendix
\section{Spectra of multiple images}

\begin{figure*}
\caption{\label{spectra}Extracted spectra of multiple images from the current sample. We mark the proeminent spectral 
features used to derive the redshift.}
\begin{minipage}{5.5cm}
\centerline{A521-1.1}
\includegraphics[height=4.5cm,angle=270]{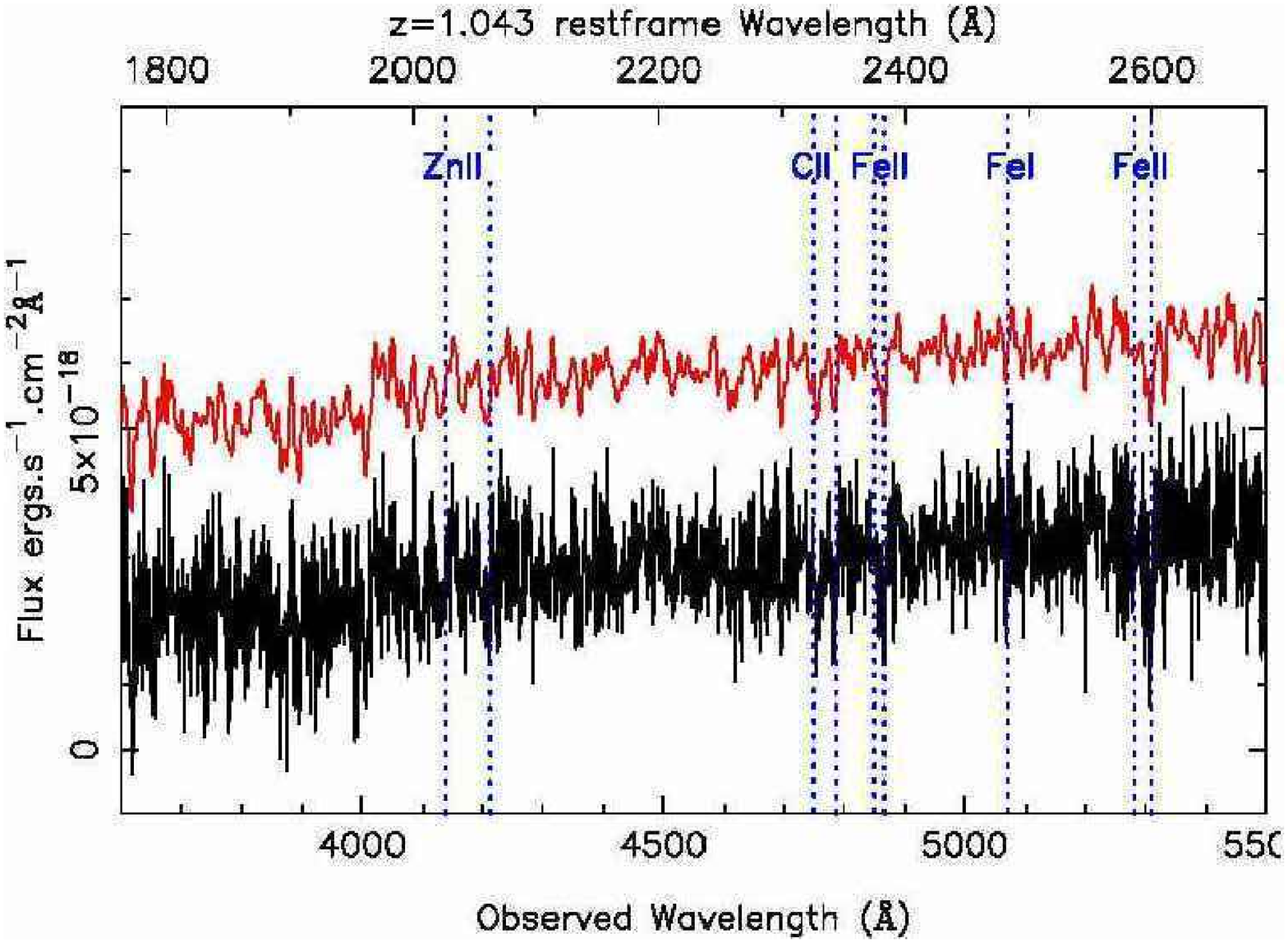}
\end{minipage}
\begin{minipage}{5.5cm}
\centerline{A521-1.2}
\includegraphics[height=4.5cm,angle=270]{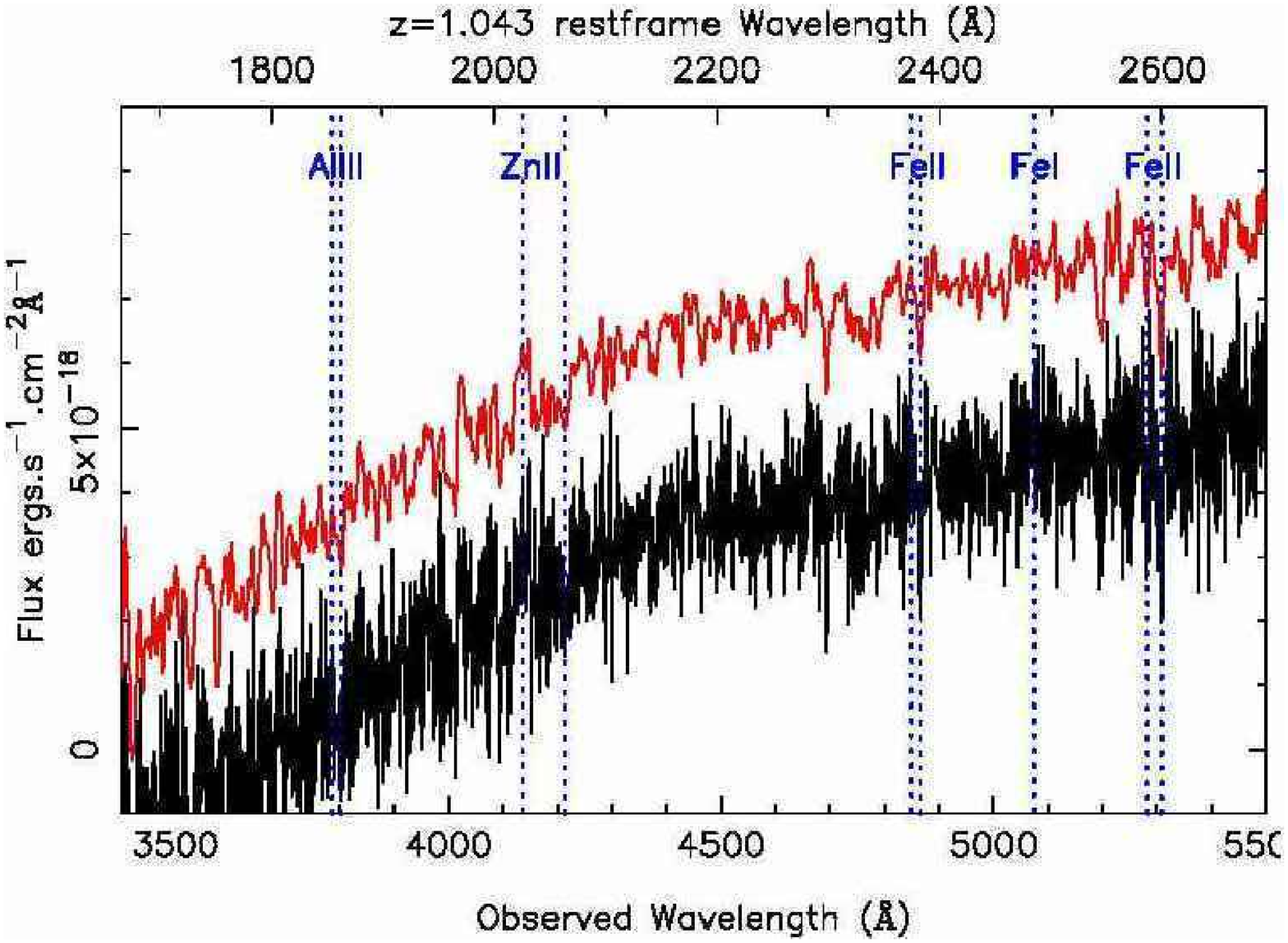}
\end{minipage}
\begin{minipage}{5.5cm}
\centerline{A521-1.3}
\includegraphics[height=4.5cm,angle=270]{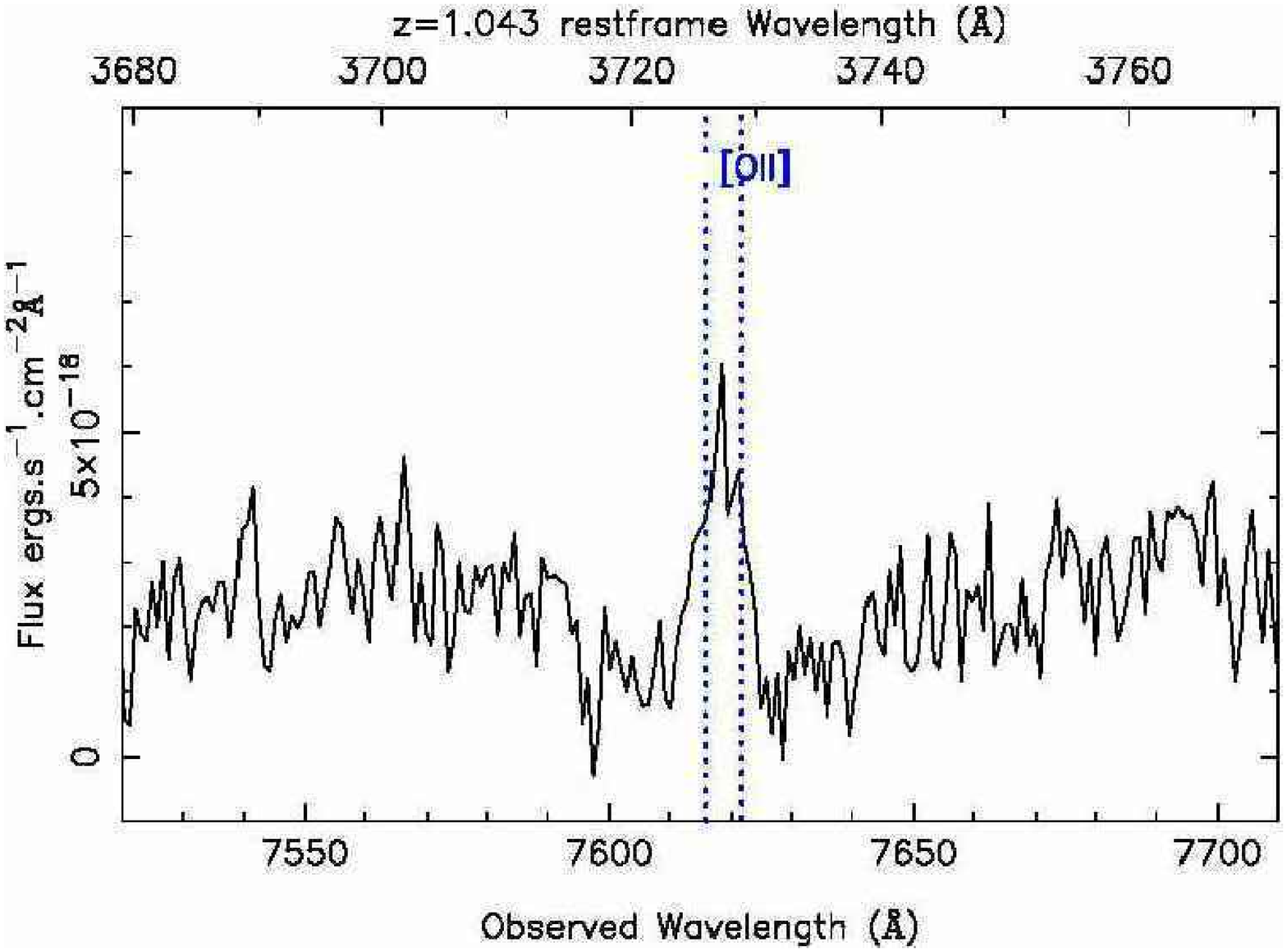}
\end{minipage}
\begin{minipage}{5.5cm}
\centerline{A611-1.2+1.3}
\includegraphics[height=4.5cm,angle=270]{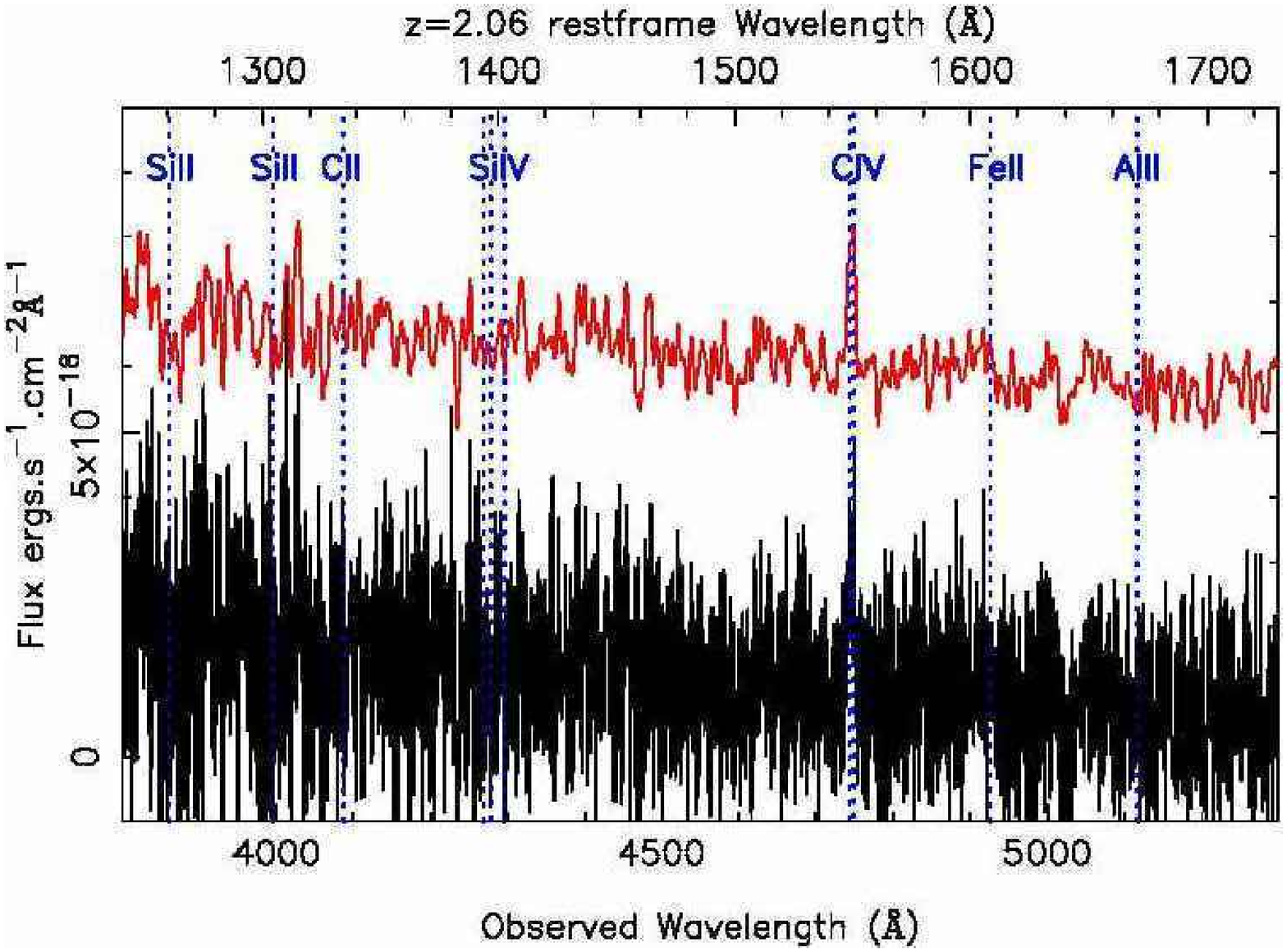}
\end{minipage}
\begin{minipage}{5.5cm}
\centerline{A611-2.1+2.2}
\includegraphics[height=4.5cm,angle=270]{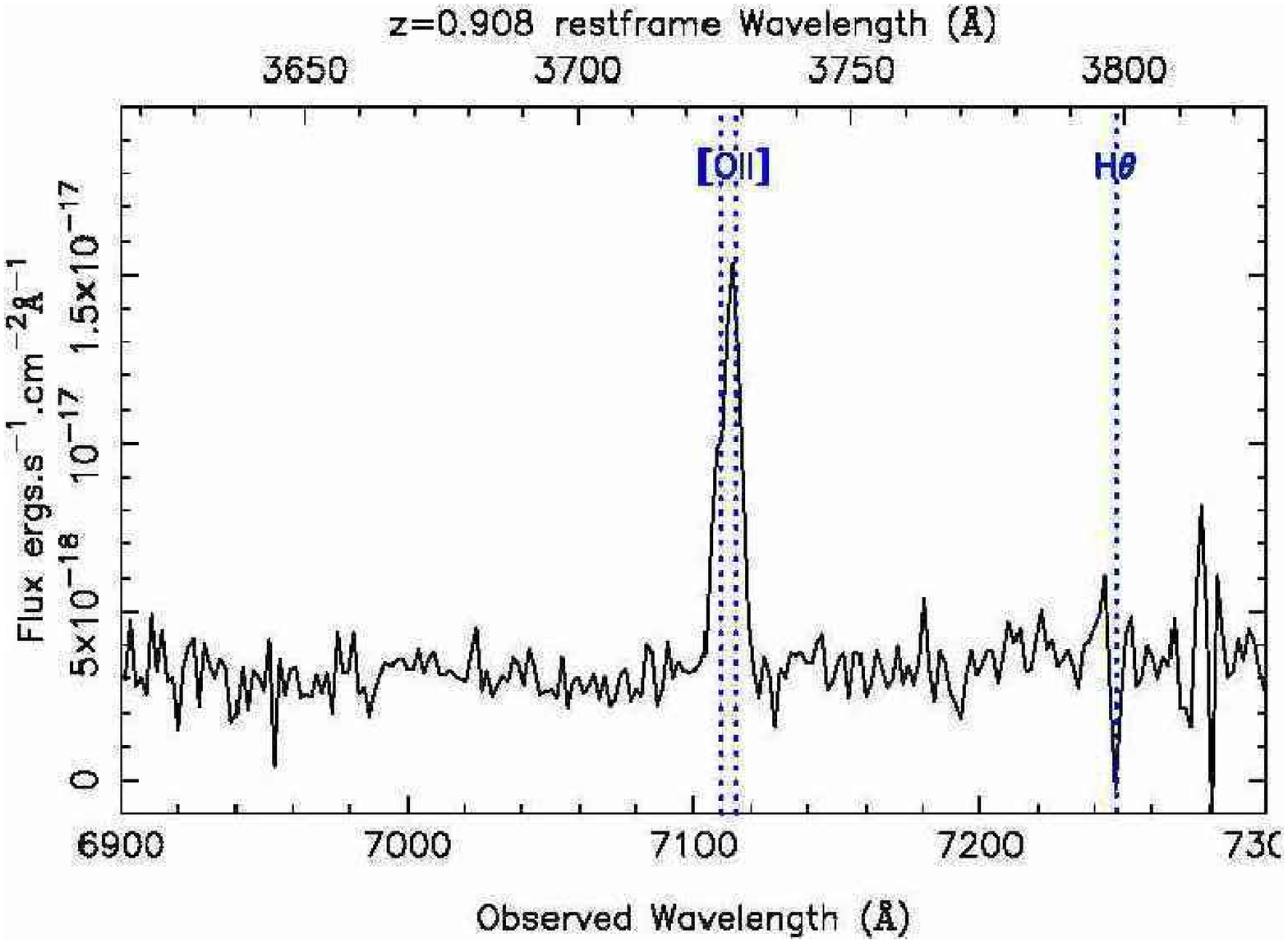}
\end{minipage}
\begin{minipage}{5.5cm}
\centerline{A611-2.3}
\includegraphics[height=4.5cm,angle=270]{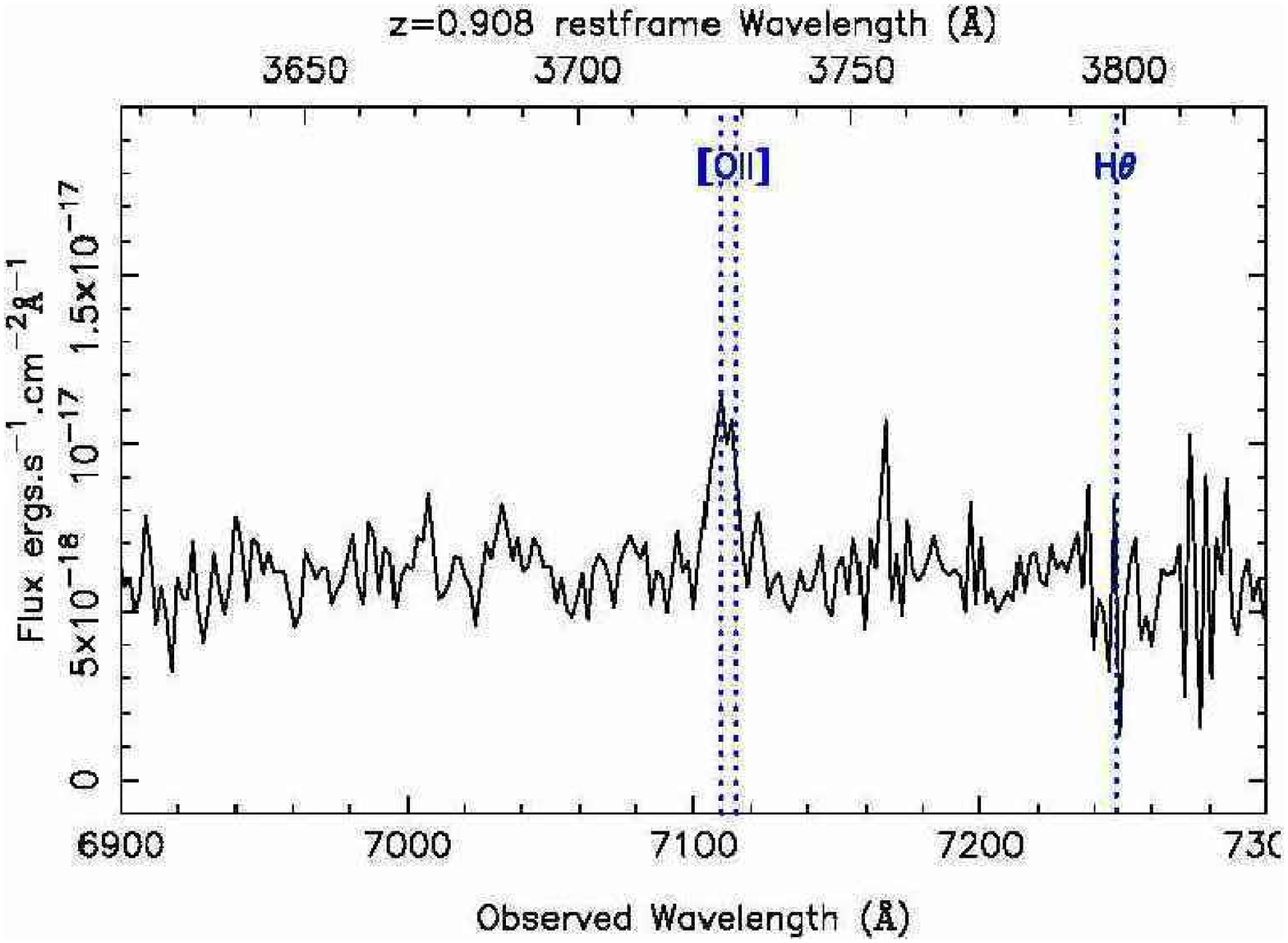}
\end{minipage}
\begin{minipage}{5.5cm}
\centerline{A611-4.1}
\includegraphics[height=4.5cm,angle=270]{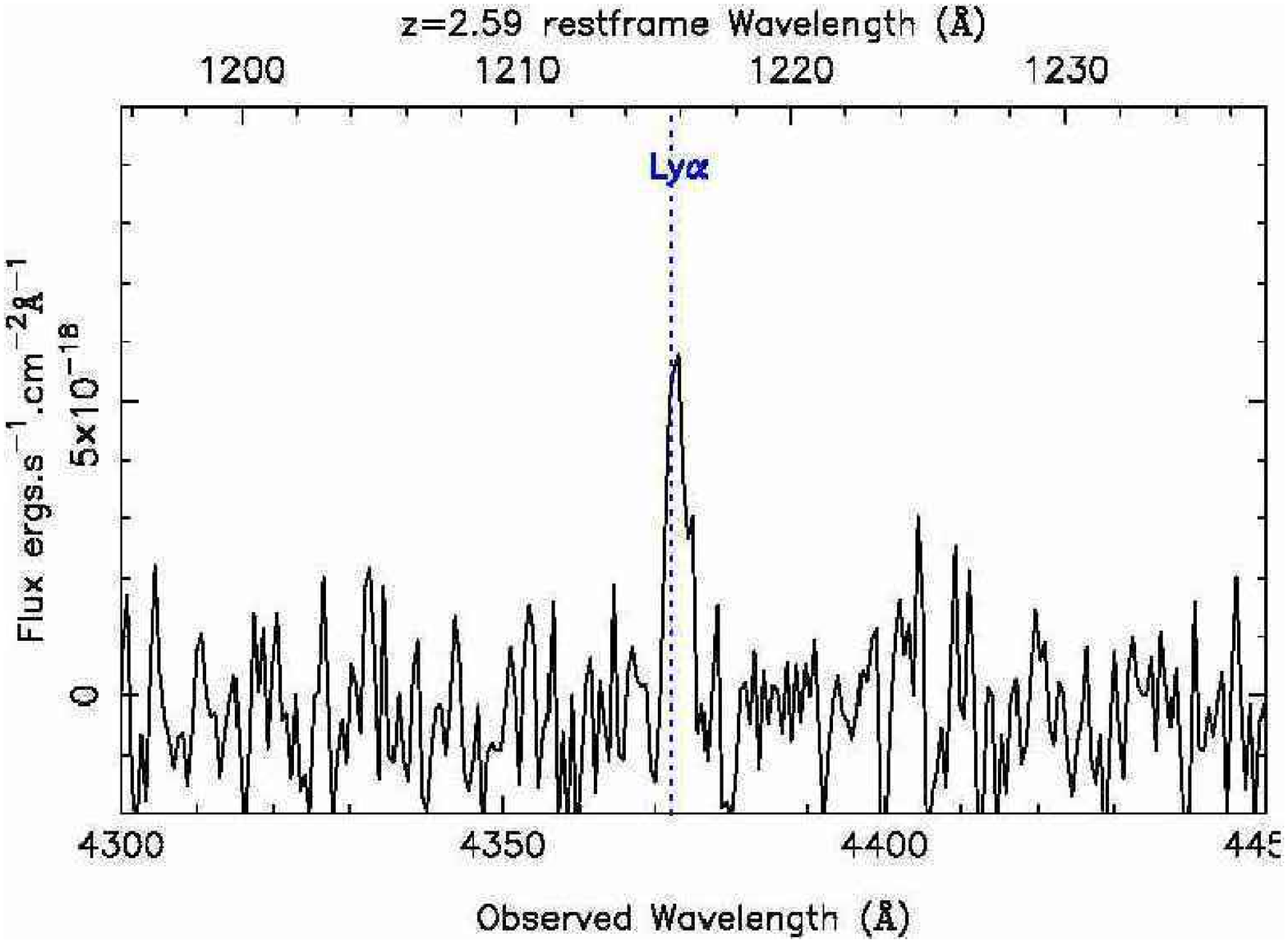}
\end{minipage}
\begin{minipage}{5.5cm}
\centerline{A773-1.1}
\includegraphics[height=4.5cm,angle=270]{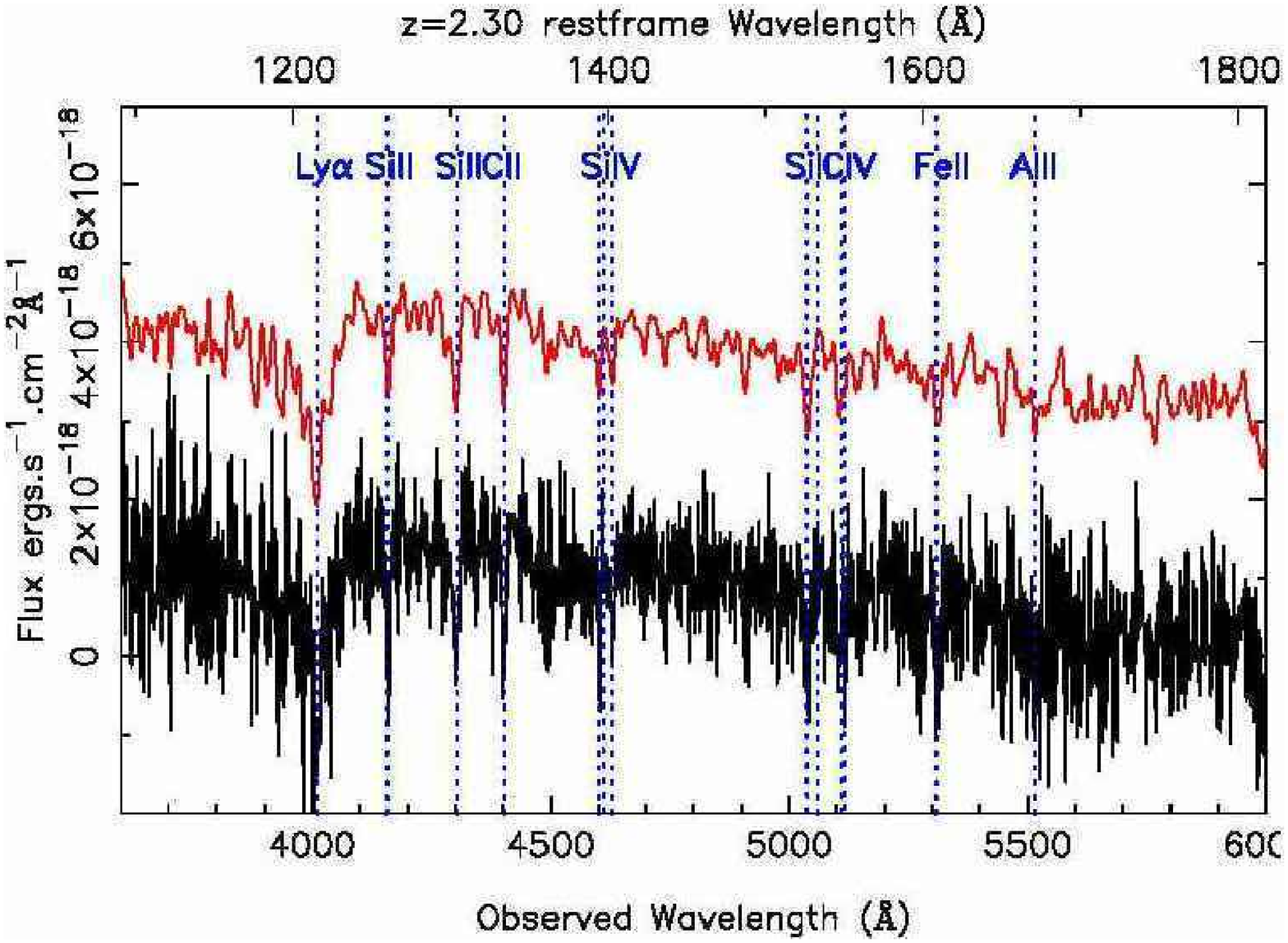}
\end{minipage}
\begin{minipage}{5.5cm}
\centerline{A773-1.2}
\includegraphics[height=4.5cm,angle=270]{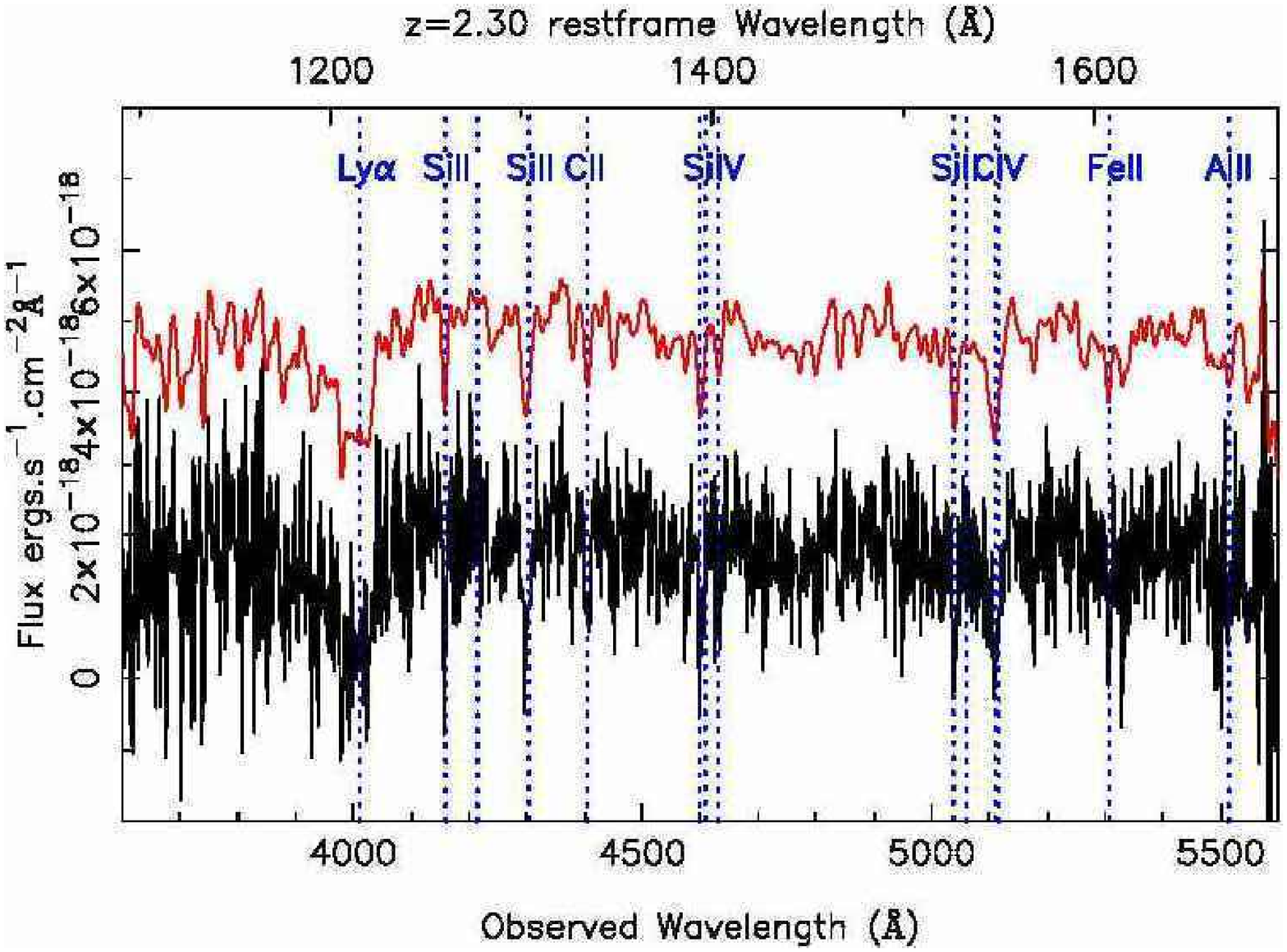}
\end{minipage}
\begin{minipage}{5.5cm}
\centerline{A773-2.1}
\includegraphics[height=4.5cm,angle=270]{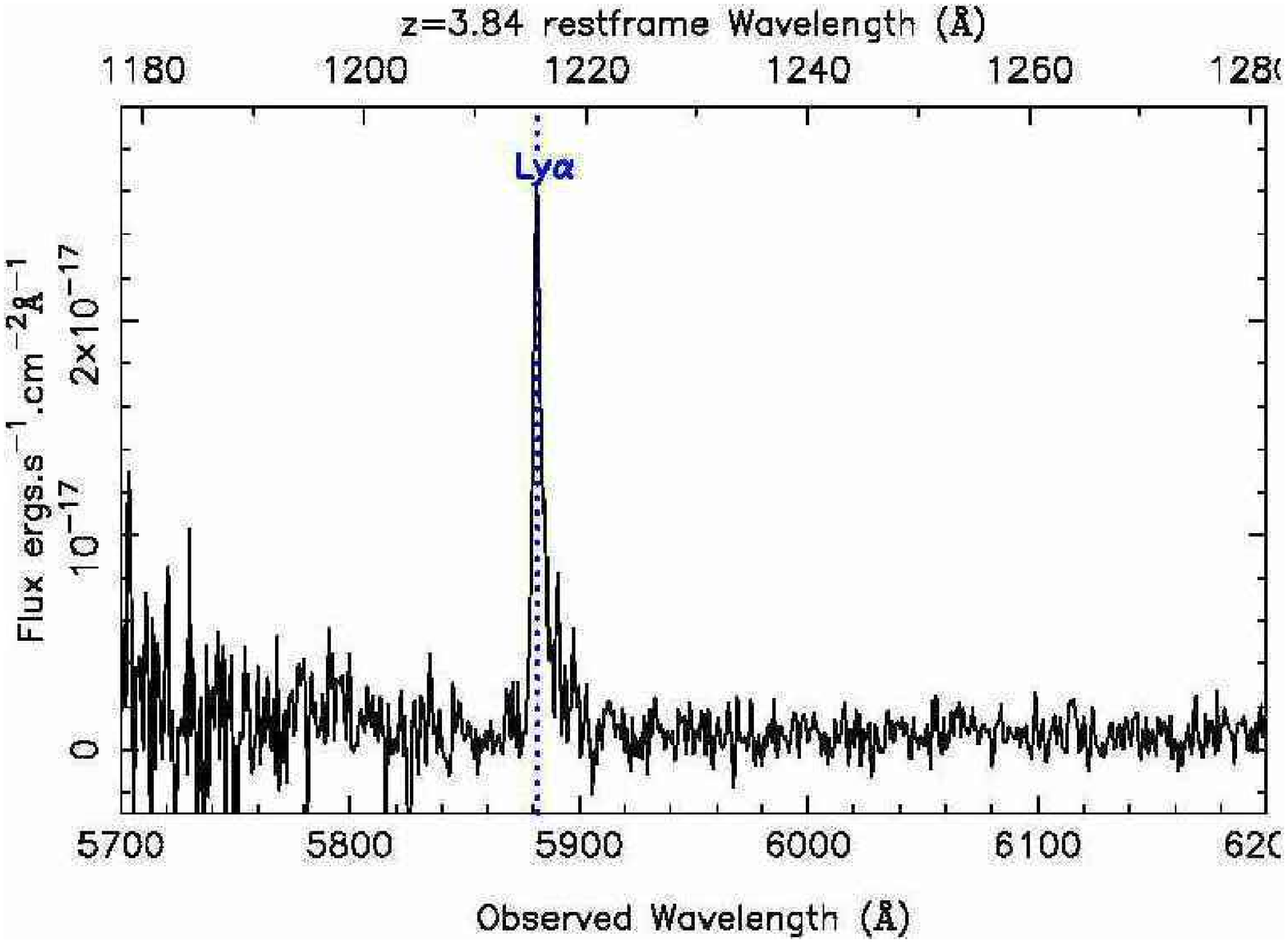}
\end{minipage}
\begin{minipage}{5.5cm}
\centerline{A773-3.1}
\includegraphics[height=4.5cm,angle=270]{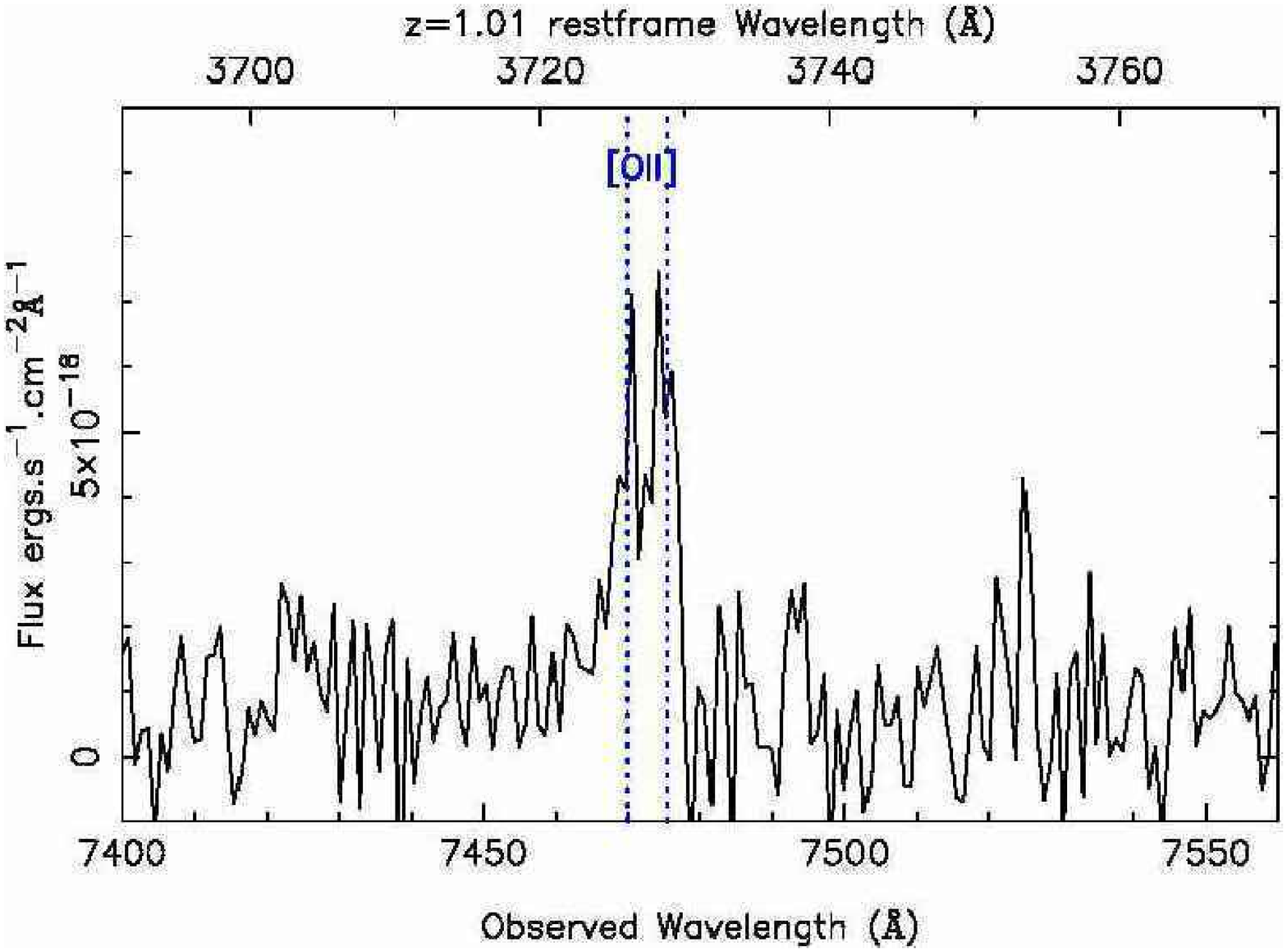}
\end{minipage}
\begin{minipage}{5.5cm}
\centerline{A868-1.1+1.2}
\includegraphics[height=4.5cm,angle=270]{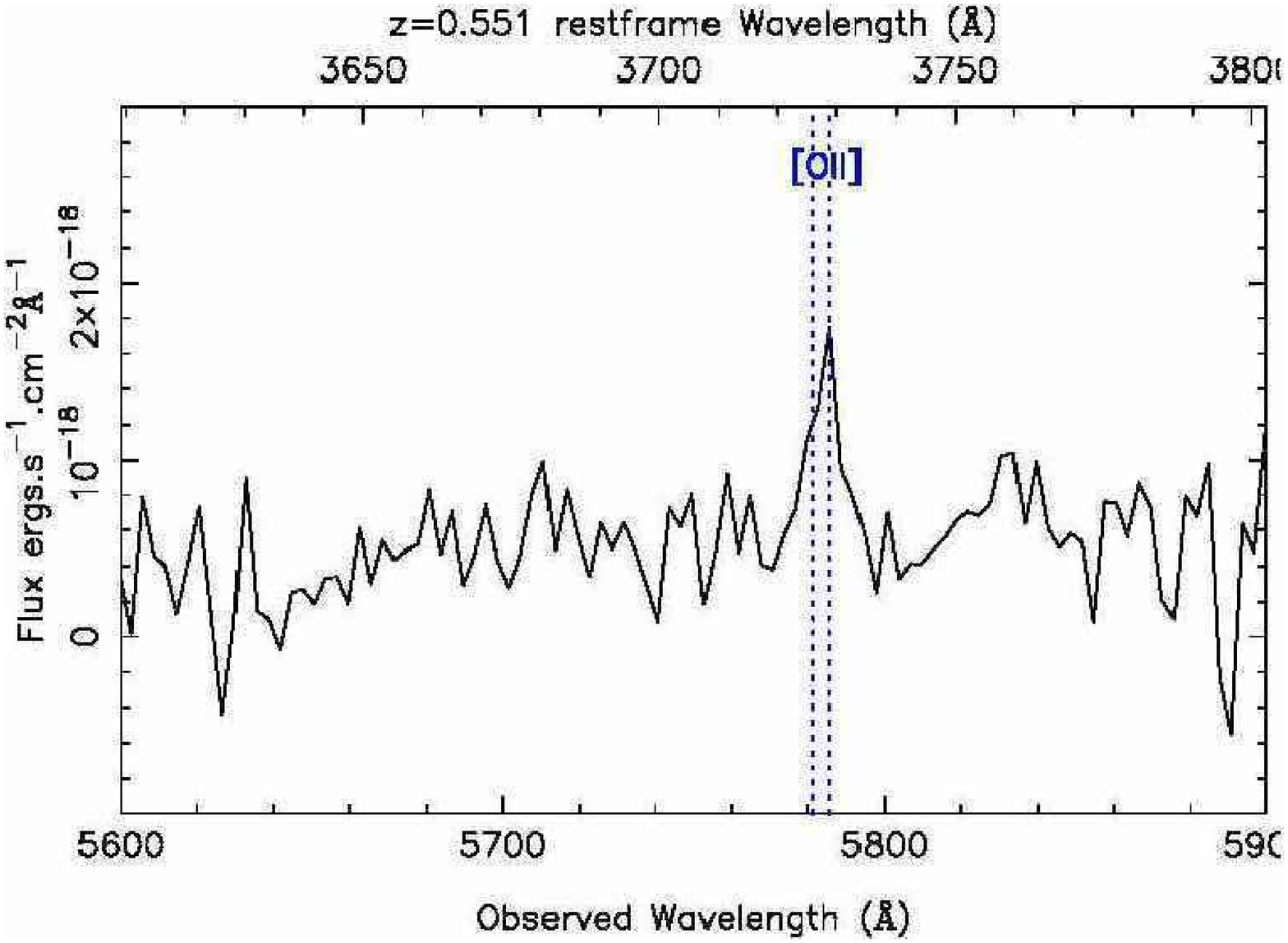}
\end{minipage}
\begin{minipage}{5.5cm}
\centerline{Z2701-1.1}
\includegraphics[height=4.5cm,angle=270]{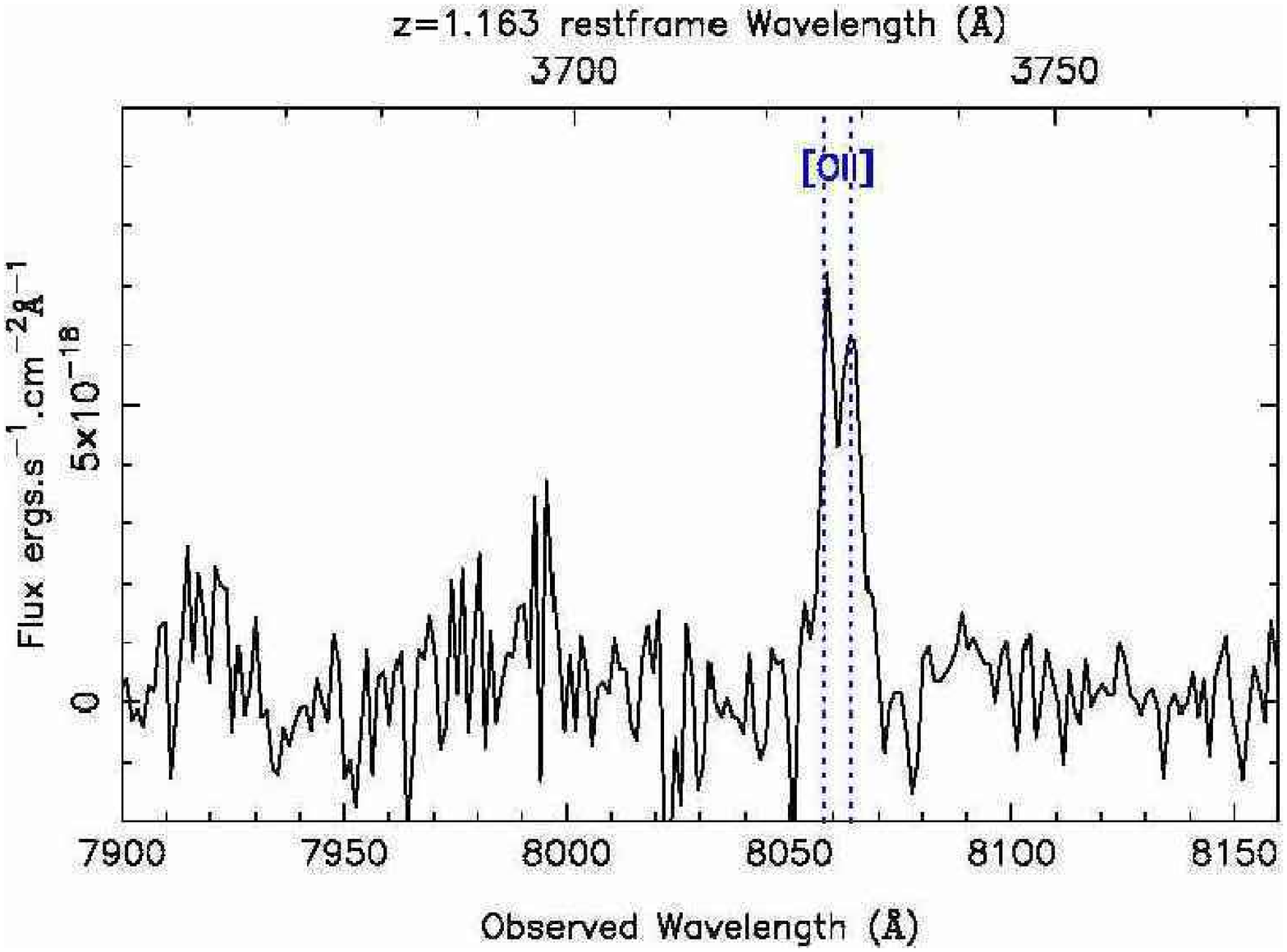}
\end{minipage}
\begin{minipage}{5.5cm}
\centerline{Z2701-1.3}
\includegraphics[height=4.5cm,angle=270]{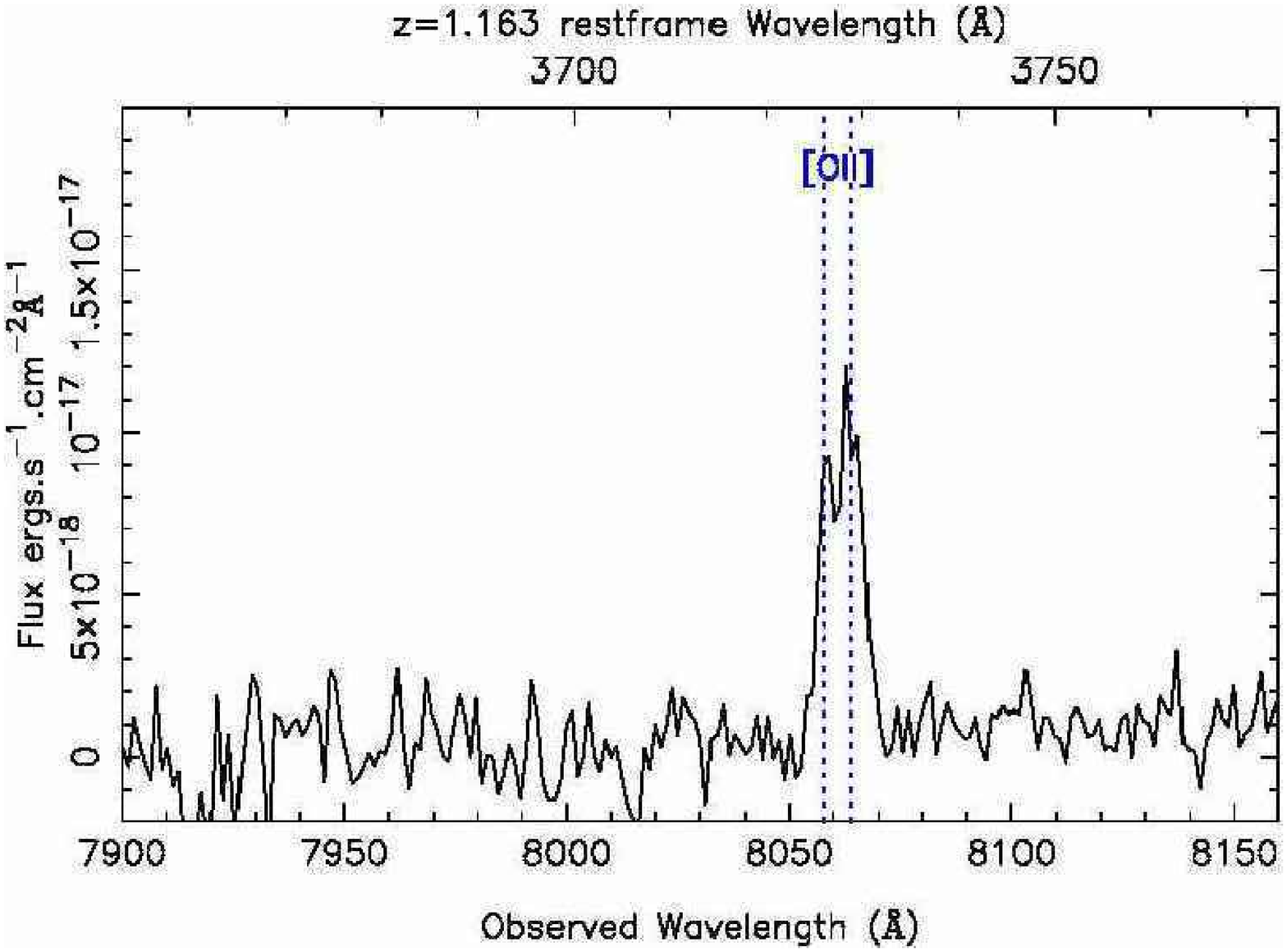}
\end{minipage}
\begin{minipage}{5.5cm}
\centerline{A1413-1.1+1.2}
\includegraphics[height=4.5cm,angle=270]{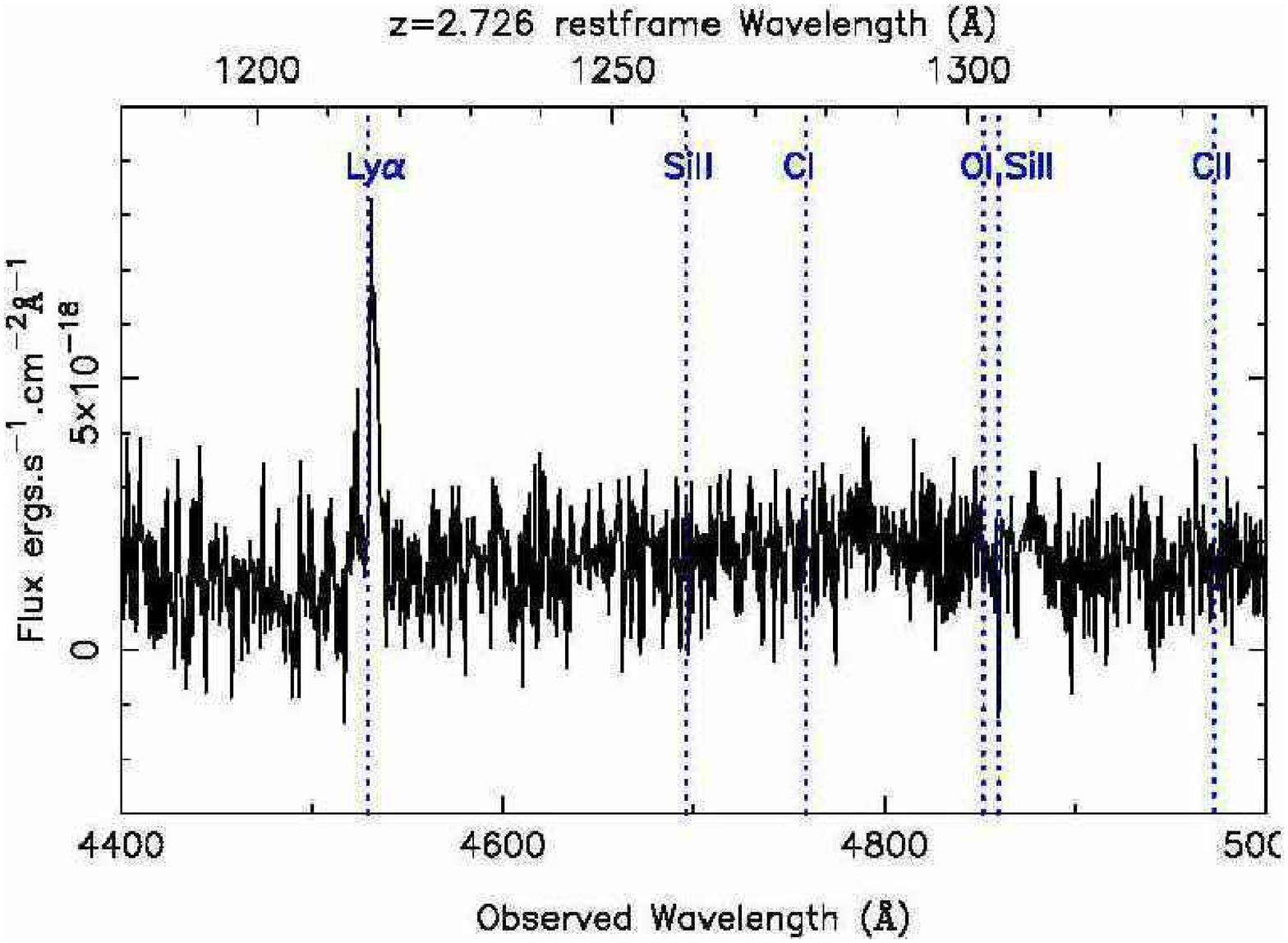}
\end{minipage}
\end{figure*}
\begin{figure*}
\begin{minipage}{5.5cm}
\centerline{A1413-2.1+2.2}
\includegraphics[height=4.5cm,angle=270]{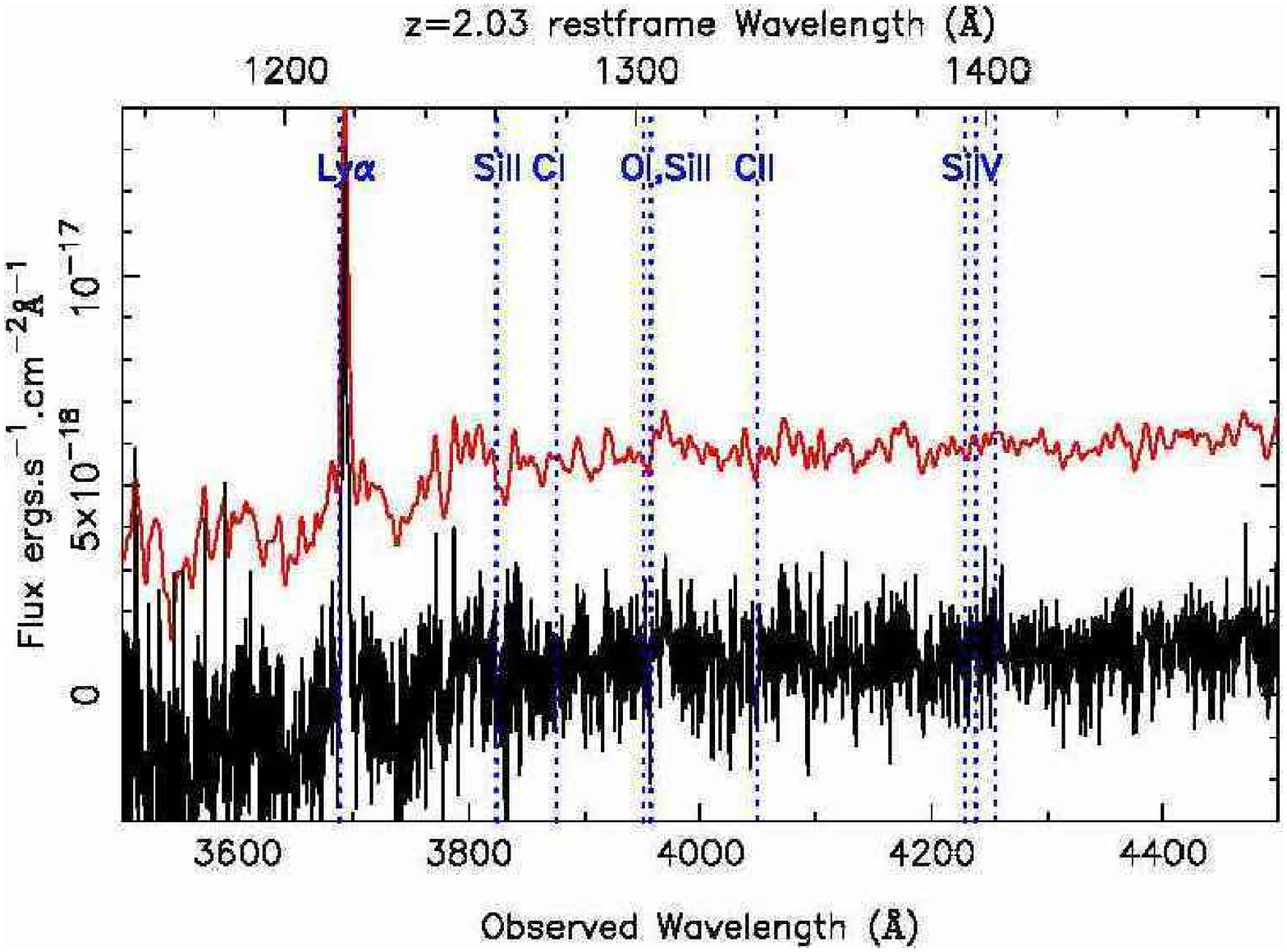}
\end{minipage}
\begin{minipage}{5.5cm}
\centerline{A1413-3.1+3.2+3.3}
\includegraphics[height=4.5cm,angle=270]{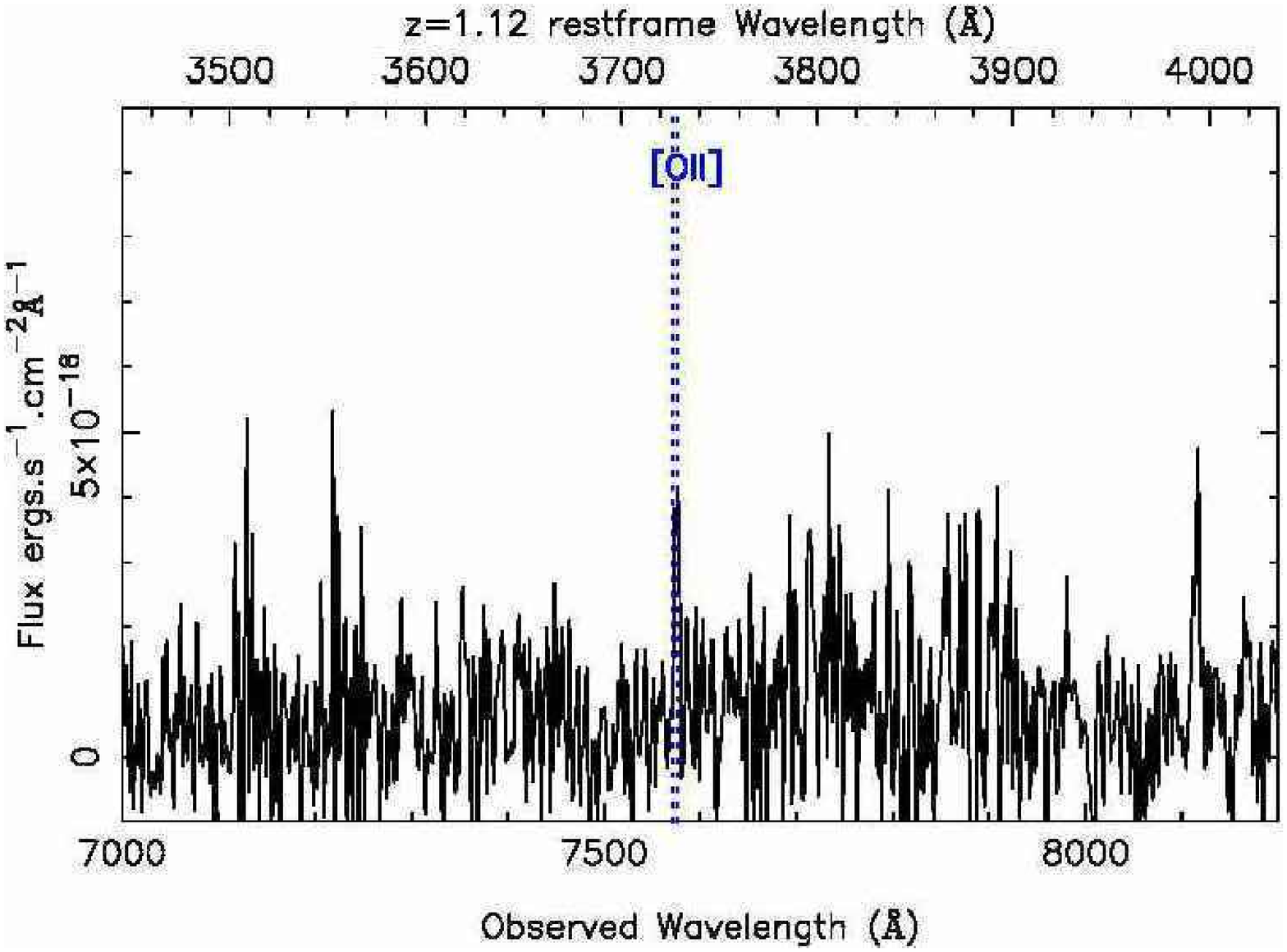}
\end{minipage}
\begin{minipage}{5.5cm}
\centerline{A1835-7.1}
\includegraphics[height=4.5cm,angle=270]{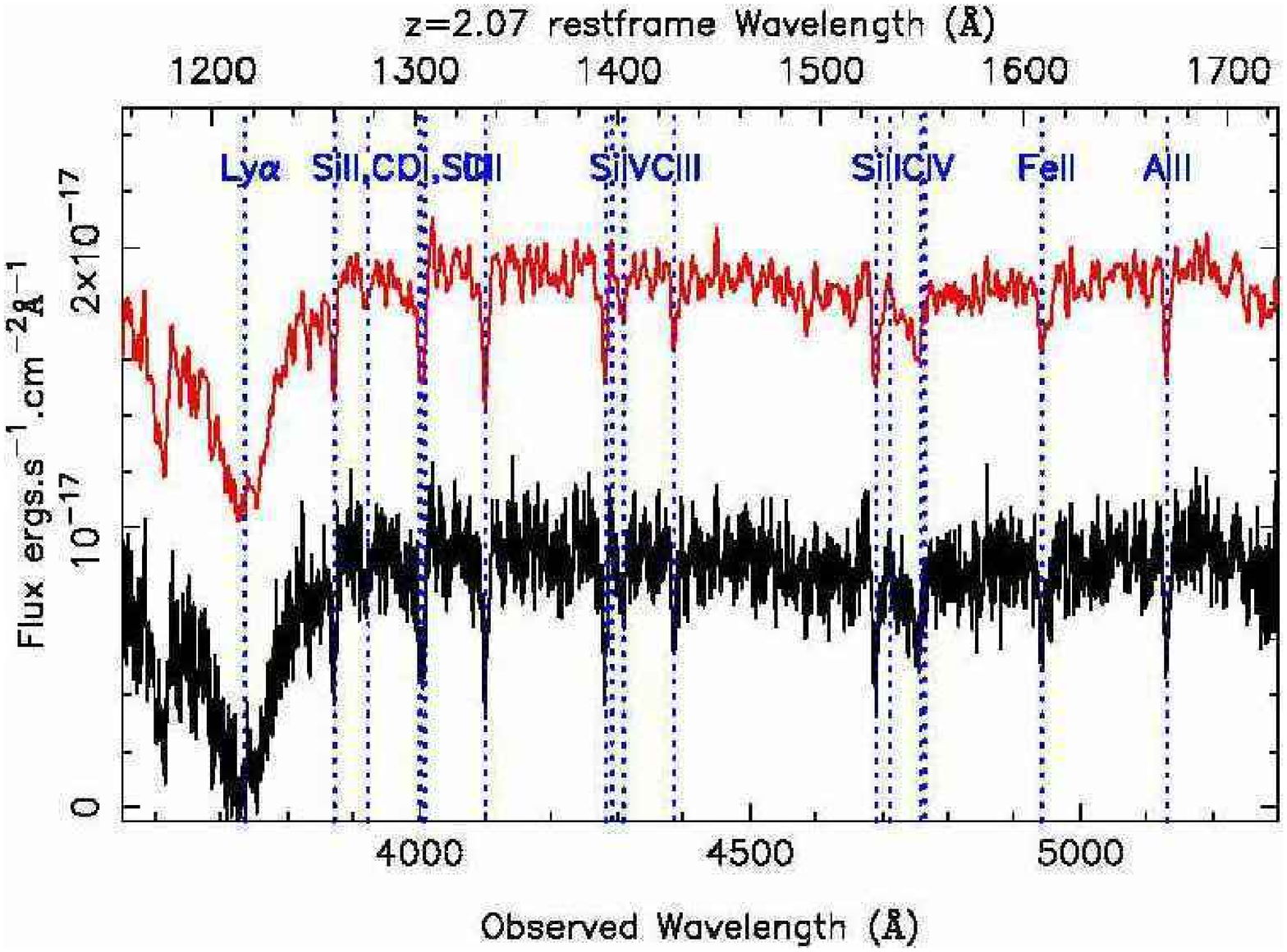}
\end{minipage}
\begin{minipage}{5.5cm}
\centerline{A2204-1.1}
\includegraphics[height=4.5cm,angle=270]{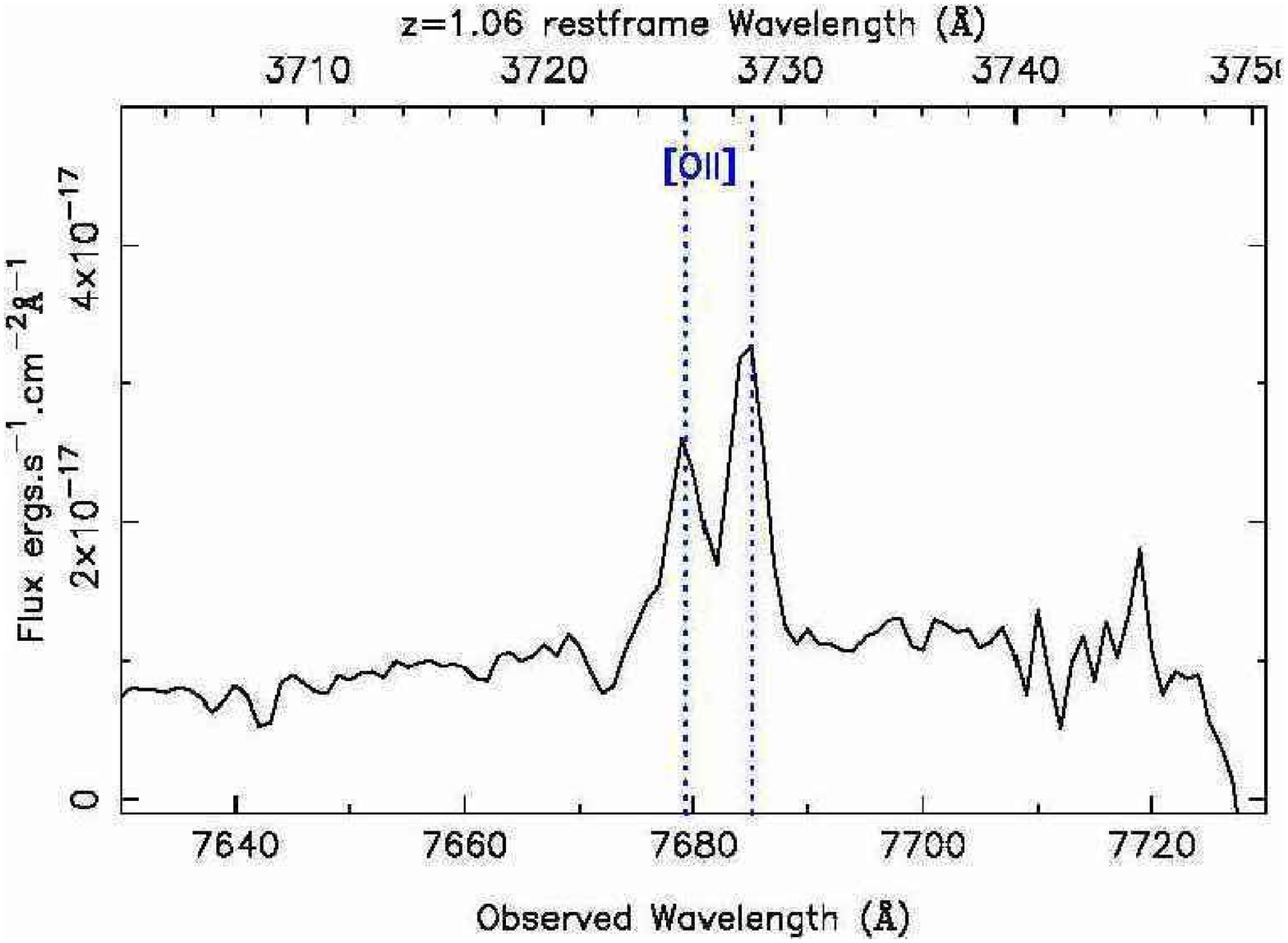}
\end{minipage}
\begin{minipage}{5.5cm}
\centerline{A2204-1.2}
\includegraphics[height=4.5cm,angle=270]{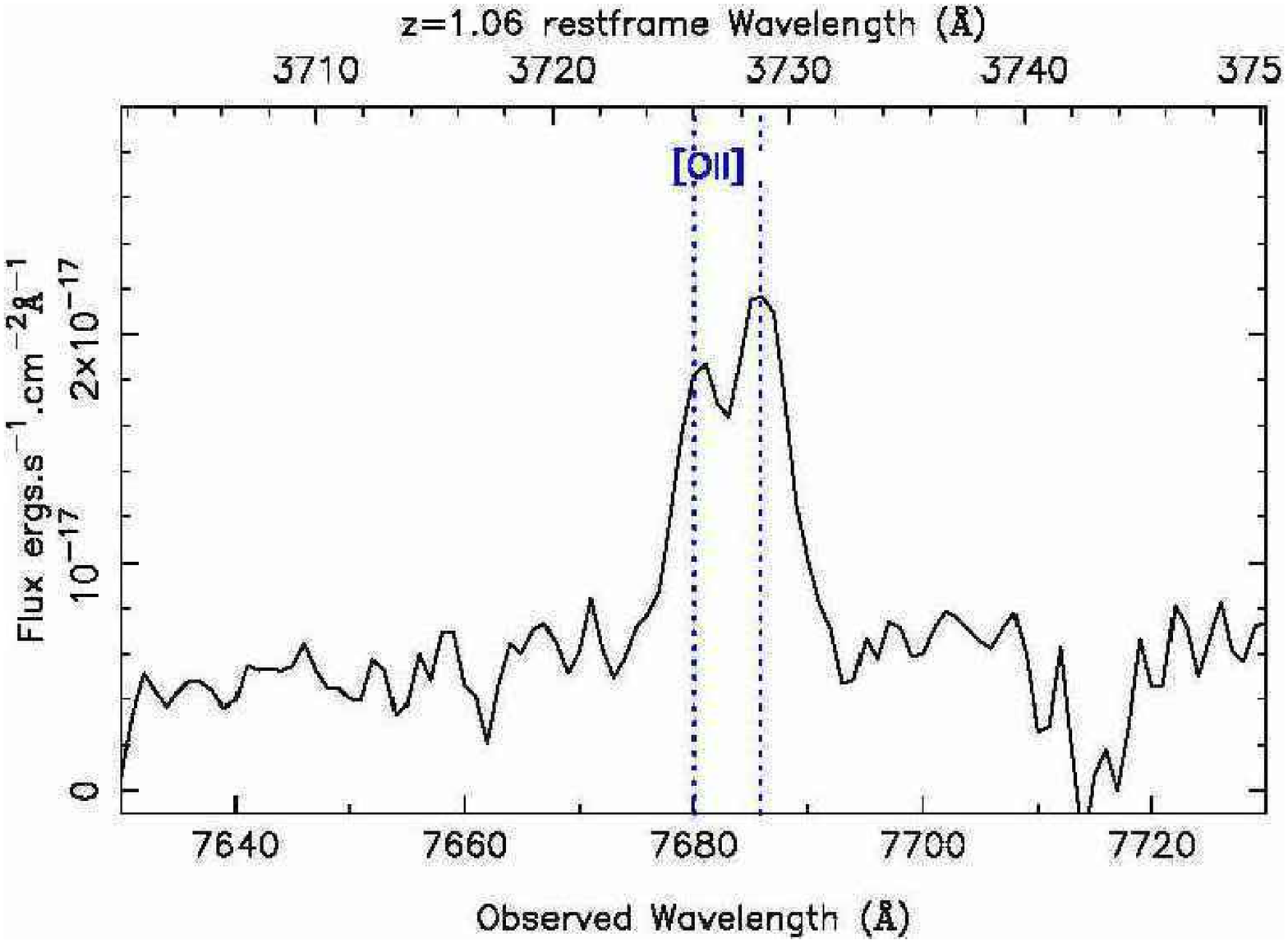}
\end{minipage}
\begin{minipage}{5.5cm}
\centerline{RXJ1720-1.1+1.2}
\includegraphics[height=4.5cm,angle=270]{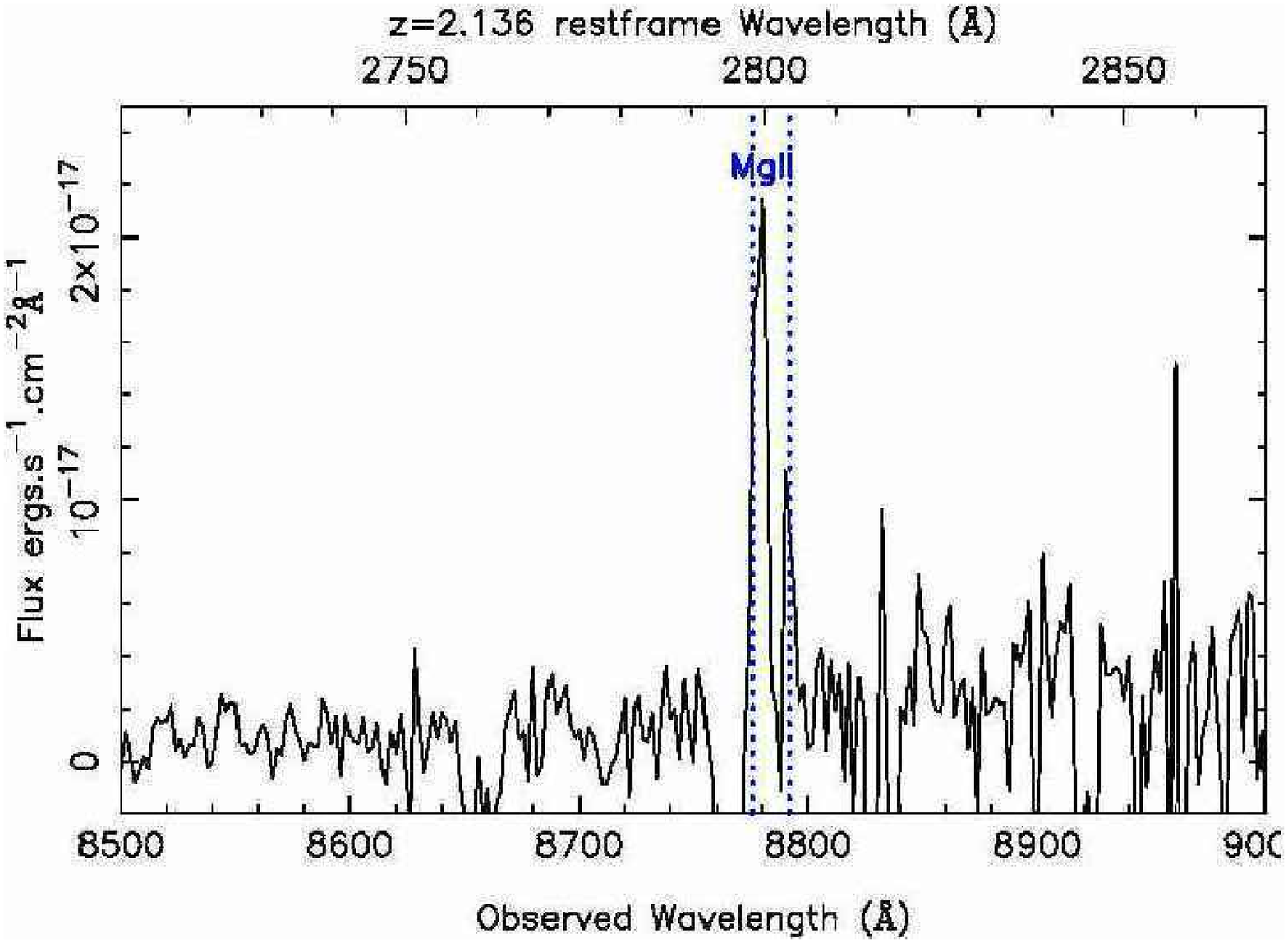}
\end{minipage}
\begin{minipage}{5.5cm}
\centerline{RXJ2129-1.1}
\includegraphics[height=4.5cm,angle=270]{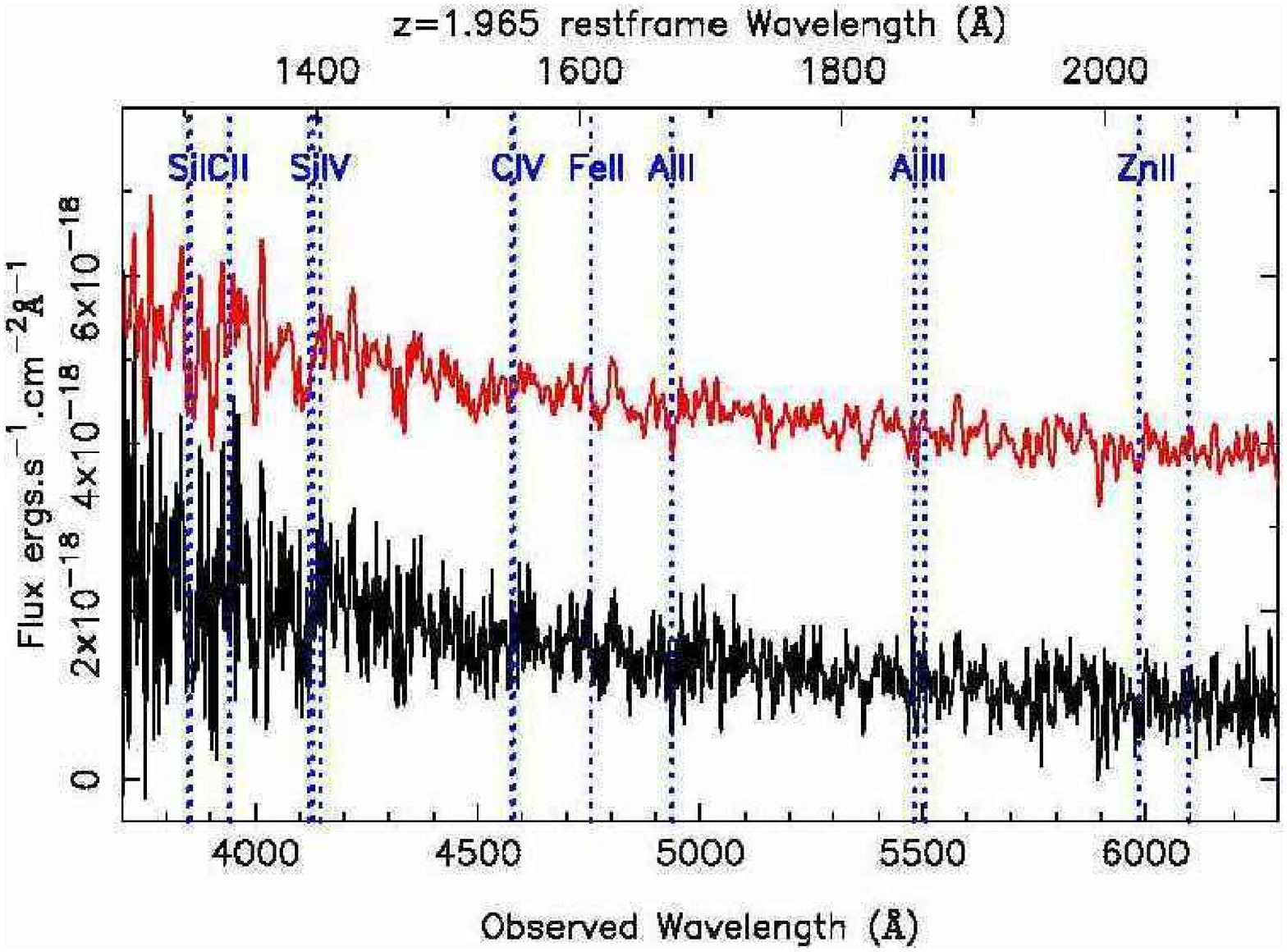}
\end{minipage}
\end{figure*}

\clearpage

\section{Best fit parameters of other mass models}

\begin{table*}
  \caption{\label{best2} Best-fit parameters of the mass models. For
    each mass component, we give the centre, ellipticity, orientation,
    core and cut radii, as well as central velocity dispersion of the
    dPIE profile. The following column gives the image plane RMS of
    this model} 
\begin{tabular}{llrrrrrrrllllllll}
Cluster & Comp. & $\Delta\alpha$ & $\Delta\delta$ & $e$ & $\theta$ & r$_{\rm core}$ & r$_{cut}$ & $\sigma_0$ & rms \\
 & & [\arcsec] & [\arcsec] & & [deg] & [kpc] & [kpc] & [km\ s$^{-1}$] & [\arcsec] \\
\hline 
A383 & DM1 & [-0.3] & [0.5] & 0.15$\pm$0.05 & 123.7$\pm$2.4 & 285.0$\pm$38.9 & [1000.0] & 1976$\pm$132 & 0.22 &   \\
 & BCG & [0.1] & [-0.1] & [0.189] & [96.400] & [0.0] & [40.0] & 117$\pm$40 &  &   \\
 & PERT1 & [14.9] & [-16.7] & [0.125] & [-6.900] & 9.5$\pm$2.8 & 15.3$\pm$8.7 & 412$\pm$110 &  &   \\
 &  L$^*$ gal & & & & & [0.15] & 18.7$\pm$10.1 & 141$\pm$29 & & \\
A963 & DM1 & [0.0] & [0.0] & [0.209] & [85] & 23.2$\pm$2.8 & [1000.0] & 743$\pm$173 & 0.22 &   \\
 & BCG & [0.0] & [0.0] & [0.209] & [85.0] & [0.0] & 47.2$\pm$4.2 & 210$\pm$27 &  &   \\
 &  L$^*$ gal & & & & & [0.15] & [45] & [158] & & \\
A1201 & DM1 & [0.0] & [0.0] & 0.99$\pm$0.26 & 57.2$\pm$7.6 & [75.0] & [1000.0] & 1085$\pm$205 & 0.07 &   \\
 & BCG & [0.0] & [0.0] & [0.705] & [59.8] & [0.0] & 20.2$\pm$55.3 & 250$\pm$44 &  &   \\
 &  L$^*$ gal & & & & & [0.15] & [45] & [158] & & \\
A2218 & DM1 & 3.1$\pm$0.5 & 20.8$\pm$0.2 & 0.04$\pm$0.02 & 38.0$\pm$0.6 & 58.3$\pm$1.0 & 596.2$\pm$4.4 & 697$\pm$1 & 0.12  \\
 & DM2 & -16.9$\pm$0.1 & -21.7$\pm$0.6 & 0.32$\pm$0.01 & 9.2$\pm$0.5 & 119.7$\pm$2.7 & 484.1$\pm$189.4 & 992$\pm$7 &  &   \\
 & BCG & [-0.5] & [0.1] & [0.46] & [52.4] & 5.2$\pm$2.6 & 38.1$\pm$2.81 & 506$\pm$2 &  &   \\
 & PERT1 & [-16.0] & [-10.3] & [0.180] & [80.4] & 1.1$\pm$0.3 & 1.3$\pm$0.3 & 425$\pm$3 &  &   \\
 & PERT2 & [-46.1] & [-49.1] & [0.199] & [59.4] & 1.5$\pm$2.2 & 28.6$\pm$0.5 & 277$\pm$1 &  &   \\
 &  L$^*$ gal & & & & & [0.15] & [45] & [158] & & \\
A2219 & DM1 & [0.1] & [0.2] & 0.65$\pm$0.03 & 32.9$\pm$0.4 & [77.0] & [1000.0] & 854$\pm$19 & 1.13 &   \\
 & DM2 & [-39.2] & [-32.0] & [0.1] & [7.6] & [157] & [1000.0] & 781$\pm$28 &  &   \\
 & DM3 & [-22.9] & [4.5] & [0.0] & [0.0] & 7.9$\pm$1.2 & [1000.0] & 328$\pm$13 &  &   \\
 & BCG & [0.0] & [0.0] & [0.442] & [29.0] & [0.041] & 12.2$\pm$23.1 & 714$\pm$111 &  &   \\
 &  L$^*$ gal & & & & & [0.15] & 2.1$\pm$15.6 & 264$\pm$93 & & \\
A2390 & DM1 & 38.9$\pm$8.2 & 27.4$\pm$0.7 & 0.61$\pm$0.08 & 215.1$\pm$0.7 & 592.3$\pm$15.7 & [2000.0] & 2038$\pm$54 & 0.13 &   \\
 & BCG & [-0.9] & [-1.4] & 0.03$\pm$0.06 & 30.5$\pm$5.5 & 29.9$\pm$0.5 & 294.8$\pm$24.5 & 633$\pm$2 &  &   \\
 & PERT1 & [46.9] & [12.8] & 0.35$\pm$0.09 & 143.7$\pm$4.3 & [0.05] & 41.5$\pm$2.8 & 152$\pm$1 &  &   \\
 &  L$^*$ gal & & & & & [0.15] & [45] & [158] & & \\
A2667 & DM1 & 0.1$\pm$0.9 & -0.5$\pm$0.8 & 0.32$\pm$0.05 & -44.1$\pm$0.5 & 82.5$\pm$5.3 & [1298.629] & 1114$\pm$26 & 0.28 &   \\
 &  L$^*$ gal & & & & & [0.15] & [45] & 109$\pm$13 & & \\
\hline
\end{tabular}

\end{table*}

\end{document}